\documentclass[aps, 10pt, superscriptaddress, twocolumn, floatfix, pra]{revtex4-2}
\usepackage{amsmath,mathtools,amsthm,amssymb}
\usepackage{tabularx}
\usepackage{units}
\usepackage{microtype}
\usepackage{graphicx}
\usepackage{subfigure}
\usepackage{color}
\usepackage{xcolor}
\usepackage{bbold}
\usepackage[linesnumbered,ruled,vlined]{algorithm2e}
\usepackage{listings}
\usepackage[utf8]{inputenc}
\usepackage[T1]{fontenc}
\usepackage{lmodern}
\usepackage{titlesec}
\usepackage{booktabs, makecell, multirow}
\usepackage{enumitem}
\usepackage{setspace}
\usepackage{comment}
\usepackage{array}
\usepackage{graphics}
\usepackage{float}
\usepackage[normalem]{ulem}
\usepackage[toc,page]{appendix}
\usepackage{physics}
\usepackage{braket}
\usepackage{booktabs, caption}
\usepackage[colorlinks=true,linkcolor=orange,citecolor=orange,urlcolor=orange]{hyperref}

\newcolumntype{Y}{>{\raggedright\arraybackslash}X} 
\newcolumntype{C}{>{\centering\arraybackslash}X}   

\newcommand{\expect}[1]{\mathbb{E} \left[#1 \right]}
\newcommand{\supp}[1]{\mathrm{supp} \left(#1 \right)}
\newcommand{\BOO}[1]{\mathcal{O} \left(#1 \right)}
\newcommand{\BOOt}[1]{\widetilde{\mathcal{O}} \left(#1 \right)}

\newcommand{\coloneqq}{:=}
\newcommand{\lind}{{\mathcal{L}}}

\newcommand{\unital}{{\mathcal{U}}}

\newcommand{\ie}{\textit{i.e.}\ }

\newcommand{\prob}[1]{\mathrm{Prob}\left[{#1}\right]}

\newcommand{\fu}{Dahlem Center for Complex Quantum Systems, Freie Universit\"{a}t Berlin, 14195 Berlin, Germany}

\newtheorem{theorem}{Theorem}

\newtheorem{lemma}{Lemma}
\newtheorem{corollary}{Corollary}

\newtheorem{define}{Definition}

\newtheorem{task}{Task}

\newtheorem*{thm*}{Theorem}
\newtheorem*{prop*}{Proposition}
\newtheorem*{lemma*}{Lemma}
\newtheorem*{cor*}{Corollary}
\newtheorem*{cj*}{Conjecture}
\newtheorem*{Def*}{Definition}

\newtheorem{thm}{Theorem}

\newtheorem{rem}[thm]{Remark}

\newtheorem{assumption}{Assumption}

\usepackage{etoolbox}

\makeatletter
\def\thm@space@setup{%
  \thm@preskip=6pt plus 2pt minus 2pt  
  \thm@postskip=6pt plus 2pt minus 2pt 
}
\makeatother
\titlespacing*{\subsection}{10pt}{9pt}{16pt}
\titlespacing*{\subsubsection}{0pt}{9pt}{9pt}
\titlespacing*{\section}{0pt}{16pt}{16pt}

\newcommand{\JRR}[1]{}

\newcommand{\TG}[1]{}

\begin{document}

\title{\bf Stability of digital and analog quantum simulations under noise}

\date{\today}

\author{Jayant Rao}
\email[]{jayant.rao@fu-berlin.de}
\affiliation{\fu}

\author{Jens Eisert}
\email[]{jense@zedat.fu-berlin.de}
\affiliation{\fu}

\author{Tommaso Guaita}
\email[]{tommaso.guaita@fu-berlin.de}
\affiliation{\fu}

\begin{abstract}
Quantum simulation is a central application of near-term quantum devices, pursued in both analog and digital architectures. A key challenge for both paradigms is the effect of imperfections and noise on predictive power. In this work, we present a rigorous and physically transparent comparison of the stability of digital and analog quantum simulators under a variety of perturbative noise models. We provide rigorous worst- and average-case error bounds for noisy quantum simulation of local observables. We find that the two paradigms show comparable scaling in the worst case, while exhibiting different forms of enhanced error cancellation on average. We further analyze Gaussian and Brownian noise processes, deriving concentration bounds that capture typical deviations beyond worst-case guarantees. These results provide a unified framework for quantifying the robustness of noisy quantum simulations and identify regimes where digital methods have intrinsic advantages and when we can see similar behavior. 

\end{abstract}
\maketitle
\tableofcontents
\newpage

Quantum simulation stands out as one of the most compelling applications of quantum technologies—and quite possibly the first to achieve practical utility \cite{CiracZollerSimulation,Fauseweh,QSim,Trotzky}. Among the various approaches, analog quantum simulation has seen especially remarkable progress over the past two decades. In this paradigm, the Hamiltonian of a strongly interacting quantum system is faithfully engineered and controlled in the laboratory. This has become feasible at impressive system sizes, particularly in platforms based on ultra-cold atoms in optical lattices and optical tweezers \cite{BlochSimulation,GreinerSpeedup}, as well as 
in systems involving trapped ions and superconducting circuits \cite{SuperconductingSimulation,Dimitris}. In the dynamical mode of \emph{analog simulation}, the time evolution of the system is monitored in real time. These techniques have enabled the exploration of a vast range of physical phenomena relevant to strongly correlated quantum matter in condensed matter and materials science
\cite{MBL2D,Dimitris,MonroeTimeCrystal,NearTermMaterialsSimulation}.
In contrast, \emph{digital quantum simulation} takes a gate-based approach: the Hamiltonian evolution is discretized into stroboscopic time steps and implemented via quantum circuits—much like in a universal quantum computer \cite{QSim,Lloyd}. 

Despite the enormous advances, it is crucial to remember that quantum simulation is only as useful as its predictive power. The primary challenge—shared by both analog and digital approaches—is quantum noise and incomplete knowledge of the system. If small perturbations or imperfections lead to drastically different outcomes, then even a sophisticated quantum simulation may become little more than a physical curiosity, offering no real advantage over classical methods.
To address this, techniques such as Hamiltonian and Liouvillian learning have been developed to improve model accuracy and compensate for imperfections
\cite{TangHamiltonianLearning2,AnshuSampleEfficientHamiltonianLearning,cole_identifying_2005,schirmer_two-qubit_2009,hangleiterRobustlyLearningHamiltonian2024a}. Yet, it is clear that this is not enough to address all types of experimental errors and noise that can appear in practical quantum simulators, which would ultimately require quantum error correction. Evidently, the robustness of quantum simulation -- whether digital or analog -- is the critical issue that will determine its possibility of success both in the short and the medium term.

 Indeed, recently understanding the stability of analog quantum simulators has become a new question of research. Several works have in particular investigated the perturbative regime \cite{Kashyap_2025,AnalogStability}. This has also been expanded to stability of long ranged systems, in particular for Gibbs states \cite{möbus2025stabilitythermalequilibriumlongrange}. The question of how stable quantum simulation with Trotterized unitaries is with regards to noise and errors has also received some recent attention 
 (see
Refs.~\cite{Knee_2015,xu2025exponentiallydecayingquantumsimulation}). This body of work has been complemented by a refined understanding of the impact of errors on quantum circuits in the absence of quantum error correction \cite{FG20,Nonunital,NonUnitalSampling}.

Against this backdrop, a central and still unresolved question emerges:
Which approach -- digital or analog -- is more robust to imperfections?
How can we make a fair and meaningful comparison between the two? One might expect analog simulations to be inherently more resilient, as certain errors may partially cancel out over continuous evolution -- unlike in digital simulations, where gate errors tend to accumulate. Indeed, some recent evidence seems to support this intuition~\cite{preskillstochasticcancellation}.

In this work, we take up this fundamental question to offer a comprehensive and mathematically rigorous, but at the same time physically grounded answer. Concretely, we consider a model of quantum simulation where a local lattice Hamiltonian $H$ is given and the task is to estimate the expectation value of a local observable $O$ on a time-evolved state under this Hamiltonian. We address the case where this is done in \emph{analog mode}, by directly implementing the Hamiltonian, and the case where it is done in \emph{digital mode}, by decomposing the time evolution in to a circuit of local gates using a suitable Suzuki-Trotter product formula~\cite{Suz91}. In both cases we assume the presence of local perturbations of magnitude $\delta$ in the system and analyse the corresponding robustness of the final outcome of the simulation. In the analog case, these perturbations are modeled as deviations of norm $\delta$ in the local terms of the implemented system's Hamiltonian compared the exact Hamiltonian which we want to simulate. In the digital case, we consider two different models of unitary perturbations which appear at the level of the circuit gates.
For all these different scenarios, we systematically analyze the error that the perturbations induce on the expectation value of a local observable. We provide upper bounds on the magnitude of this error and compare them to each other and to similar results obtained in the literature specifically for analog systems~\cite{AnalogStability}.

While these worst-case bounds provide rigorous guarantees, in many cases they do not fully capture the more intricate structure of the problem. In particular, the experimental perturbations that we are considering most likely entail some form of randomness, which commonly implies concentration effects.
Indeed, several recent analyses have highlighted the important role of error cancellation in analog quantum simulators, which makes them more stable to and unaffected by errors than the worst case bounds would suggest~\cite{preskillstochasticcancellation,Poggi_2020}. 
In digital simulation, it has similarly been observed that Trotter discretization errors might scale far better than worst case bounds would suggest~\cite{Chen_2024,PhysRevLett.129.270502}. 
For these reasons, 
on top of the worst case analysis, we also consider the stochastic behaviour of errors under different realizations of the random perturbations. We provide in particular results on the concentration of these random errors around their typical values, which are in some cases significantly better than the worst case ones. Our work confirms previous findings for analog simulation, extending the results to a much wider class of random perturbations.
We further complement these results by showing that error cancellation effects also apply in digital quantum simulators under several models of circuit-level errors.

The work is structured as follows. In Section~\ref{sec:prelim-notation} we establish the required conventions and notations, introducing the simulation tasks we consider, the analog and digital methods of solving them and the error models which we assume they may be subject to. In the following two sections we introduce the main results of this work, first the ones concerning worst-case errors in Section~\ref{sec:worst-case} and then the ones concerning average errors under stochastic noise models in Section~\ref{sec:average-case}. In both cases, we discuss the results that apply specifically to the digital and the analog simulation modes separately. For each we present the formal statements of our findings and discuss their implication, while we mostly postpone their rigorous mathematical proofs to the appendices.

\begin{figure*}[t]
    \centering
    \includegraphics[width=.75\linewidth]{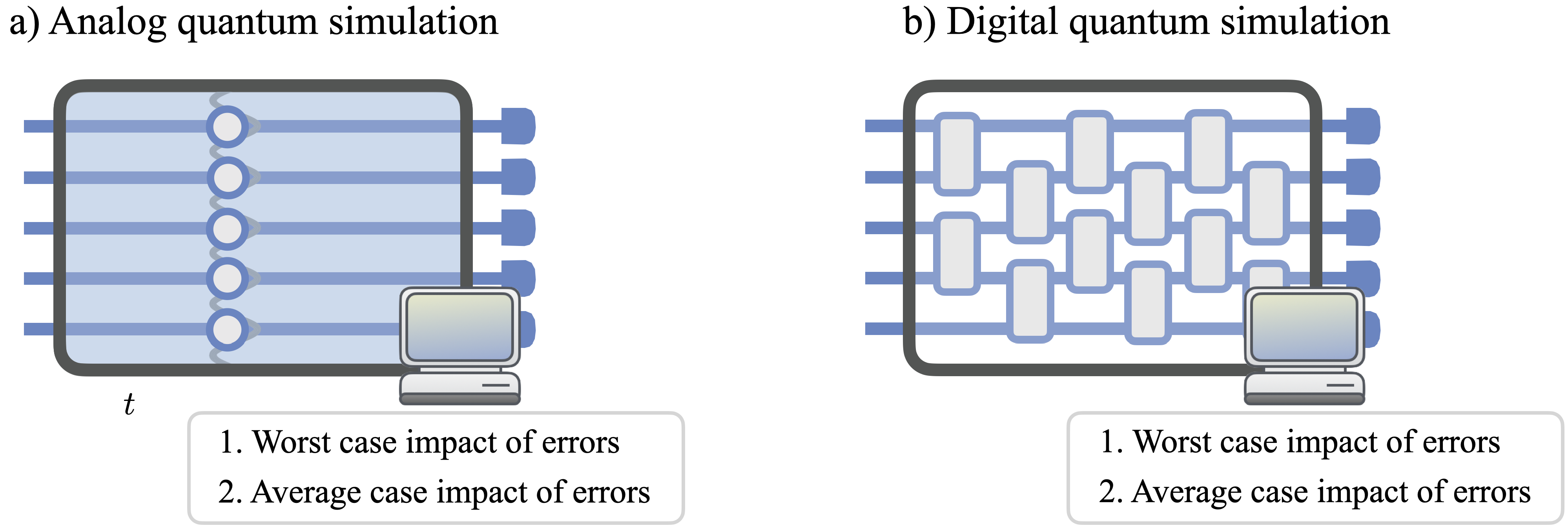}
    \caption{In this work, (a) analog quantum simulation 
    provided by precisely controlled quantum systems naturally evolving in time $t>0$
    is 
    comprehensively and rigorously compared 
    to digital quantum simulation run on non-quantum error corrected quantum circuits with respect to the 1. worst case and 2. average case impact of natural errors.}
\label{figure}
    \end{figure*}

\section{Preliminaries and notation} \label{sec:prelim-notation}

\subsection{Analog and digital quantum simulation}
We consider a $d$-dimensional hypercubic lattice $\mathbb{Z}^d$. To each lattice site is associated a local system, which we assume for simplicity to be a qubit. We are interested in the simulation of time evolution under local Hamiltonians defined on this lattice system. In particular, we consider Hamiltonians that can be expressed as a sum of local terms, that is, of the form 
\begin{align}
    H = \sum_{\gamma\in\Gamma} H_\gamma,
\end{align}
where the operators $H_\gamma$ satisfy the following assumptions on the geometric locality of their support. We assume that there exists a global constant $R$ and that each term $H_\gamma$ can be associated to a lattice site $x$ such that
\begin{equation}
    \supp{H_\gamma}\subseteq B_R(x)
\end{equation}
where $B_R(x)$ is the ball centered on $x$ and of radius $R$ in the $l_1$ metric on the lattice.

The specific task that we focus on is estimating the expectation values of geometrically local observables, that is, observables whose support is also contained in a region of constant diameter of the lattice.

\begin{task}[Dynamical quantum simulation]
    Given a local observable $O$, a local Hamiltonian $H$, a time $t$ and some initial state $\rho$, compute the expectation value of the observable after time evolution
    \begin{align}
        \braket{O(t)} = \tr{U^\dag (t) O U(t) \rho}. \label{eq:O-task1}
    \end{align}
    Here $U(t)=e^{-itH}$ is the time evolution unitary.
\end{task}

This task can be approached on quantum devices in two conceptually different ways, which we refer to as \emph{analog} and \emph{digital} simulation. In the case of analog simulation, we assume that the Hamiltonian is implemented natively on the simulator device, i.e., that there exists a way of encoding the Hamiltonian $H$ such that the unitary evolution $U(t) = e^{- iH t}$ is the time evolution operator of the device.

In the case of digital simulation, instead we first decompose the time evolution unitary $U(t)$ at time $t$ into an approximate circuit representation composed of a product of local unitary gates, i.e., gates that only act on a local patch of the lattice of radius $R$. We then implement this circuit on a digital quantum computing device. Here the crucial step is the choice of decomposition which provides the discretization of the time evolution into a circuit form. In our analysis we consider a specific class of discretizations based on the Suzuki-Trotter formulas. This is one of the most commonly used approaches and contains a large class of methods, including product formulas of arbitrary even order $p$. In general, a product unitary of this class takes the following form.

\begin{define}[Product unitary]
    \label{trotter1}
    Given a Hamiltonian $H = \sum_{\gamma\in\Gamma} H_\gamma$,  a corresponding $p$-th order Suzuki-Trotter product unitary with Trotter number $n$ is of the form
    \begin{align}
        U^{(p)}_n (t) = \prod_{j = 1}^n \prod_{v=1}^{\Upsilon_p} \prod_{\gamma\in\Gamma}  e^{-i\frac{t}{n}a_{v,\gamma} H_{\pi_\upsilon (\gamma)}}, \label{eq:product-unitary-def}
    \end{align}
    where $a_{v, \gamma}$ are constants associated to higher order product formulas. The index $\upsilon$ runs over the stages of the given formula (whose number $\Upsilon_p$ depends on $p$). The permutation $\pi_\upsilon (\gamma)$ is chosen for every $\upsilon$.
\end{define}
The main parameter which appears in these product unitaries is the Trotter number $n$. Generally a larger Trotter number leads to a better approximation of the exact time evolution unitary $U(t)$ at time $t$. In Appendix~\ref{app:digital-worst-case} we provide a more detailed review of how these product formulas are constructed and of how well they approximate the exact time evolution as a function of $n$ and $t$.

So far we have considered systems and Hamiltonians in the thermodynamic limit, \ie defined on lattices of infinite size. Of course, when simulating them on a physical system, they will necessarily need to be truncated to a finite size for the implementation to be possible. We will, therefore, always consider analog and digital simulations which are actually run on a system truncated to a finite distance $l$ from the support of the local observable $O$. The idea is that, for the systems that we consider, taking a large but finite $l$ is enough to obtain a sufficiently good approximation of the full evolution of $O(t)$, due to the existence of a Lieb-Robinson light cone in the system's dynamics (see Appendix~\ref{app:notation-preliminaries} for a more detailed review of the corresponding results). The truncation to distance $l$ is performed more concretely as follows.

Given an observable $O$ and a truncation length $l>0$, we consider the subregion of the lattice
\begin{align}
    \Omega_l = \{y \; | \; d(x,y) \leq l, \forall x \in \supp{O}\}
\end{align}
 within a distance $l$ of the initial support of $O$,
where $d(\cdot,\cdot)$ is again the natural $l_1$ distance on the lattice (intuitively, the number of steps one needs to go from one site to the other). The corresponding truncated Hamiltonian is then the one where only the local terms are retained that have a non-trivial support on $\Omega_l$.

\begin{define}[Truncated Hamiltonian]
    The truncated Hamiltonian associated with $l > 0$ and a corresponding local observables $O$ is
    \begin{align}
        H_l = \sum_{\gamma \in \Theta_l} H_\gamma
    \end{align}
    where $\Theta_l=\{\gamma\in\Gamma \; | \; \supp{H_\gamma}\cap\Omega_l\neq\emptyset\}$.
\end{define}

Once a truncation length has been specified, then we assume that the analog and digital implementation of the simulation will take place on the correspondingly 
reduced system and taking into account the truncated Hamiltonian. In particular we assume that the analog simulator will implement the truncated evolution
\begin{align}
    U_l(t)=e^{-itH_l}\,.
\end{align}
For what concerns the digital simulator, we assume it will implement a product formula, which we denote by $U_{l,n}^{(p)}(t)$, which has the same form as~\ref{eq:product-unitary-def} but where the product now runs over $\gamma\in\Theta_l$. 

\subsection{Meaningful error models}\label{subsec:error-models}

In the discussion above, we have introduced the ideal notion of analog and digital quantum simulators. However, real-world implementation of these protocols will necessarily be affected by a certain amount of experimental imperfections. Here, we will now discuss some ways in which these perturbations can be modeled and parametrized. This is a key step to then introduce the concept of stability under perturbations. The noise models which we will analyze represent a sufficiently large range of practical scenarios, while remaining sufficiently simple to be able to establish rigorous mathematical proofs for our novel results. It is nonetheless worth noting, that several directions exists to consider more general and exhaustive models: this represents an open and challenging question for future research.

In the setting of analog simulation, the main error model that we consider is the one where the Hamiltonian $H'$ that is implemented in the physical simulator system is not exactly the one that should be simulated but a slightly perturbed one. These perturbations could in principle also be time-dependent. Then, more specifically, we assume that the time-dependent Hamiltonian $H'(t)=\sum_\gamma H_\gamma'(t)$ is implemented, where the terms $H_\gamma'(t)$ have exactly the same local support as the ideal ones $H_\gamma$, but can be perturbed by up to a distance $\delta$ in operator norm, \ie, for all $t$,
\begin{align}
    \norm{H_\gamma - H_\gamma'(t)} \leq \delta.
\end{align}
Note that here and in what follows $\norm{\,\cdot\,}$ denotes the operator norm, unless otherwise specified, so the largest singular value. We denote the imperfect evolution implemented with this perturbed Hamiltonian as $V(t)$.

In order to discuss the behaviour of average errors in analog quantum simulation, we would like to introduce a more specific error model, where the perturbations are explicitly drawn from a well-defined random ensemble. For this, we will assume that the perturbed evolution is explicit defined by 
\begin{align}
    \frac{d}{dt} V(t) = -i\sum_\gamma\left(H_\gamma+\delta L_\gamma(t)\right)\, V(t), \label{eq:analog-stoc-errors-def}
\end{align}
where the possibly time-dependent operators $t\mapsto L_\gamma(t)$ 
are drawn from an ensemble of random Hermitian matrix processes independently for each $\gamma$. We will later specify further the precise processes that we consider, however 
we will always assume that the perturbation $L_\gamma(t)$ 
at time $t$ has support on the same region of the lattice as the Hamiltonian term $H_\gamma$ which it perturbs and that it has mean zero.

In the setting of digital simulation, we consider a model where experimental imperfection perturb the implementation of each individual gate. In particular, we consider unitary errors: we assume that the perturbed gates remain unitary, although slightly different from the ideal unitary we intend to implement. Here, there are different approaches with which one can parametrize the magnitude of this perturbation. 

In the simplest case, we can assume that the implemented unitary deviated from the ideal one by up to a distance $\delta$ in operator norm. That is, we assume that each gate $U_{j,\upsilon,\gamma}=e^{-i\frac{t}{n}a_{v,\gamma} H_{\pi_\upsilon (\gamma)}}$, appearing at the step labeled by $j,\upsilon,\gamma$ of the ideal product unitary~\eqref{eq:product-unitary-def}, is replaced in the practical implementation by a unitary gate $V_{j,\upsilon,\gamma}$ such that 
\begin{equation}
    \norm{V_{j,\upsilon,\gamma}-U_{j,\upsilon,\gamma}}\leq \delta\,.
\end{equation}
Another model that one may consider, is one where the perturbation of magnitude $\delta$ occurs at the level of the Hamiltonian generating the gate $U_{j,\upsilon,\gamma}$. As the gate also depends on a rotation angle that scales as $\frac{t}{n}$, the total error on the unitary gate will in this case be parametrized as
\begin{equation}
    \norm{V_{j,\upsilon,\gamma}-U_{j,\upsilon,\gamma}}\leq \delta\frac{t}{n}\,.
\end{equation}
The latter error models represents a case where the gate error depends linearly on the rotation angle of the corresponding gate, while the former represents a case where the error magnitude is totally independent of the considered gate. Clearly, these two models represents two extreme cases of the different possible gate-dependencies of the error models present in practical scenarios. We will derive and discuss our results for both these choices, indicating as $V^{(p)}_{l,n}(t)$ the perturbed product formula, that is the product of the perturbed gates $V_{j,\upsilon,\gamma}$.

Finally, in order to discuss the behaviour of average errors in digital quantum simulation, we would like to introduce a more specific error model, where the perturbations are explicitly drawn from a well-defined random ensemble. For this, we will assume that the perturbed version of the product unitary~\eqref{eq:product-unitary-def} takes the explicit form
\begin{align}
    V^{(p)}_{l,n}(t) = \prod_{j=1}^n \prod_{\upsilon=1}^{\Upsilon_p} \prod_{\gamma\in\Theta_l} e^{-i\frac{t}{n} \left( a_{v,\gamma} H_{\pi_\upsilon (\gamma)} -i \delta L_{\gamma, \upsilon, j}\right)}\label{eq:noisy-product-unitary-def}
\end{align}
in the first error model and, for the second error model,
\begin{align}
    V^{(p)}_{l,n}(t) = \prod_{j=1}^n \prod_{\upsilon=1}^{\Upsilon_p} \prod_{\gamma\in\Theta_l} e^{-i\frac{t}{n} a_{v,\gamma} H_{\pi_\upsilon (\gamma)}+ \delta L_{\gamma, \upsilon, j}}, \label{eq:noisy-product-unitary-def2}
\end{align}
where the operators $L_{\gamma, \upsilon, j}$ have support on the same region of the lattice as the Hamiltonian terms $H_{\pi_\upsilon (\gamma)}$ which they perturb and are independently drawn from an ensemble of Hermitian matrices. The only assumptions we make on this ensemble are that $\expect{L_{\gamma, \upsilon, j}}=0$ and $\norm{L_{\gamma, \upsilon, j}}\leq 1$.

\subsection{Stability}
In this work, we discuss stability as a notion of perturbative robustness. In particular, we are interested in determining bounds on the final error committed on the expectation value~\eqref{eq:O-task1} and analyzing how the behave as a function of the strength $\delta$ of the perturbations present in the simulator device, as defined in the previous section on error models. 
To be more specific, we define 
\begin{align}
    \Delta(\rho) &\coloneqq
    \abs{\Tr{O(t) \rho} - \Tr{O'(t) \rho}}
\end{align}
as the deviation of the observed expectation value $\braket{O(t)}$ from its exact value, for a given initial state $\rho$. 
Here, $O(t)=U^\dag(t)OU(t)$ is the evolved observable that we ideally want to measure, while $O'(t)$ represents instead the perturbed dynamics that is actually implemented in the physical simulator. So the precise definition of $O'(t)$ depends on the analog or digital context that we are considering. In the analog case $O(t)=V^\dag(t)OV(t)$, while in the digital case $O(t)={V^{(p)}_{l,n}}^\dag(t) \, O\, V^{(p)}_{l,n}(t)$, where $V(t)$ and $V^{(p)}_{l,n}(t)$ are the perturbed implementations  at time $t$ defined in the previous section.
As we often want to avoid a dependence on the initial state $\rho$, we shall also consider the maximal deviation over all possible initial state, that is
\begin{align}
\nonumber
    \Delta &\coloneqq \sup_{\rho} \abs{\Tr{O(t) \rho} - \Tr{O'(t) \rho}} \\
    &= \norm{O(t) - O'(t)}\,, \label{eq:def-Delta}
\end{align}
which clearly provides an upper bound to the state-dependent one.

In what follows we will determine how severely the error $\Delta$ is affected by the magnitude of the perturbations $\delta$ and compare this scaling among the various modes (digital and analog) and error models considered. We will, therefore, prove bounds of the form
\begin{align}
    \Delta \leq h(\delta,t), \label{eq:stability-bound-Delta}
\end{align}
for suitable functions $h$ of $\delta$ and $t$, where we are specifically interested in the asymptotic scalings for small errors $\delta$ and large times $t$. In the analysis of worst case errors we will be interested in bounds of the form~\eqref{eq:stability-bound-Delta} which hold for any possible perturbation of magnitude $\delta$ within the model considered. In the case of stochastic perturbation models, we are instead interested in the typical behaviour of $\Delta$, that is ranges of values in which $\Delta$ is guaranteed to lie \emph{with high probability} over the considered ensemble of random perturbations.
In the following sections, we will present our stability results, first in the setting of worst case errors and then for stochastic perturbation models.

\subsection{Comparison to Previous Work}
In general, some of the questions introduced here have been tackled previously, especially in the setting of analog quantum simulation for  specific unitary and dissipative dynamics and error models \cite{AnalogStability, Kashyap_2025}. Furthermore, a first average case result was discussed in Ref.~\cite{preskillstochasticcancellation}. Our work summarizes these analog case discussions and extends them to more general time-dependent noise models. This will reveal, especially in the average case setting, different error scalings depending on the time correlations of the stochastic noise.

More importantly, our work also provides a comprehensive comparison to the digital simulation model. While this topic has also been addressed in Ref.~\cite{trivedi2025noiserobustnessproblemtosimulatormappings} which appeared shortly before our article, we focus on a different family of discretization strategies (namely Suzuki-Trotter product formulas of arbitrary order) and, especially, we establish the first results for the average case treatment of digital noise models. Additionally, we will provide an analysis of the optimal choice of Trotter number $n$ and system size and its relation to the presence of noise.

\renewcommand{\arraystretch}{1.8}

\begin{table*}[t]
\centering
\renewcommand{\arraystretch}{1.35}
\setlength{\tabcolsep}{8pt}
\begin{tabular}{@{} l l l l @{}}
\toprule
\multirow{2}{*}{\textbf{Error measure}}
  & \multirow{2}{*}{\textbf{Analog simulator}}
  & \multicolumn{2}{c}{\textbf{Digital simulator}} \\
\cmidrule(lr){3-4}
  &  & \textbf{Model M1} & \textbf{Model M2} \\
\midrule
\makecell[tl]{\textbf{Worst case $\Delta$}\\[2pt]
  {\scriptsize Thms.~\ref{thm:wcas},~\ref{wcds},~\ref{wccds}}}
  & $\mathcal{O}\!\left(\delta\, t^{d+1}\right)$
  & $\BOOt{\delta\, t^{d+1}}$
  & $\BOOt{\delta^{\tfrac{p}{p+1}}\, t^{d+1}}$ \\
\midrule
\makecell[tl]{\textbf{Average case $\Delta(\rho)$}\\[2pt]
  {\scriptsize Thms.~\ref{acac},~\ref{ito-acas}, \ref{fixed-states-ads},~\ref{fixed-states-2}}}
  & \makecell[tl]{%
      Time-independent: $\BOOt{\delta\, t^{\tfrac{d}{2}+1}}$ \\
      Finite $\lambda$: $\BOOt{\sqrt{\lambda}\,\delta\, t^{\tfrac{d+1}{2}}}$ \\
      White noise: $\BOOt{\delta\, t^{\tfrac{d+1}{2}}}$}
  & $\BOOt{\dfrac{\delta\, t^{\tfrac{d}{2}+1}}{\sqrt{n}}}$
  & $\BOOt{\delta^{\tfrac{2p}{2p+1}}\, t^{\tfrac{2(d+1)}{3}}}$ \\
\midrule
\makecell[tl]{\textbf{Average case $\Delta$}\\[2pt]
  {\scriptsize Thms.~\ref{acas},~\ref{acdc},~\ref{accds}}}
  & $\mathcal{O}\!\left(\delta\, t^{d+1}\right)$
  & $\BOOt{\dfrac{\delta\, t^{d+1}}{\sqrt{n}}}$
  & $\BOOt{\delta^{\tfrac{2p}{2p+1}}\, t^{d+\tfrac{2}{3}}}$ \\
\midrule
\makecell[tl]{\textbf{Lindblad-type $\Delta(\rho)$}\\[2pt]
  {\scriptsize Thms.~\ref{pertlind}, ~\ref{ito-acdc}}}
  & $\BOOt{\delta\, t^{\tfrac{d+1}{2}}}$
  & \multicolumn{2}{c}{$\BOOt{\delta\, t^{\tfrac{d+1}{2}}}$} \\
\bottomrule
\end{tabular}
\caption{Summary of main results. \textbf{M1} is the error model of Eq.~\eqref{eq:noisy-product-unitary-def}; \textbf{M2} is the error model of Eq.~\eqref{eq:noisy-product-unitary-def2}. The notation $\BOOt{\,\cdot\,}$ hides polylogarithmic factors, i.e.\ $\BOOt{f}=\BOO{f \cdot \mathrm{polylog}(\cdot)}$.}
\label{tab:main-results}
\end{table*}
\section{Results for worst case errors} \label{sec:worst-case}
In this section, we want to derive upper bounds on the error
\begin{align}
    \Delta \coloneqq \sup_{\rho} \abs{\Tr{O(t) \rho} - \Tr{O'(t) \rho}},
\end{align}
which hold for arbitrary perturbations of magnitude $\delta$ within the error models that we 
have introduced in the previous section. We will first discuss the error models for analog quantum simulation and then the ones for digital quantum simulation. We will then conclude the section with a comparison of the results in the two cases.

\subsection{Analog quantum simulation}
For errors in analog quantum simulators, we consider the model  where the local Hamiltonian terms $H_\gamma$ are replaced by perturbed terms $H_\gamma'$ with $\norm{H_\gamma-H_\gamma'}\leq \delta$. For this model it is straightforward to observe that $\Delta$ will scale at most as $\BOO{\delta t^{d+1}}$. This bound has been derived in references \cite{AnalogStability,Harley_2024}, where it was in particular discussed how the system's local nature implies that the bound does not depend explicitly on the system size. We present here a version of this result, derived with the notation and framework of our present work, which in particular also includes time-dependent perturbations.

\begin{theorem}[Upper bound for worst case errors in analog simulators]
\label{thm:wcas}
    Consider a perturbed analog time evolution $V(t)$ defined by the time-dependent local Hamiltonian 
    \begin{align}
        H'(s)=\sum_\gamma H_\gamma'(s),
    \end{align}
    where $\norm{H'_\gamma(s) - H_\gamma} \leq \delta$ for all $s<t$ and all $\gamma\in\Gamma$.
    Then, the error on the time-evolution of a local observable $O$ is at most
    \begin{align}
        \Delta \leq \BOO{t^{d+1} \delta}.
    \end{align}
\end{theorem}
The full proof of the theorem is presented in Appendix~\ref{app:analog-worst-case}. 
The main ingredient of the proof is Duhamel's formula
    \begin{align}
        e^{iHt} - e^{iH't} = \int_0^t ds e^{i(t-s)H} (H-H') e^{isH'}\,,
    \end{align}
which we use to relate the error on the Hamiltonian to the one on the time evolution. We further exploit the Lieb-Robison light cone of the system's dynamics to observe that the dominant contribution to the final error is given by the perturbations occurring inside such light cone. 

Note that we have formally stated here the result for evolutions on the infinite lattice. However, it is straightforward to see from the proof technique that the same result applies to any analog simulation implemented on a system truncated at any length $l > vt - \frac{1}{\mu} \log(\delta t^{d+1})$, where $v$, $\mu$ are suitable Lieb-Robinson constants of the system (see Remark~\ref{rem:wcas} in the Appendix for more details).

\subsection{Digital
quantum simulation}
For digital quantum simulation with Suzuki-Trotter formulas, we consider stability under two different error models introduced in Section~\ref{subsec:error-models}. In both cases, each gate $U_{j,\upsilon,\gamma}$ in the product unitary is replaced by a perturbed gate $V_{j,\upsilon,\gamma}$. This structure leads to three distinct contributions to the final error $\Delta$ on the observable expectation value. First, we have the error stemming from the fact that we simulate the evolution of the infinite lattice Hamiltonian on a finite system truncated to length $l$. Then we have a discretization error given by representing this evolution by a product formula with Trotter number $n$. Finally, we have the perturbation error coming from implementing the perturbed gates $V_{j,\upsilon,\gamma}$ instead of the ideal ones $U_{j,\upsilon,\gamma}$. 

Each of these terms depends on the choice of parameters $l$ and $n$ in the product unitary and trade-offs between the various contributions evidently play a role. For instance, increasing the total number of gates in the product unitary may reduce the discretization error but will at the same time increase the error contribution coming from faulty gate implementations. The main observation of this analysis is thus that an optimal scaling of the total error can be achieved only by a careful choice of the implementation parameters $l$ and $n$. The ideal choice is the one that achieves a balance in the trade-offs between the error terms, giving the same scaling with respect to $t$ and $\delta$ in all terms. 

The precise nature of the trade-offs and the ideal scaling that can be achieved by balancing them depends on the choice of model with which we describe the perturbations in the system. In the first model, the strength of the perturbation is parametrized by assuming that noisy gate $V_{j,\upsilon,\gamma}$ is at most within distance $\delta\frac{t}{n}$ of the ideal gate $U_{j,\upsilon,\gamma}$. Here, we observe that the optimal scaling of $\Delta$ in digital quantum simulators reproduces the same result that we derived for analog quantum simulation, up to logarithmic factors. 

\begin{theorem}[Upper bound for worst case errors in digital simulators with gate-dependent perturbations]
\label{wcds}
    Consider a perturbed Suzuki-Trotter product unitary of even order $p=2k$, which takes the 
    form
    \begin{align}
       V^{(p)}_{l,n} (t)  = \prod_{j=1}^n \prod_{\upsilon=1}^\Upsilon \prod_{\gamma\in\Theta_l} V_{\gamma, j, \upsilon}\,,
    \end{align}
    where each local gate is a perturbed version of the exact gate, satisfying $ \norm{ V_{\gamma, j, \upsilon} - e^{-i\frac{t}{n}a_{\upsilon,j} H_\gamma}} \leq \delta\frac{t}
    {n}$. The product unitary has Trotter number $n$ and is implemented on a system of size $l$. 
    Then, the error on the time-evolution of a local observable $O$ is at most
\begin{align}
        \Delta \leq \BOO{\delta t^{d+1} \log^d(\frac{1}{\delta t^{d+1}})},
    \end{align}
    if the \textit{optimal choices} of $n_{\rm opt} \geq t/\delta^{\frac{1}{p}}$ and $l_{\rm opt}\geq vt-\frac{1}{\mu}\log(\delta t^{d+1})$ are made, for suitable constants $v$, $\mu$.
\end{theorem}

In the second model, the actual strength of the perturbation is parametrized by assuming a distance between the perturbed and exact gate of up to a constant value $\delta$. Here, we observe a similar scaling of $\Delta$ with respect to $t$, but a slightly worse scaling with respect to $\delta$ compared to the bounds in Theorem \ref{thm:wcas} and Theorem \ref{wcds}. For large $p$, we, however, see that 
this difference in the scaling vanishes.
\begin{theorem}[Upper bound for worst case errors in digital simulators with constant gate perturbations]
\label{wccds}
     Consider a perturbed Suzuki-Trotter product unitary of even order $p=2k$, which takes the form
    \begin{align}
       V^{(p)}_{l,n} (t)  = \prod_{j=1}^n \prod_{\upsilon=1}^\Upsilon \prod_{\gamma\in\Theta_l} V_{\gamma, j, \upsilon}\,,
    \end{align}
    where each local gate is a perturbed version of 
    the exact gate, satisfying $ \norm{ V_{\gamma, j, \upsilon} - e^{-i\frac{t}{n}a_{\upsilon,j} H_\gamma}} \leq \delta$. 
    The product unitary has Trotter number $n$ and is implemented on a system of size $l$.
    Then, the error on the time-evolution of a local observable $O$ is at most
\begin{align}
        \Delta \leq \BOO{\delta^{\frac{p}{p+1}} t^{d+1} \log^d(\frac{1}{\delta^{\frac{p}{p+1}}t^{d+1}})},
    \end{align}
    if the \textit{optimal choices} of $n_{\rm opt} = t/\delta^{\frac{1}{p+1}}$ and $l_{\rm opt}\geq vt-\frac{1}{\mu}{\log}(\delta^{\frac{p}{p+1}} t^{d+1})$ are made, for suitable constants $v$, $\mu$.
\end{theorem}

The proofs of both theorems are presented in Appendix~\ref{app:digital-worst-case}. The core idea is to separately derive bounds for the three error contributions described above. The truncation error due to the finite system size can be bounded using an instance of the Lieb-Robinson theorem, the discretization error is bounded by standard results in Trotter theory and then, finally, the gate error term is bounded by applying a telescopic product identity. The total error scaling is then derived as the one that balances these terms, making them scale equally. 

\subsection{Comparison of digital vs.\ \ analog simulation}

From our analysis of worst case errors in digital and analog quantum simulators we can conclude that very similar stability bounds apply in both cases, giving a comparable polynomial scaling of the error in $t$ and $\delta$. In the case of digital simulation for this optimal scaling to be achieved, the employed product unitary needs to be tailored to the considered setting by making an appropriate choice of Trotter number. 

A behaviour that we observe specifically for digital simulation is that the precise scaling in $\delta$ of the final error depends to some degree on the considered error model. The case that most naturally reproduces the analog simulation results is the one where errors of magnitude $\delta$ apply to the Hamiltonian generating the gate. This is not surprising, as this corresponds directly to the analog setting, except with a time dependent Hamiltonian which changes between each gate. Note, however, that this assumption implies that gate errors depend explicitly on the rotation angle of the given gate, with gates with a smaller rotation angle incurring proportionally smaller errors. This may not always necessarily capture the experimental reality. If we instead consider an opposite limit, where the unitary gates incur an error of magnitude $\delta$ independently of their rotation angle, then the digital error scaling deviates from the analog one, acquiring a slightly worse exponent for $\delta$ which may be compensated only by using higher order Suzuki-Trotter formulas.

\section{Results for average errors}\label{sec:average-case}
In this section, we consider the expectation value error
\begin{align}
    \Delta(\rho) \coloneqq \abs{\Tr{O(t) \rho} - \Tr{O'(t) \rho}},
\end{align}
as a random variable, where the randomness is given by 
different realizations of the simulator perturbations, which are sampled according to the random error models which we introduced in detail in Section~\ref{subsec:error-models}. We will 
make different kinds of statements about this random variable: 
we will analyze its mean value $\expect{\Delta(\rho)}$ (which can be significantly lower than the worst case value of $\Delta(\rho)$) 
and we will provide some concentration bounds to show that the typical value of $\Delta(\rho)$ fluctuates away from the mean value only with very low probability.

One subtlety that should be noted is that it is important to consider
here the state-dependent error $\Delta(\rho)$. In the deterministic setting, $\Delta=\sup_\rho \Delta(\rho)$ is clearly the best way for controlling the error given arbitrary inputs states. In the average case, however, $\expect{\Delta}$ is only an upper bound to the quantity we are interested in (as discussed also in Ref.~\cite{Chen_2024}). Indeed, by the convexity of the sup,
\begin{align}
    \sup_\rho \expect{\Delta(\rho)}\leq \expect{\sup_\rho \Delta(\rho)} = \expect{\Delta}\,.
\end{align}
This upper bound is often not optimal, as we will see in what follows, and it is therefore more useful to directly analyze $\expect{\Delta(\rho)}$.

\subsection{Analog quantum simulation}\label{sec:average-analog}
For analog quantum simulations, we consider here the error model where the local Hamiltonian terms $H_\gamma$  are replaced by randomly perturbed terms $H'_\gamma=H_\gamma+\delta L_\gamma(t)$ as in Eq.~\eqref{eq:analog-stoc-errors-def}.
To analyze this case further, we consider a model of stochastic time-dependent perturbations given by
\begin{align}
    L_\gamma(t)= \sum_{a=1}^m \xi_{\gamma,a}(t) X_{\gamma,a}\,, \label{eq:perturbation-stoc-process}
\end{align}
where $t\mapsto \xi_{\gamma,a}(t)$ are independent Gaussian noise processes with $\expect{\xi_{\gamma,a}(t)}=0$ and $\expect{\xi_{\gamma,a}(t) \xi_{\gamma',b}(s)}= \delta_{\gamma,\gamma'}\delta_{a,b} \, D(t-s)$~\cite{wassner2025holonomicquantumcomputationscalable}. Here, $X_{\gamma,a}$ is an arbitrary set of Hermitian operators, with support on the same lattice sites as $H_\gamma$ and $\norm{X_\gamma}=1$. The function $D(t-s)$ is called the time correlation function and is often assumed to be of the form
\begin{align}
    D(t-s) = e^{-\frac{(t-s)^2}{2\lambda^2}}, \label{eq:noise-time-correlation}
\end{align}
where $\lambda$ is the time correlation length. To simplify the presentation, we also make this choice here. Note however, that our results are readily extended to arbitrary correlation functions. In the following we compute the average value of $\Delta(\rho)$ for a finite correlation length $0 <\lambda < \infty $ and in the limit $\lambda \rightarrow \infty$, which corresponds to processes that are perfectly correlated, \ie constant in time. We further find that with high probability over the random perturbations, the value of $\Delta(\rho)$ is close to this average. 
\begin{theorem}[Average case bounds for errors in analog simulators with Gaussian perturbations]\label{acac}
    Consider a perturbed analog time evolution given by the Hamiltonian
    \begin{align}
        H'(t)=\sum_{\gamma\in\Gamma} \left(H_\gamma + \delta  \sum_{a=1}^m \xi_{\gamma,a}(t) X_{\gamma,a}\right)\,,
    \end{align}
    where $t\mapsto \xi_{\gamma,a}(t)$ are uncorrelated Gaussian noise processes with time correlation function given by~\eqref{eq:noise-time-correlation}. Assume that the initial state is a given pure state $\rho=\ket{\psi}\!\bra{\psi}$.
    Then, the error on the time-evolution of a local observable $O$ is, on average over the noise realizations,
    \begin{align}
        \expect{\Delta (\rho)} \leq \BOO{\sqrt{\lambda}\,\delta t^{\frac{d+1}{2}}\log^{\frac{d}{2}}\left(\frac{1}{\delta t^{\frac{d+1}{2}}}\right)}\,.
    \end{align}
    Additionally, 
    \begin{align}
        \prob{   \Delta(\rho)\geq  \BOO{s \, \sqrt{\lambda} \delta  t^{\frac{d+1}{2}} \log^{\frac{d}{2}}\left(\frac{1}{ \delta t^{\frac{d+1}{2}}}\right)}} \leq 2 e^{-s^2}.
    \end{align}
    In the case the infinite correlation length (\ie $\lambda \rightarrow +\infty$) we instead have
    \begin{align}
       \expect{\Delta(\rho)} \leq \BOO{\delta t^{\frac{d}{2} +1}\log^{\frac{d}{2}}\left(\frac{1}{\delta t^{\frac{d}{2}+1}}\right)}\,.
    \end{align}
    Likewise:
    \begin{align}
        \prob{   \Delta(\rho)\geq  \BOO{s \, \delta  t^{\frac{d}{2}+1} \log^{\frac{d}{2}}\left(\frac{1}{ \delta t^{\frac{d+1}{2}}}\right)}} \leq 2 e^{-s^2}.
    \end{align}
\end{theorem}
A full proof of the 
theorem 
is presented in Appendix~\ref{app:analog-average-case}. It is based on the derivation of perturbation theory results for evolutions under stochastic Schr\"odinger equations. 
We note that the case $\lambda \rightarrow \infty$ essentially corresponds to Gaussianly distributed time-independent perturbations. This is the scenario considered in Ref.~\cite{preskillstochasticcancellation}, for which we find the same error scaling. So in particular our results generalize the previously known ones to arbitrary time-dependent Gaussian perturbations. We observe that finite time correlation lengths in general correspond to a better scaling of the error $\Delta$ with respect to time. 

Taking the $\lambda \rightarrow 0$ limit, on the other hand, is not as straightforward as just inserting $0$ into the above expressions. This limit is known as the white noise limit, where the perturbation processes become completely uncorrelated in time, and needs different mathematical tools to be treated.
We, in particular, use the Ito stochastic calculus formalism for this analysis. As this is very technical, we defer most of the analysis to appendix~\ref{app:analog-average-case} and just briefly state here the result. We find that, if the analog simulator perturbations are of the form~\eqref{eq:perturbation-stoc-process} where $t\mapsto \xi_{\gamma,a}(t)$ are uncorrelated white noise processes, then the expected error is upper-bounded, with high probability over the noise realizations, by $\Delta(\rho)\leq  \BOO{\delta t^{\frac{d+1}{2}}}$ (up to logarithmic factors). More precisely, we have the following statement.
\begin{theorem}[Average case bounds for errors in analog simulators with white noise perturbations] \label{ito-acas}
    Consider a perturbed analog time evolution $\ket{\psi'_t}$ given by evolution under white noise perturbations. Assume that the initial state is a given pure state $\rho=\ket{\psi}\!\bra{\psi}$.
    Then, the error on the time-evolution of a local observable $O$ is, on average over the noise realizations,
    \begin{align}
        \expect{\Delta(\rho)}
        &\leq  \BOO{\delta t^{\frac{d+1}{2}} \log^{\frac{d}{2}}\left(\frac{1}{ \delta t^{\frac{d+1}{2}}}\right)}.
    \end{align}
    Additionally, 
    \begin{align}
        \prob{   \Delta(\rho)\geq  \BOO{s \, \delta t^{\frac{d+1}{2}} \log^{\frac{d}{2}}\left(\frac{1}{ \delta t^{\frac{d+1}{2}}}\right)}} \leq 2 e^{-s^2}.
    \end{align}
\end{theorem}

Finally, we observe that, as hinted to above, the quantity $\expect{\Delta}$ is not the most useful metric to analyze. Indeed, considering a setting analogous to Theorem~\ref{thm:wcas} with time-independent perturbations, we find the following bound, proven in Appendix~\ref{app:analog-average-case}.
\begin{theorem}[Upper bound for average case errors in analog simulators]
    \label{acas}
    Consider a perturbed analog time evolution of the form
    \begin{align}
        V (t) = e^{-it\sum_\gamma\left(H_\gamma+\delta L_\gamma\right)},
    \end{align}
    where $L_\gamma$ are a set of independent random variables drawn from a ensemble of Hermitian operators with $\norm{L_{\gamma}} \leq 1$ and $\expect{L_\gamma}=0$.
    Then, the expected worst-case error on measuring $O$ is
    \begin{align}
        \expect{\Delta} \leq \BOO{t^{d+1} \delta}.
    \end{align}
\end{theorem}
We thus see that this only gives an average-case bound that scales identically to the worst-case bound.

\subsection{Digital 
quantum simulation}
In the case of digital simulation, the crucial technical ingredient for our analysis is a generalized version of the well-known Hoeffding inequality applied to vector valued random variables. This allows us to derive the fact (discussed more in detail in Lemma~\ref{randomsum} of the Appendix) that
\begin{align}
    \label{khitchine-tr}
    \expect{\norm{\sum_{j=1}^n \sum_{\upsilon=1}^\Upsilon \sum_{\gamma\in\Theta_l} L_{\gamma,\upsilon, j}\ket{\psi}}_2} \leq \BOO{\sqrt{n\abs{\Theta_l}} },
\end{align}
for mean-zero random perturbations $L_{\gamma,\upsilon, j}$. Note that the same quantity in the worst case will necessarily scale as $\BOO{n\abs{\Theta_l}}$. This hints at a possibly different behavior in the average case, compared to what we saw before. In the following, we seek to exploit this fact to our advantage. 

We first consider the error model where the perturbed gates appearing in the Suzuki-Trotter circuit are of the form~\eqref{eq:noisy-product-unitary-def}. 
Here, we see that this different scaling in $n$ has the consequence that, unlike in Theorem~\ref{wcds}, we no longer need to carefully choose the Trotter number $n$ to balance the different error contributions. Instead, the average error behaves in a way more similar to the one of noiseless Suzuki-Trotter formulas. That is, it is possible to always decrease the error by arbitrarily increasing $n$. This is true not only for the average error but also in general for the quantity $\Delta(\rho)$, with high probability over the random perturbations. These results are summarized in the following theorem, for which we provide a full proof in Appendix~\ref{app:digital-average-case}.

\begin{theorem}[Average case errors in digital simulators with gate-dependent perturbations]\label{fixed-states-ads}
    Consider a perturbed Suzuki-Trotter product unitary of order $p =2k$, which takes the form
    \begin{align}
        V_{l,n}^{(p)}(t) = \prod_{j=1}^n \prod_{\upsilon =1}^\Upsilon \prod_{\gamma}^{\abs{\Theta_l}} e^{i\frac{t}{n} (H_\gamma a_{\gamma, \upsilon} + \delta L_{\gamma, \upsilon, j})},
    \end{align}
    where $L_{\gamma, \upsilon, j}$ are random perturbations, drawn independently from a distribution of Hermitian operators with bounded norm $\norm{L_{\gamma, \upsilon, j}}\leq 1$ and vanishing mean $\expect{L_{\gamma, \upsilon, j}}=0$. The product unitary has Trotter number $n$ and is implemented on a system of size $l$. Assume that the initial state is a given pure state $\rho=\ket{\psi}\!\bra{\psi}$.
    
    Then, for any $\varepsilon>0$, there exists a choice of $n\geq \BOO{\frac{t^{d+2}}{\varepsilon^2}\log^d\!\left(\frac{1}{\varepsilon}\right)}$ and $l\geq vt-\frac{1}{\mu}\log\BOO{\varepsilon}$ such that the error on time evolution of a local observable $O$ is on average
    \begin{align}
        \expect{\Delta(\rho)} \leq \varepsilon\,.
    \end{align}
    Here, $v$ and $\mu$ are suitable constants. Additionally, for the same choices, we have
    \begin{align}
        \prob{\Delta(\rho) > s\,\varepsilon} \leq 2e^{-s^2}\,.
    \end{align}
\end{theorem}

In the case of the second error model, where  the perturbed gates appearing in the Suzuki-Trotter circuit are of the form~\eqref{eq:noisy-product-unitary-def2}, we find that we again need to choose an optimal Trotter number $n$ to balance the various error contributions. This leads to a fundamental limit also on the average error that can be achieved in the presence of noise. At this optimal point, we find that, with high probability over the random perturbations, $\Delta(\rho)\leq \BOO{ \delta^{\frac{2p}{2p+1}} t^{\frac{2}{3}(d+1)}}$ (up to logarithmic factors). This can be more formally stated as in the following theorem, also proven in Appendix~\ref{app:digital-average-case}.
\begin{theorem}[Average case errors in digital simulators with constant gate perturbations] \label{fixed-states-2}
    Consider a perturbed Suzuki-Trotter product unitary of order $p =2k$, which takes the form
    \begin{align}
        V_{l,n}^{(p)}(t) = \prod_{j=1}^n \prod_{\upsilon =1}^\Upsilon \prod_{\gamma}^{\abs{\Theta_l}} e^{i\frac{t}{n} H_\gamma a_{\gamma, \upsilon} + i\delta L_{\gamma, \upsilon, j}}\,,
    \end{align}
    where $L_{\gamma, \upsilon, j}$ are random perturbations, drawn independently from a distribution of Hermitian operators with bounded norm $\norm{L_{\gamma, \upsilon, j}}\leq 1$ and vanishing mean $\expect{L_{\gamma, \upsilon, j}}=0$. The product unitary has Trotter number $n$ and is implemented on a system of size $l$. Assume that the initial state is a given pure state $\rho=\ket{\psi}\!\bra{\psi}$.
    
    Then, the error on time evolution of a local observable $O$ is on average
    \begin{align}
        \expect{\Delta(\rho)}  \leq \BOO{ \delta^{\frac{2p}{2p+1}} t^{\frac{2}{3}(d+1)}\log^d\left(\frac{1}{\delta^{\frac{2p}{2p+1}} t^{\frac{2}{3}(d+1)}}\right)}\,,
    \end{align}
    if the optimal choices $n_{\rm opt}=\delta^{-\frac{2}{2p+1}}\,t^{\frac{d+4}{3}}$ and $l_{\rm opt}=vt-\frac{1}{\mu}\log\left( \delta^{\frac{2p}{2p+1}} t^{\frac{2}{3}(d+1)}\right)$ are made, for suitable constants $\mu$ and $v$. Additionally, for the same choices, we have
\begin{align}
    &\mathrm{Prob}\!\left[\Delta(\rho) > \mathcal{O}\!\left(s \; \delta^{\frac{2p}{2p+1}} t^{\frac{2}{3}(d+1)}\log^d\!\left(\frac{1}{\delta^{\frac{2p}{2p+1}} t^{\frac{2}{3}(d+1)}}\right)\!\right)\!\right]\nonumber\\
    &\hspace{70mm}\leq 2e^{-s^2}\,.
\end{align}
\end{theorem}

To conclude we discuss also in this case the role of the quantity $\expect{\Delta}$. As before, we find that this is only a loose upper bound on the average error that can be obtained for a fixed input state. Indeed, it is possible to prove the following scalings, both of which are worse than the ones discussed above. Under the assumptions of Theorem~\ref{fixed-states-ads}, that is in the case of the gate-dependent error model, we find
\begin{align}
    \expect{\Delta} \leq \varepsilon\,,
\end{align}
provided that one chooses \begin{align}
n\geq\BOO{\frac{t^{2d+2}}{\varepsilon^2} \,\log^{2d}\!\left(\frac{1}{\varepsilon}\right)}
\end{align}
and $l\geq vt-\frac{1}{\mu}\log\BOO{\varepsilon}$. Under the assumptions of Theorem~\ref{fixed-states-2}, that is in the case of the constant gate error model, we instead find that the best achievable scaling is
\begin{align}
    \expect{\Delta} \leq \BOO{\delta^{\frac{2p}{2p+1}} t^{d+\frac{2}{3}}} \log^d\left(\frac{1}{\delta^{\frac{2p}{2p+1}} t^{d+\frac{2}{3}}}\right) \,.
\end{align}
In both cases the value of $\Delta$ concentrates around this average scaling with high probability. A full proof of these statements can be found in Theorems~\ref{acdc} and~\ref{accds} in the Appendix.

\subsection{Comparison of digital vs.\  analog simulation}
From our analysis of average case errors in digital and analog
quantum simulators, we can conclude that in both settings the expected error will, with high probability, show an improved scaling in comparison to the worst case. In the analog setting, in particular, we see that the improved dependence of the average error on the evolution time $t$, which had been previously observed for time-independent perturbations~\cite{preskillstochasticcancellation}, also applies in several cases of time-dependent random perturbations, including white noise and noise with finite time correlations. 

In the case of digital simulation, we see a possibly even larger improvement. Indeed, we see that, for one of the error models that we consider, the average case behaviour of Suzuki-Trotter formulas reproduces the one of the noiseless case. That is, the error can be arbitrarily reduced by choosing a larger Trotter number $n$. 

\subsection{Lindbladians and Brownian random walks}
We conclude by commenting on a further pair of error models that are of relevance. In the analog setting, we have shown the behaviour of errors under white noise perturbations. It is well-known that the mixed state evolution of the averaged density matrix in such cases follows a Lindbladian evolution. This allows us to say something also about the stability of analog simulation under this non-unitary noise model. In fact we are able to prove the following result, as shown in Appendix~\ref{app:analog-average-case}.
\begin{theorem}[Lindbladian perturbations] \label{pertlind}
Consider the noisy evolution give by the Lindbladian:
\begin{align}
    \lind[\rho_t] \!= \!-i[H, \rho_t] + \delta^2 \sum_a L_a \rho_t L_a^\dag -\! \frac{\delta^2}{2} \left\{L_a^\dag L_a, \rho_t\right\}. \label{eq:Lindblad-perturbation}
\end{align}
Let $\rho'(t) = e^{\lind t} (\rho)$ and $\rho(t)=e^{-iHt}\rho e^{iHt}$, then
\begin{align}
      \norm{\rho'(t)-\rho(t)}_1 \leq\BOO{\delta t^{\frac{d+1}{2}} \log^{\frac{d}{2}} (\frac{1}{\delta t^{\frac{d+1}{2}}})}
\end{align}
\end{theorem}

Lastly, consider the perturbed digital simulation model of the form
\begin{align}   
    \label{eq:noisy-trotter-circuit-def3}
    V_{l, n}^{(p)} = \prod_{\gamma, j, \upsilon} e^{i\frac{t}{n} a_{\gamma, \upsilon} H_{\pi_\upsilon(\gamma)} + i\sqrt{\frac{t}{n}} \delta L_{\gamma, \upsilon, j} }.
\end{align}
In the $n \rightarrow \infty$ limit, the averaged density matrix
\begin{align}
    \rho(t) = \lim_{n \rightarrow \infty} \expect{V_{l, n}^{(p)} (t) \ketbra{\psi(0)}{\psi(0)} V_{l, n}^{(p), \dag} (t)},
\end{align}
obeys a Lindblad type noise model of the form~\eqref{eq:Lindblad-perturbation}, where the Lindblad generator contains only Hermitian jump operators. The proof that this converges in distribution uses a variation of the central limit theorem, known as \textit{Donsker's Theorem} \cite{stochasticcal, legall} and fundamentally shows how a rescaled random walk converges to a Wiener process. Thus, for large $n$, this model behaves similarly to the process described in Theorem~\ref{pertlind}. At finite $n$ this gives an error behaviour described in the following theorem.
 \begin{theorem}[Discrete-Ito perturbations]
    \label{ito-acdc}
        Given $V_{l,n}^{(p)}$ as in Equation \eqref{eq:noisy-trotter-circuit-def3}, let $\rho=\ket{\psi}\bra{\psi}$ be a pure initial state and let $\ket{\psi'(t)} =  V_{l, n}^{(p)} (t) \ket{\psi(0)}$, then the error:
        \begin{align}
            \Delta(\rho) \leq \BOO{\delta t^{\frac{d+1}{2}} \log^{\frac{d}{2}}(\frac{1}{\delta t^{\frac{d+1}{2}}})}.
        \end{align}
     \end{theorem}
The proof of this theorem is presented in Appendix~\ref{app:digital-average-case}

\section{Outlook and conclusion}

Recent years have enjoyed a rapid progress in the field of quantum simulation, both in the digital and the analog realm. There are good reasons to believe that quantum simulation may be the first technology-ready application of the quantum technologies. This development poses pressing questions on how to compare the two scenarios fairly. This work is meant to be a substantial contribution 
along these lines. Concretely,
in this work, we provide a comprehensive theoretical analysis of the behavior of analog and digital quantum simulation under noise. We provide in both cases deterministic worst-case bounds and average case statements under stochastic error models. This allows for a global comparison of the performance of analog and digital methods, supported by drastically improved bounds on how unitary errors accumulate. This allows us to derive actionable advice on the best practical implementations in the presence of different forms of noise. 

In the analog setting, we provide stochastic error bounds for measuring local observables using different assumptions on the underlying noise model. These results provide a greater insight into the stability of analog quantum simulation, as well as allow to design further techniques for noise suppression in analog devices. We believe further considering quantum noises of the form \cite{Benoist2022deviationbounds}, will complete the picture we considered in this work.

In the digital setting, we believe that this work will have impact on our general understanding of Trotter products as a theoretical tool of approximating unitaries~\cite{Childs_2021, Chen_2024}: it answers questions about robustness of this method in general, while also providing concrete recipes for optimal choices of Trotter number $n$ and system size $l$. We believe that this work can also help to improve techniques for filtering and control \cite{bouten2006introductionquantumfiltering,Bouten_2009, bouten2015trotterkatotheoremquantummarkov, Bouten_2008}.

Our analysis may finally help to design new tools for quantum error mitigation \cite{ErrorMitigationObstructions, ErrorMitigationObstructionsOld}, benchmarking \cite{BenchmarkingReview}, and help with aspects of quantum control theory in order to design more noise resilient quantum devices \cite{hu2025universaldynamicsgloballycontrolled, Dutkiewicz_2024}. We believe that it is such kind of technical work that will help driving the theory of quantum simulation forward.
 
\section{Acknowledgements}
The authors thank Lennart Bittel for inspiring this project and many helpful comments. 
The authors further thank Onno Pfohl, Paul Faehrmann, Jose Carrasco, Jonas Fuksa, Antonio Anna Mele, and Gregory A.~L White for fruitful discussions and helpful comments. This work has been  supported by the BMFTR (DAQC, MUNIQC-Atoms, QuSol, PasQuops, hybrid++), the Munich Quantum Valley (K-4 and K-8), the Quantum Flagship (PasQuans2, Millenion), QuantERA (HQCC), the Clusters of Excellence MATH+ and ML4Q, the DFG (CRC183), the Einstein Foundation (Einstein Research Unit on Quantum Devices), Berlin Quantum, and the ERC (DebuQC).

\bibliography{references,BigReferences68}

@article{QSim, author={I. M. Georgescu and S. Ashhab and F. Nori}, 
  doi={10.1103/RevModPhys.86.153},
pages=153,
  title ={Quantum simulation}, journal={Rev. Mod. Phys.},volume= 86, 
year=2014}

@article{Fauseweh,
title={Quantum many-body simulations on digital quantum computers: State-of-the-art and future challenges},
author={B. Fauseweh},
doi={10.1038/s41467-024-46402-9},
journal={Nature Comm.}, volume=15, pages=2123 ,year=2024}

@misc{Appendix,
note={See the supplemental material for further details of the proof of Observation 2, 
which includes Refs.\ \cite{Continuous,GaussianQuantumInfo,AreaReview}.
}}

@article{AnshuSampleEfficientHamiltonianLearning,
title={Sample-efficient learning of interacting quantum systems},
author={Anurag Anshu and Srinivasan Arunachalam and Tomotaka Kuwahara and Mehdi Soleimanifar},
doi={10.1038/s41567-021-01232-0},
journal={Nature Phys.}, volume=17, pages={931} , year=2021}

@article{cole_identifying_2005,
  title = {Identifying an experimental two-state {{Hamiltonian}} to arbitrary accuracy},
  author = {Cole, Jared H. and Schirmer, Sonia G. and Greentree, Andrew D. and Wellard, Cameron J. and Oi, Daniel K. L. and Hollenberg, Lloyd C. L.},
  year = {2005},
  month = jun,
  journal = {Phys. Rev. A},
  volume = {71},
  number = {6},
  pages = {062312},
  publisher = {{American Physical Society}},
  doi = {10.1103/PhysRevA.71.062312},
  urldate = {2021-10-05}
}

@article{hangleiterRobustlyLearningHamiltonian2024a,
  title = {Robustly Learning the {{Hamiltonian}} Dynamics of a Superconducting Quantum Processor},
  author = {Hangleiter, Dominik and Roth, Ingo and Fuksa, Jon{\'a}{\v s} and Eisert, Jens and Roushan, Pedram},
  year = {2024},
  month = nov,
  journal = {Nature Communications},
  volume = {15},
  number = {1},
  pages = {9595},
  publisher = {Nature Publishing Group},
  issn = {2041-1723},
  doi = {10.1038/s41467-024-52629-3},
  urldate = {2025-03-07},
  copyright = {2024 The Author(s)},
  langid = {english}
}

@article{schirmer_two-qubit_2009,
  title = {Two-qubit {{Hamiltonian}} tomography by {{Bayesian}} analysis of noisy data},
  author = {Schirmer, Sonia G. and Oi, Daniel K. L.},
  year = {2009},
  month = aug,
  journal = {Phys. Rev. A},
  volume = {80},
  number = {2},
  pages = {022333},
  publisher = {{American Physical Society}},
  doi = {10.1103/PhysRevA.80.022333},
  urldate = {2021-10-19}
}

@Article{TangHamiltonianLearning2,
title={{Structure learning of Hamiltonians from real-time evolution}},
author={Ainesh Bakshi and Allen Liu and Ankur Moitra and Ewin Tang},
archiveprefix={arxiv},
eprint={2405.00082},
journal = ""
}

@Misc{r,
  title                     = {{The Complexity of Translationally-Invariant Spin Chains with Low Local Dimension}},

  Author                   = {Bausch, Johannes and Cubitt, Toby and Ozols, Maris},
  Note                     = {arXiv:1605.01718}
}

@article{GreinerSpeedup,
journal={Nature}, volume=551, 
doi={10.1038/nature24622},
pages={579-584}, year=2017,
      title ={Probing many-body dynamics on a 51-atom quantum simulator},
    author={H. Bernien and S. Schwartz and A. Keesling and H. Levine and A. Omran and H. Pichler and S. Choi and  A. S. Zibrov and M. Endres and M. Greiner and V. Vuletic and M.  Lukin}
    }

@Article{BlochSimulation,
  title                     = {Quantum simulations with ultracold quantum gases},
  Author                   = {I. Bloch and J. Dalibard and S. Nascimbene},
  doi={10.1038/nphys2259},
  Journal                  = {Nature Phys.},
  Year                     = {2012},
  Pages                    = {267},
  Volume                   = {8}
}

@Article{Stability,
  Author                   = {S. Bravyi and M. Hastings and S. Michalakis},
  Journal                  = {J. Math. Phys.},
  
  Year                     = {2010},
  Pages                    = {093512},
  Volume                   = {51}
}

@article{BenchmarkingReview,
title={Quantum certification and benchmarking},
doi={10.1038/s42254-020-0186-4},
Author={J. Eisert and D. Hangleiter and N. Walk and I. Roth and D. Markham and R. Parekh and U. Chabaud and E. Kashefi},
journal={Nature Rev. Phys.},
volume=2, pages={382-390},
year={2020}
}

@Article{CiracZollerSimulation,
  title                     = {Goals and opportunities in quantum simulation},
  Author                   = {J. I. Cirac and P. Zoller},
  doi={10.1038/nphys2275},  
  Journal                  = {Nature Phys.},
  Year                     = {2012},
  Pages                    = {264},
  Volume                   = {8}
}

@Article{SuperconductingSimulation,
  title                     = {On-chip quantum simulation with superconducting circuits},
  Author                   = {A. A. Houck and H. E. Tuereci and J. Koch},
  Journal                  = {Nature Phys.},
  Year                     = {2012},
doi={10.1038/nphys2251},
  Pages                    = {292},
  Volume                   = {8}
}

@InCollection{LRreviewchapter,
  title                     = {{Lieb-Robinson bounds and the simulation of time-evolution of local observables in lattice systems}},
  Author                   = {Kliesch, M. and Gogolin, C. and Eisert, J.},
  Booktitle                = {Many-Electron Approaches in Physics, Chemistry and Mathematics},
  Publisher                = {Springer},
  Year                     = {2014},
  Editor                   = {Bach, V. and Delle Site, L.},
  Pages                    = {301},
  Series                   = {Mathematical Physics Studies}
}

@Article{Lloyd,
  title                     = {Universal quantum simulators},
  Author                   = {S. Lloyd},
  doi={10.1126/science.273.5278.1073},
  Journal                  = {Science},
  Year                     = {1996},
  Pages                    = {1073},
  Volume                   = {273}
}

@Article{Suz91,
  title                     = {General theory of fractal path integrals with applications to many-body theories and statistical physics},
  Author                   = {Suzuki, Masuo},
  Journal                  = {J. Math. Phys.},
  Year                     = {1991},
  optNumber                   = {2},
  Pages                    = {400-407},
  Volume                   = {32},

  Doi                      = {10.1063/1.529425},
  opturl                      = {http://scitation.aip.org/content/aip/journal/jmp/32/2/10.1063/1.529425}
}

@article{FG20,
	title = {Limitations of optimization algorithms on noisy quantum devices},
	author = {Stilck Franca, D. and Garc\'{i}a-Patr\'{o}n, R.},
	journal = {Nature Phys.},
	doi={10.1038/s41567-021-01356-3},
	volume = {17},
	pages=
	{1221},
	year = {2020}
}

@article{SpeedLimits,
    title={Speed limits and locality in many-body quantum dynamics},
    author={Chi-Fang (Anthony) Chen and  Andrew Lucas and  Chao Yin},
    year= 2023,
    journal={Rept. Prog. Phys.},
    volume= 86, 
    pages=116001,
    doi={10.1088/1361-6633/acfaae}
}

@article{MBL2D,
   title ={Exploring the many-body localization transition in two dimensions},
    author={J.-Y. Choi and S. Hild and J. Zeier and P. Schau{\ss} and A. Rubio-Abadal and T. Yefsah and V. Khemani  and  D. A. Huse and I. Bloch and C. Gross},
    journal={Science},
    doi={10.1126/science.aaf8834},
    volume= 352, 
    pages=1547,
    year=2016}

@Article{onorati_mixing_2017,
  title                     = {Mixing {properties} of {stochastic} {quantum} {Hamiltonians}},
  Author                   = {Onorati, E. and Buerschaper, O. and Kliesch, M. and Brown, W. and Werner, A. H. and Eisert, J.},
  Journal                  = {Commun. Math. Phys.},
  Year                     = {2017},

  Month                    = nov,
  optNumber                   = {3},
  Pages                    = {905--947},
  Volume                   = {355},
archiveprefix = {arXiv},
eprint = {1606.01914},
  Doi                      = {10.1007/s00220-017-2950-6},
  optissn                     = {0010-3616, 1432-0916},
  opturl                      = {https://link.springer.com/article/10.1007/s00220-017-2950-6},
  opturldate                  = {2018-03-26}
}

@article{AnalogStability,
title={Quantum advantage and stability to errors in analogue quantum simulators},
author={Rahul Trivedi and Adrian Franco Rubio and J. Ignacio Cirac},
journal={Nature Comm.}, volume=15, pages=6507, year=2024,
doi={10.1038/s41467-024-50750-x}}

@Article{Trotzky,
  title                     = {Probing the relaxation towards equilibrium in an isolated strongly correlated one-dimensional {B}ose gas},
  Author                   = {S. Trotzky and Y.-A. Chen and A. Flesch and I. P. McCulloch and U. Schollw\"ock and J. Eisert and I. Bloch},
  Journal                  = {Nature Phys.},
  Year                     = {2012},
  Pages                    = {325-330},
  Volume                   = {8},
  eprint = {1101.2659},
  archiveprefix = {arXiv},
  Doi                      = {doi:10.1038/nphys2232}
}

@article{Dimitris,
journal={Science},
  title ={Spectral signatures of many-body localization with interacting photons},
 author={P. Roushan and C. Neill and J. Tangpanitanon and V.M. Bastidas and A. Megrant and R. Barends and Y. Chen and Z. Chen and B. Chiaro and A. Dunsworth and A. Fowler and B. Foxen and M. Giustina and E. Jeffrey and J. Kelly and E. Lucero and J. Mutus and M. Neeley and C. Quintana  and  Sank and A. Vainsencher and J. Wenner and T. White and H. Neven  and  G. Angelakis and J. Martinis},
volume=358, pages={1175-1179},
DOI={10.1126/science.aao1401},
year=2017,
 }

@article{Nonunital,
    year = {2024},
    eprint = {2403.13927},
    archivePrefix = {arXiv},
    title={Noise-induced shallow circuits and absence of barren plateaus},
    author={A. A. Mele and A. Angrisani and S. Ghosh and S. Khatri and J. Eisert and D. Stilck França and Y. Quek},
    journal={}
}

@article{NonUnitalSampling,
    eprint = {2306.16659},
    archivePrefix = {arXiv},
    year=2023,
    title={Effect of non-unital noise on random circuit sampling},
    author={B. Fefferman and S. Ghosh and M. Gullans and K. Kuroiwa and K. Sharma},
    journal={}
}

@article{ErrorMitigationObstructions,
    year=2024,
    title={Exponentially tighter bounds on limitations of quantum error 
    mitigation},
    volume=20, 
    pages={1648},
    doi={10.1038/s41567-024-02536-7},
    journal={Nature Phys.},
    author={Y. Quek and D. Stilck França and S. Khatri and J. J. Meyer
    and J. Eisert}
}

@article{ErrorMitigationObstructionsOld,
     note={arXiv:2210.11505},year=2022,
     journal={npj Quant. Inf.},
     volume=8, pages=114,
     doi={10.1038/s41534-022-00618-z},
    title={Fundamental limits of quantum error mitigation},
    author={R. Takagi and S. Endo and S. Minagawa and M. Gu}}

@article{NearTermMaterialsSimulation,
title={Towards near-term quantum simulation of materials},
author={Laura Clinton and Toby Cubitt and Brian Flynn and Filippo Maria Gambetta and Joel Klassen and Ashley Montanaro and Stephen Piddock and Raul A. Santos and Evan Sheridan},
doi={10.1038/s41467-023-43479-6},
journal={Nature Comm.},
volume={15},
pages={211},
year={2024}
}

@article{MonroeTimeCrystal,
author={J. Smith and A. Lee and P.  Richerme and B. Neyenhuis and P. W.
Hess and P. Hauke and M. Heyl  and  D. Huse and C. Monroe}, year= 2016,
title={Many-body localization in a quantum simulator with programmable random disorder}, 
doi={10.1038/nphys3783},
journal={Nature Phys.}, volume=12, pages={907-911}}

@PREAMBLE{
 "\providecommand{\noopsort}[1]{}" 
 # "\providecommand{\singleletter}[1]{#1}%" 
}

@inproceedings{preskillstochasticcancellation,
  author =	{Cai, Yiyi and Tong, Yu and Preskill, John},
  title =	{{Stochastic error cancellation in analog quantum simulation}},
  booktitle =	{19th Conference on the Theory of Quantum Computation, Communication and Cryptography (TQC 2024)},
  pages =	{2:1--2:15},
  series =	{Leibniz International Proceedings in Informatics (LIPIcs)},
  ISBN =	{978-3-95977-328-7},
  ISSN =	{1868-8969},
  year =	{2024},
  volume =	{310},
  editor =	{Magniez, Fr\'{e}d\'{e}ric and Grilo, Alex Bredariol},
  publisher =	{Schloss Dagstuhl -- Leibniz-Zentrum f{\"u}r Informatik},
  address =	{Dagstuhl, Germany},
  URL =		{https://drops.dagstuhl.de/entities/document/10.4230/LIPIcs.TQC.2024.2},
  URN =		{urn:nbn:de:0030-drops-206720},
  doi =		{10.4230/LIPIcs.TQC.2024.2}
}

@article{Schur1911,
author = {Schur, J.},
journal = {Journal f\"ur die reine und angewandte Mathematik},
pages = {1-28},
title = {Bemerkungen zur {Theorie} der beschr\"ankten {Bilinearformen} mit unendlich vielen {Ver\"anderlichen}},
url = {http://eudml.org/doc/149352},
volume = {140},
year = {1911},
}

@article{random,
  title = {User-Friendly Tail Bounds for Sums of Random Matrices},
  author = {Tropp, Joel A.},
  journal = {Found. Comput. Math.},
  volume = {12},
  issue = {4},
  pages = {389-434},
  year = {2012},
  doi = {10.1007/s10208-011-9099-z},
  url = {https://doi.org/10.1007/s10208-011-9099-z}
}

@article{random3,
  title = {THE EXPECTED NORM OF A SUM OF INDEPENDENT RANDOM MATRICES: AN ELEMENTARY APPROACH},
  author = {J.A.Tropp},
  year = {2015},
  journal = {},
  url = {https://arxiv.org/pdf/1506.04711},
  eprint={1506.04711},
  archivePrefix={arXiv},
}

@article{_Anthony_Chen_2023,
   title={Speed limits and locality in many-body quantum dynamics},
   volume={86},
   ISSN={1361-6633},
   url={http://dx.doi.org/10.1088/1361-6633/acfaae},
   DOI={10.1088/1361-6633/acfaae},
   number={11},
   journal={Rep. Prog. Phys.},
   publisher={IOP Publishing},
   author={(Anthony) Chen, Chi-Fang and Lucas, Andrew and Yin, Chao},
   year={2023},
   month=sep, pages={116001} }

@article{Childs_2021,
   title={{Theory of Trotter error with commutator scaling}},
   volume={11},
   DOI={10.1103/physrevx.11.011020},
   pages=011020,
   journal={Phys. Rev. X},
   publisher={American Physical Society (APS)},
   author={Childs, Andrew M. and Su, Yuan and Tran, Minh C. and Wiebe, Nathan and Zhu, Shuchen},
   year={2021},
   month=feb }

@article{chen2021concentrationotocliebrobinsonvelocity,
      title={{Concentration of OTOC and Lieb-Robinson velocity in random Hamiltonians}}, 
      author={Chi-Fang Chen},
      year={2021},
      eprint={2103.09186},
      archivePrefix={arXiv},
      optprimaryClass={quant-ph},
      url={https://arxiv.org/abs/2103.09186}, 
      journal={}
}

@article{Chen_2024,
   title={Average-Case Speedup for Product Formulas},
   volume={405},
pages= 32,
   DOI={10.1007/s00220-023-04912-5},
   journal={Comm. Math. Phys.},
   publisher={Springer Science and Business Media LLC},
   author={Chen, Chi-Fang and Brandão, Fernando G. S. L.},
   year={2024},
   month=feb }

@article{Harley_2024,
   title={Going beyond gadgets: the importance of scalability for analogue quantum simulators},
   volume={15},
pages=6527,
   DOI={10.1038/s41467-024-50744-9},
   journal={Nature Comm.},
   publisher={Springer Science and Business Media LLC},
   author={Harley, Dylan and Datta, Ishaun and Klausen, Frederik Ravn and Bluhm, Andreas and França, Daniel Stilck and Werner, Albert H. and Christandl, Matthias},
   year={2024},
   month=aug }

@misc{xu2025exponentiallydecayingquantumsimulation,
      title={Exponentially Decaying Quantum Simulation Error with Noisy Devices}, 
      author={Jue Xu and Chu Zhao and Junyu Fan and Qi Zhao},
      year={2025},
      eprint={2504.10247},
      archivePrefix={arXiv},
      optprimaryClass={quant-ph},
      url={https://arxiv.org/abs/2504.10247}, 
}

@article{Knee_2015,
   title={{Optimal Trotterization in universal quantum simulators under faulty control}},
   volume={91},
   DOI={10.1103/physreva.91.052327},
   pages=052327,
   journal={Phys. Rev. A},
   publisher={American Physical Society (APS)},
   author={Knee, George C. and Munro, William J.},
   year={2015},
   month=may }

@article{PhysRevLett.129.270502,
  title = {Hamiltonian Simulation with Random Inputs},
  author = {Zhao, Qi and Zhou, You and Shaw, Alexander F. and Li, Tongyang and Childs, Andrew M.},
  journal = {Phys. Rev. Lett.},
  volume = {129},
  issue = {27},
  pages = {270502},
  numpages = {7},
  year = {2022},
  month = {Dec},
  publisher = {American Physical Society},
  doi = {10.1103/PhysRevLett.129.270502},
  url = {https://link.aps.org/doi/10.1103/PhysRevLett.129.270502}
}

@article{Poggi_2020,
   title={Quantifying the Sensitivity to Errors in Analog Quantum Simulation},
   volume={1},
pages=020308,
   DOI={10.1103/prxquantum.1.020308},
   number={2},
   journal={PRX Quantum},
   publisher={American Physical Society (APS)},
   author={Poggi, Pablo M. and Lysne, Nathan K. and Kuper, Kevin W. and Deutsch, Ivan H. and Jessen, Poul S.},
   year={2020},
   month=nov }

@article{wassner2025holonomicquantumcomputationscalable,
      title={Holonomic quantum computation: a scalable adiabatic architecture}, 
      author={Clara Wassner and Tommaso Guaita and Jens Eisert and Jose Carrasco},
      year={2025},
      eprint={2502.17188},
      archivePrefix={arXiv},
      optprimaryClass={quant-ph},
      url={https://arxiv.org/abs/2502.17188}, 
      journal={}
}

@misc{möbus2025stabilitythermalequilibriumlongrange,
      title={Stability of thermal equilibrium in long-range quantum systems}, 
      author={Tim Möbus and Jorge Sánchez-Segovia and Álvaro M. Alhambra and Ángela Capel},
      year={2025},
      eprint={2506.16451},
      archivePrefix={arXiv},
      optprimaryClass={quant-ph},
      url={https://arxiv.org/abs/2506.16451}, 
}

@article{Kashyap_2025,
   title={Accuracy Guarantees and Quantum Advantage in Analog Open Quantum Simulation with and without Noise},
   volume={15},
pages=021017,
   DOI={10.1103/physrevx.15.021017},
   journal={Phys. Rev. X},
   publisher={American Physical Society (APS)},
   author={Kashyap, Vikram and Styliaris, Georgios and Mouradian, Sara and Cirac, J. Ignacio and Trivedi, Rahul},
   year={2025},
   month=apr }

@book{stochasticcal,
  title = {Brownian Motion and Stochastic Calculus},
  author = {Ioannis Karatzas, Steven E. Shreve},
  publisher = {Springer New York, NY},
  doi = {https://doi.org/10.1007/978-1-4612-0949-2},
}

@book{legall,
  title = {Brownian Motion, Martingales, and Stochastic Calculus},
  author = {Jean-François Le Gall},
  publisher = {Springer Cham},
  doi = {https://doi.org/10.1007/978-3-319-31089-3},
}

@article{klieschlocality,
  title = {{Quasilocality and efficient simulation of Markovian quantum dynamics}},
  author = {Barthel, Thomas and Kliesch, Martin},
  journal = {Phys. Rev. Lett.},
  volume = {108},
  issue = {23},
  pages = {230504},
  numpages = {5},
  year = {2012},
  month = {Jun},
  publisher = {American Physical Society},
  doi = {10.1103/PhysRevLett.108.230504},
  url = {https://link.aps.org/doi/10.1103/PhysRevLett.108.230504}
}

@article{tropp2015introductionmatrixconcentrationinequalities,
      title={An Introduction to Matrix Concentration Inequalities}, 
      author={Joel A. Tropp},
      year={2015},
      eprint={1501.01571},
      archivePrefix={arXiv},
      optprimaryClass={math.PR},
      url={https://arxiv.org/abs/1501.01571}, 
      journal={}
}

@misc{hu2025universaldynamicsgloballycontrolled,
      title={Universal Dynamics with Globally Controlled Analog Quantum Simulators}, 
      author={Hong-Ye Hu and Abigail McClain Gomez and Liyuan Chen and Aaron Trowbridge and Andy J. Goldschmidt and Zachary Manchester and Frederic T. Chong and Arthur Jaffe and Susanne F. Yelin},
      year={2025},
      eprint={2508.19075},
      archivePrefix={arXiv},
      optprimaryClass={quant-ph},
      url={https://arxiv.org/abs/2508.19075}, 
}

@article{Dutkiewicz_2024,
   title={{The advantage of quantum control in many-body Hamiltonian learning}},
   volume={8},
   ISSN={2521-327X},
   url={http://dx.doi.org/10.22331/q-2024-11-26-1537},
   DOI={10.22331/q-2024-11-26-1537},
   journal={Quantum},
   publisher={Verein zur Forderung des Open Access Publizierens in den Quantenwissenschaften},
   author={Dutkiewicz, Alicja and O'Brien, Thomas E. and Schuster, Thomas},
   year={2024},
   month=nov, pages={1537} }

@article{pinelis,
 ISSN = {00911798, 2168894X},
 URL = {http://www.jstor.org/stable/2244912},
 author = {Iosif Pinelis},
 journal = {The Annals of Probability},
 number = {4},
 pages = {1679--1706},
 publisher = {Institute of Mathematical Statistics},
 title = {{Optimum} {Bounds} for the {Distributions} of {Martingales} in {Banach} {Spaces}},
 urldate = {2025-09-25},
 volume = {22},
 year = {1994}
}

@book{watanabe,
 author = {Nobuyuki Ikeda and Shinzo Watanabe},
 title = {Stochastic Differential Equations and Diffusion Processes, Second Edition},
 urldate = {2025-09-30},
 year = {1981},
 publisher = {North Holland Publishing Company}
}

@article{Benoist2022deviationbounds,
  doi = {10.22331/q-2022-08-04-772},
  url = {https://doi.org/10.22331/q-2022-08-04-772},
  title = {Deviation bounds and concentration inequalities for quantum noises},
  author = {Benoist, Tristan and H{\"{a}}nggli, Lisa and Rouz{\'{e}}, Cambyse},
  journal = {{Quantum}},
  issn = {2521-327X},
  publisher = {{Verein zur F{\"{o}}rderung des Open Access Publizierens in den Quantenwissenschaften}},
  volume = {6},
  pages = {772},
  month = aug,
  year = {2022}
}

@misc{ito-concentration,
      title={Some new concentration inequalities for the It\^o stochastic integral}, 
      author={Nguyen Tien Dung},
      year={2024},
      eprint={2310.18699},
      archivePrefix={arXiv},
      optprimaryClass={math.PR},
      url={https://arxiv.org/abs/2310.18699}, 
}

@misc{bouten2006introductionquantumfiltering,
      title={An introduction to quantum filtering}, 
      author={Luc Bouten and Ramon van Handel and Matthew James},
      year={2006},
      eprint={math/0601741},
      archivePrefix={arXiv},
      primaryClass={math.OC},
      url={https://arxiv.org/abs/math/0601741}, 
}

@article{Bouten_2009,
   title={A Discrete Invitation to Quantum Filtering and Feedback Control},
   volume={51},
   ISSN={1095-7200},
   url={http://dx.doi.org/10.1137/060671504},
   DOI={10.1137/060671504},
   number={2},
   journal={SIAM Review},
   publisher={Society for Industrial & Applied Mathematics (SIAM)},
   author={Bouten, Luc and van Handel, Ramon and James, Matthew R.},
   year={2009},
   month=may, pages={239–316} }

@misc{bouten2015trotterkatotheoremquantummarkov,
      title={A Trotter-Kato Theorem for Quantum Markov Limits}, 
      author={Luc Bouten and Rolf Gohm and John Gough and Hendra Nurdin},
      year={2015},
      eprint={1409.2260},
      archivePrefix={arXiv},
      primaryClass={math-ph},
      url={https://arxiv.org/abs/1409.2260}, 
}

@article{Bouten_2008,
   title={Discrete approximation of quantum stochastic models},
   volume={49},
   ISSN={1089-7658},
   url={http://dx.doi.org/10.1063/1.3001109},
   DOI={10.1063/1.3001109},
   number={10},
   journal={Journal of Mathematical Physics},
   publisher={AIP Publishing},
   author={Bouten, Luc and Van Handel, Ramon},
   year={2008},
   month=oct }

@misc{trivedi2025noiserobustnessproblemtosimulatormappings,
      title={Noise robustness of problem-to-simulator mappings for quantum many-body physics}, 
      author={Rahul Trivedi and J. Ignacio Cirac},
      year={2025},
      eprint={2509.17579},
      archivePrefix={arXiv},
      primaryClass={quant-ph},
      url={https://arxiv.org/abs/2509.17579}, 
}

@article{bogachev1998gaussian,
  title={Gaussian measures, volume 62 of Mathematical Surveys and Monographs},
  author={Bogachev, Vladimir I},
  journal={American Mathematical Society, Providence, RI},
  volume={348},
  pages={355},
  year={1998}
}

\clearpage
\onecolumngrid
\appendix

\section{Notation and preliminary results} \label{app:notation-preliminaries}
In this appendix we review and discuss in more detail the precise assumptions that we make on the considered systems and the notation that we use to represent them. We then introduce some preliminary technical results from the literature that we will need to prove the main theorems of our work.
\subsection{Notation and assumptions}
We consider a hypercubic lattice $\mathbb{Z}^d$, in $d$ spatial dimensions. On this lattice $\mathbb{Z}^d$ we will use the $l^1$ distance, which we indicate as $d(\cdot, \cdot)$. With respect to this metric, we denote the ball of radius $R$ and center $x$ as $B_R(x)$. It contains a number of sites (i.e., a volume) of 
\begin{equation}
    \abs{B_R(x)}=\Lambda_d \, R^d \,,
\end{equation}
where $\Lambda_d = \frac{2^d}{d!}$.
With this lattice, we associate a Hilbert space given by
\begin{align}
    \mathcal{H} = \bigotimes_{x \in \mathbb{Z}^d} \mathcal{H}_x,
\end{align}
where $\mathcal{H}_x$ are local Hilbert spaces associated to each lattice site. In what follows we assume these local systems to be qubits (i.e., $\mathcal{H}_x=\mathbb{C}^2$), however, 
it should be straightforward to generalize all our results to arbitrary finite-dimensional local Hilbert spaces. When we consider linear operators on $\mathcal{H}$, we will say that an operator is supported on a certain set of lattice sites, if it has non-trivial support on the factors of the tensor product associated to these sites and acts trivially like the identity on all others. 

Let $H$ be a geometrically local Hamiltonian on 
this lattice, i.e., a Hermitian operator on $\mathcal{H}$ which can be written as
\begin{align}
    H = \sum_{\gamma \in \Gamma} H_\gamma,
\end{align}
where $\Gamma$ is a set of indices labeling local Hamiltonian terms $H_\gamma$. Each term $H_\gamma$ is local in the sense that it is supported on a set of sites restricted to a geometrically local region of constant size. This assumption is made more precise as follows.
\begin{assumption}[Geometric locality] \label{ass:locality}
    We consider local Hamiltonians of the form $H = \sum_{\gamma \in \Gamma} H_\gamma$ which satisfy the following properties:
    \begin{itemize}
        \item $\norm{H_{\gamma}}\leq 1$ for all $\gamma\in\Gamma$.
        \item There exists a constant $R$ such that each term $H_\gamma$ can be associated to a lattice site $x$ in a way that $\supp{H_\gamma}$ is contained in the ball of center $x$ and radius $R$.
        \item The mapping of the previous point associates at most a constant number $P$ of terms $H_\gamma$ to each lattice site $x$. Without loss of generality, by appropriately regrouping the terms and renormalising the Hamiltonian by a constant, we can always assume $P=1$, which we will do in what follows.
    \end{itemize}
\end{assumption}

The main task we are interested in is to 
simulate time evolution under these local Hamiltonians, 
that is under the time evolution operator
\begin{equation}
    U(t)\coloneqq e^{-iHt}\,.
\end{equation}
In particular, we are interested in the dynamics of the expectation value of local observables
\begin{align}
        \braket{O(t)} = \tr{O(t) \rho} = \tr{ U^\dag(t) \,O\, U(t) \rho}\,,
\end{align}
where $O$ is a local observable  according to 
the following definition.

\begin{define}[Local observable] \label{def:local-observable}
    A local observable $O$ is a Hermitian operator on $\mathcal{H}$ such that its support is contained in 
    a ball $B_{R_O}(x_O)$ of constant radius $R_O$ and 
    center $x_O$. 
\end{define}

Up to now we have introduced operators that are defined on the whole lattice $\mathbb{Z}^d$, however, in what follows we will also need to consider operators truncated to act on systems of a finite size, as physical implementations to simulate $U(t)$ will necessarily be realised on a finite system. In particular we would like to consider systems truncated up to a fixed maximal distance $l$ from the observable $O$, as this allows for a straightforward application of Lieb-Robinson bounds to the time evolution of observables. More precisely, for any $l>0$ let $\Omega_l\subset\mathbb{Z}^d$ be the set of all lattice sites within distance $l$ of $\supp{O}$,
defined as
\begin{equation}
    \Omega_l=\{x \;|\; d(x,\supp{O})<l\}\,
\end{equation}
where we identify $x$ with $\{x\}$ in this definition.
From the definition of $\Omega_l$, it is clear that it contains a number of sites $\abs{\Omega_l} \leq \Lambda_d \left( R_O + l\right)^d$.
We now restrict the Hamiltonian to those terms that have support overlapping with $\Omega_l$:
\begin{define}[Truncated Hamiltonian] \label{def:trunctated-Hamiltonian}
    Given a local Hamiltonian $H = \sum_{\gamma \in \Gamma} H_\gamma$, the truncated Hamiltonian $H_l$ associated with $l > 0$ and a local observables $O$ contains only the terms with (partial) support in $\Omega_l$, that is
    \begin{align}
        H_l = \sum_{\gamma \in \Theta_l} H_\gamma \hspace{10mm}\mbox{where}\hspace{5mm}\Theta_l=\{\gamma \, | \, \supp{H_\gamma}\cap\Omega_l\neq\emptyset\} \,.
    \end{align}
    We will refer to the corresponding truncated time evolution unitary as $U_l(t)=e^{-iH_lt}$. 
\end{define} 

\subsection{Locality in quantum systems}
Here, we introduce some results concerning what happens when the system introduced above is truncated to a finite system size. Firstly, we will often need to estimate the number of terms of a local Hamiltonian satisfying Assumption~\ref{ass:locality} which are relevant for the truncated Hamiltonian $H_l$.
\begin{lemma}[Truncated Hamiltonian terms]
    \label{localterms}
   For any truncation length $l>0$, the truncated Hamiltonian of 
   Definition~\ref{def:trunctated-Hamiltonian} contains a number of local Hamiltonian terms bounded by
    \begin{align}
        \abs{\Theta_l} \leq \Lambda_d \left( R_O + l + R\right)^d . 
    \end{align}
    Furthermore, the truncated Hamiltonian $H_l$ and the corresponding evolution $U_l(t)=e^{-iH_lt}$ have non-trivial support on a number of lattice sites bounded by
    \begin{equation}
        \abs{\supp{H_l}}\leq \Lambda_d \left( 2R+L+R_O\right)^d.
    \end{equation}
    If $l$ is large enough (in particular, $l \geq 2R+R_O$), then this clearly reduces to
    \begin{align}
        \abs{\Theta_l} &\leq 2^d \Lambda_d l^d, \\
        \abs{\supp{H_l}} &\leq 2^d \Lambda_d l^d.
    \end{align}
    
\end{lemma}
\begin{proof}
    The number of local terms in the truncated Hamiltonian is simply equal to the number of all terms $H_\gamma$, such that $\supp{H_\gamma} \cap \Omega_l \neq \emptyset$.
    Using Assumption \ref{ass:locality}, we can associate to each $x \in \mathbb{Z}^d$ at most one $H_{\gamma}$ with $\supp{H_\gamma} \subset B_R (x)$. We see that this $H_\gamma$ can satisfy $\supp{H_\gamma} \cap \Omega_l \neq \emptyset$ only if $x$ is within at most distance $R+L+R_O$ of the site $x_O$ introduced in Definition~\ref{def:local-observable}. Then the number of terms $\abs{\Theta_l}$ is upper bounded by the number of sites within such range, that is,
    \begin{align}
        \abs{\Theta_l} &= \abs{B_{R+L+R_O}(x_O)}  
        \leq \Lambda_d \left( R+L+R_O\right)^d.
        \nonumber
    \end{align}
    By applying Assumption \ref{ass:locality} again, we conclude that for any site $x$ within this range, the corresponding local Hamiltonian term $H_\gamma$ must have a support that extends at most to a distance $2R+L+R_O$ from $x_O$. Therefore, the total support of $H_l$ must be contained in $B_{2R+L+R_O}(x_O)$, that is
    \begin{equation}
        \abs{\supp{H_l}}\leq  \Lambda_d \left( 2R+L+R_O\right)^d.
    \end{equation}
\end{proof}
When we consider the dynamics of local observables, truncating the system can have a limited impact, if the truncation length is large enough. This can be made more precise by the Lieb-Robinson bound. We restate it here in the form that we will use in what follows.

\begin{lemma}[Truncation lemma (see Proposition 4.3 in \cite{SpeedLimits})]
    \label{lem:trunc}
    For any local operator $O$ with support $\supp{O}$, and for any $l \geq 0$, there exist positive constants $\mu,v$ that depend only on the lattice such that
    \begin{align}
        \label{lr-unit}
        \norm{U_l^\dag(t) O U_l(t) - U^\dag(t) O U(t)} \leq \abs{\supp{O}} \norm{O} \min \left( e^{-\mu l} \left( e^{\mu v t} - 1\right),1 \right).
    \end{align}
    Assuming that $H$ is defined according to Assumption \ref{ass:locality}, then $v = e \Lambda_d R^{d+1} $, $\mu = \frac{1}{R}$ (see Refs.\ \cite{klieschlocality,LRreviewchapter}).
\end{lemma}

\subsection{Concentration inequalities for random matrices and stochastic processes}

Here, we collect some useful results concerning norm bounds and concentration inequalities for random matrices and stochastic processes. We begin by introducing a result that generalizes the well-known Hoeffding inequality to sums of vector valued random variables.

\begin{lemma}[Pinelis' lemma~\cite{Pinelis}]
    \label{lem:pinelis}
    Let $X_1, \dots, X_T \in \mathbb{C}^D$ be a collection of $T$ independently distributed vector valued random variables. Consider another set of $T$ random variables $Y_1, \dots, Y_T \in \mathbb{C}^D$ defined as functions of $X_1, \dots, X_T$ and assume that, for every $t=1,\dots,T$, we have $\norm{Y_t}_2 \leq M$ and that
    \begin{align}
        &\mathbb{E}_{X_{t},\dots,X_T} \left[Y_t\right] = 0\,, \label{eq:pinelis-partial expectation}
    \end{align}
    where $\mathbb{E}_{X_{t},\dots,X_T}$ means taking the expectation value only over the random variables $X_t,\dots,X_T$. Then
    \begin{align}
        \prob{\norm{\sum_{t=1}^T Y_t}_2 > s} \leq 2 e^{-\frac{s^2}{2 TM^2}}, 
    \end{align}
\end{lemma}
\begin{proof}
    Consider the stochastic process $Z_t=\sum_{s=1}^tY_{s}$. The assumption~\eqref{eq:pinelis-partial expectation} implies that $Z_t$ is a martingale with respect to the random variables $X_1,\dots,X_t$. Indeed, from~\eqref{eq:pinelis-partial expectation} we can conclude that
    \begin{align}
        \expect{Z_{t+1}\!-\!Z_{t}\;|\; X_{1}=x_1, \dots,X_{t}=x_{t}}=\expect{Y_{t+1}\;|\; X_{1}=x_1, \dots,X_{t}=x_{t}}=0\,,
    \end{align}
    which implies that $Z_t$ has independent increments and is thus a martingale. We also have that $\sum_{t=1}^T\norm{Y_t}_2^2\leq M^2T$.
    These observations allow us to apply Theorem 3.5 of Ref.~\cite{Pinelis}, from which the lemma's statement follows.
\end{proof}

Note that this result applies specifically to the vector $2$-norm, while it does not apply to the operator norm for matrix valued random variables. When dealing with the operator norm we will instead use a slightly weaker result: 
\begin{lemma}[Matrix Azuma inequality~\cite{random}]
    \label{lem:azuma}
    Let $X_1, \dots, X_T$ be a collection of $T$ independently distributed $D\times D$ Hermitian random matrices. Consider another set of $T$  Hermitian $D\times D$ random matrices $Y_1, \dots, Y_T$ defined as functions of $X_1, \dots, X_T$ and assume that, for every $t=1,\dots,T$, we have $\norm{Y_t} \leq M$ and that
    \begin{align}
        &\mathbb{E}_{X_{t},\dots,X_T} \left[Y_t\right] = 0\,,
    \end{align}
    where $\mathbb{E}_{X_{t},\dots,X_T}$ means taking the expectation value only over the random variables $X_t,\dots,X_T$. Then 
\begin{align}
    \prob{\norm{\sum_{t=1}^T Y_t} > s} \leq 2D \, e^{-\frac{s^2}{ 8TM^2}}.
    \end{align}
\end{lemma}
\begin{proof}
    The same logic applies as in 
    the previous Lemma, except that we 
    now use the results from Ref.~\cite{random} (Section 7.2) which apply to the operator norm.  
\end{proof}

Let us now move to stochastic noise processes. Each instance of such a noise process is a square integrable real valued function on the time interval $[0,T]$, that is it belongs to the space $E \coloneqq L^2([0,T])$. On this space we define the standard norm $\|\xi\|_E^2=\int_0^T|\xi(t)|^2 \, dt$. Stochastic processes are then just stochastic distributions over this space of functions.

We consider a specific family of stochastic processes, namely centered Gaussian processes. These are characterized by their first and second moments
\begin{align}
    \expect{\xi(t)} = 0\,,  \hspace{20mm}  \expect{\xi(t) \xi(s)} =  D(t-s). 
\end{align} 
Given such a Gaussian process, we can define its corresponding \emph{covariance operator} $C$. This is a linear operator on the function space $E$ defined by
\begin{align}
    (C f) (t) := \int_0^T \!ds \, D(t-s) f(s), \label{eq:covariance-op}
\end{align}
for every $f\in E$. 
Through $C$, we can now define a further concept related to this Gaussian process, namely the \emph{Cameron-Martin norm} of a process $f\in E$, which is  defined as
\begin{equation}
    \norm{f}_{CM} := \norm{C^{-\frac{1}{2}} f}_E.
\end{equation}
The Cameron-Martin space $E_{CM}$, relative to the stochastic process, is the space of all function $h\in E$ such that $\|h\|_{CM}<\infty$. Note that we have for every $h \in E_{CM}$:
\begin{align}
    \norm{h}_E = \norm{C^{\frac{1}{2} }C^{-\frac{1}{2}}  h}_E  \leq \sqrt{\norm{C}_{\rm op}} \hspace{1mm} \norm{h}_{CM}, \label{eq:CM-vs-L2-norms}
\end{align}
where $\norm{C}_{\rm op}$ is the operator norm of $C$ as a linear operator on $E$.

Note that the same construction applies to the slightly more general case of a space $E$ of vector-valued processes, which we will encounter in Section~\ref{app:analog-average-case}: it suffices to define the corresponding norm $\norm{\xi}_E^2=\int_0^T\sum_\sigma\abs{\xi_\sigma(t)}^2$. In general, the operator $C$ will be defined as $(C f)_\sigma (t) := \int_0^T \!ds \,\sum_{\sigma'} D_{\sigma\sigma'}(t-s) f_{\sigma'}(s)$, where $D_{\sigma\sigma'}(t-s)=\expect{\xi_\sigma(t)\xi_{\sigma'}(s)}$. However, in what follows we will always focus on the diagonal case $D_{\sigma\sigma'}(t)=D(t)\delta_{\sigma\sigma'}$, whose operator norm is the same as the one of~\eqref{eq:covariance-op}.

With the definition of the Cameron-Martin norm, we can now introduce the notion of Lipschitz continuous functions on $E_{CM}$. These functions satisfy the following concentration property. 
\begin{lemma}[Gaussian concentration for Cameron-Martin Lipschitz functionals]
\label{lem:cm-concentration}
Let $\xi$ be a centered Gaussian stochastic process with corresponding Cameron--Martin norm $\norm{\cdot}_{CM}$.
Let $F$ be a measurable real valued function on $E$ for which there exists $L \geq 0$ such that for all $h \in E_{CM}$,
\begin{align}
    \abs{F(\xi+h) - F(\xi)} \leq L \norm{h}_{CM}
    \qquad \text{almost surely.}
    \label{eq:cm-lip}
\end{align}
Then for all $s \geq 0$,
\begin{align}
    \prob{\abs{F(\xi) - \mu} \geq s + \sigma}
    \leq  \exp\!\left(-\frac{s^2}{2 L^2}\right),
    \label{eq:cm-conc}
\end{align}
where $\mu:=\expect{F(\xi)}$ and $\sigma^2 := \expect{(F(\xi) - \mu)^2}$.
\end{lemma}
\begin{proof}
    We adapt here a statement proven in Ref.~\cite{bogachev1998gaussian} for the concentration around the median $\mathrm{med}(F(\xi))$. Let us set $\mu :=  \expect{(F(\xi))}$ and $m := \mathrm{med}(F(\xi))$. Then, Theorem 4.5.6 of Ref.~\cite{bogachev1998gaussian} states that
    \begin{align}
    \prob{\abs{F(\xi) - m} \geq s}
    \leq
    \exp\!\left(-\frac{s^2}{2 L^2}\right).
    \end{align}
    We further have that 
    \begin{align}
        \abs{\mu - m} = \abs{\expect{F(\xi) - m}} \leq \expect{\abs{F(\xi) - m}} \leq \expect{\abs{F(\xi)-\mu}} \leq \sigma,
    \end{align}
    where $\sigma = \sqrt{\expect{(F(\xi) - \mu)^2}}$. We used here that the median is defined as the minimum of $c \mapsto \expect{\abs{F(\xi) - c}}$. Then, $\abs{F(\xi) - m} \geq \abs{F(\xi) - \mu} - \abs{m-\mu}\geq  \abs{F(\xi) - \mu} - \sigma$, and so
    \begin{align}
          \prob{\abs{F(\xi) - \mu} \geq s+\sigma} \leq \prob{\abs{F(\xi) - m} \geq s} \leq  \exp\!\left(-\frac{s^2}{2 L^2}\right). 
    \end{align}
\end{proof}

\subsection{Further helpful lemmas}

We state here a series of helpful technical results, which we will repeatedly use in the following derivations.
The following expansion is convenient for dealing with expressions written as products of matrices.
\begin{lemma}[Telescope product]
    \label{telesc}
    Let $A = \prod_{i=1}^N A_i$, $B = \prod_{j=1}^N B_j$ be two products of $k \times k $ matrices.
    Then
\begin{align}
        A - B = \sum_{i= 1}^N \prod_{j = 1}^{i-1} A_j (A_i - B_i) \prod_{k = i+1}^N B_k.
    \end{align}
\end{lemma}
\begin{corollary}
\label{cor:telescopic}
    Let $A = \prod_{i=1}^N A_i$, $B = \prod_{j=1}^N B_j$ be two products of $k \times k $ unitary matrices. Then
    \begin{align}
        \norm{A - B} \leq \sum_{i= 1}^N \norm{A_i - B_i}.
    \end{align}
\end{corollary}

\begin{proof}
    Using that the spectral norm is unitarily invariant, by application of Lemma \ref{telesc} and with the triangle inequality, we arrive at
    \begin{align}
    \begin{split}
        \norm{A - B} = \norm{\prod_{i=1}^N A_i - \prod_{j=1}^N B_j} \leq &\norm{ \sum_{i= 1}^N \prod_{j = 1}^{i-1} A_j (A_i - B_i) \prod_{k = i+1}^N B_k} \\ & \leq 
        \sum_{i=1}^N \prod_{j=1}^{i-1} \norm{A_j} \norm{A_i - B_i} \prod_{k = i+1}^N \norm{B_k}  \leq \sum_{i=1}^N \norm{A_i - B_i}.
    \end{split}
    \end{align}
\end{proof}
A standard result from the theory of ordinary differential equations is Duhamel's formula:
\begin{lemma}[Duhamel's formula]
\label{duhamel}
Let $H(t), K(t)$ be two bounded, time-dependent Hamiltonians and $O$ a Hermitian observable. Let $U(t,s)$ and $V(t,s)$ be the time evolution operators, from time $s$ to time $t$, defined by the Hamiltonians $H(t), K(t)$ respectively. We further set $U(t)\equiv U(t,0)$ and $V(t)\equiv V(t,0)$. Then the following relations hold:
\begin{align}
    U(t) - V(t) = - i\int_0^t ds\;  V(t,s) (H(s) - K(s)) U(s,0).
\end{align}
As a corollary:
\begin{align}
    V^\dag(t) O V(t) - U(t)^\dag O U(t)&= i\int_0^t \!ds\;  V(s)^\dag [K(s) - H(s), U(t,s)^\dag O U(t,s)] V(s)\,. \label{duhamel-eq}
\end{align}
\end{lemma}
We now introduce a way to compute operator norms of integral operators:
\begin{lemma}[Schur test~\cite{Schur1911}]
    \label{lem:schur}
    Let $K : [0,t]\times[0,t] \to \mathbb{C}$ be measurable and define the integral operator
    \begin{align}
        (Tf)(s) \coloneqq \int_0^t K(s,s') f(s') ds',
        \qquad f \in L^2([0,t]).
    \end{align}
    Assume
    \begin{align}
        M \coloneqq \sup_{s \in [0,t]} \int_0^t \abs{K(s,s')} ds' < \infty,
        \qquad
        M' \coloneqq \sup_{s' \in [0,t]} \int_0^t \abs{K(s,s')} ds < \infty.
    \end{align}
    Then $T$ is bounded on $L^2([0,t])$ and
    \begin{align}
        \norm{T}_{\rm op} \leq \sqrt{M M'}.
    \end{align}
    In particular, if $K(s,s') = k(s-s')$ with $\abs{k}$ even, then $M = M'$ and
    \begin{align}
        \norm{T}_{\rm op} \leq M = \sup_{s \in [0,t]} \int_0^t \abs{k(s-s')} ds'.
    \end{align}
\end{lemma}

The difference in the expectation value of a given operator on two different states can be related to the Hilbert space distance of the states in the following way.
\begin{lemma}[Hilbert Schmidt distance]
    \label{onlypure}
    Consider an observable $O$ and two states $\rho$, $\rho'$. These could be pure states or mixed states obtained by evolving the same initial state with different unitary evolutions, i.e., $\rho = U\rho_0 U^\dag =  \sum_k p_k \ketbra{\Psi_k}{\Psi_k}$ and $\rho' = U'\rho_0 U'^\dag =  \sum_k p_k \ketbra{\Psi_k'}{\Psi_k'}$. Then
    \begin{align}
         \abs{\tr{O \rho} - \tr{O\rho'}} &\leq 2\norm{O} \;\sup_{k} \norm{\ket{\Psi_k} -  \ket{\Psi'_k}}_2,
    \end{align}
    where $\norm{\,\cdot\,}_2$ is the canonical norm of Hilbert space vectors. 
\end{lemma}
\begin{proof}
Consider the case where the inputs are pure
\begin{align}
    \abs{\tr{O \rho} - \tr{O\rho'}}  &\leq \norm{O} \norm{\ketbra{\Psi}{\Psi} - \ketbra{\Psi'}{\Psi'}}_{1} \\
    \nonumber
    &\leq  \norm{O} \: 2\,\sqrt{1-|\braket{\Psi|\Psi'}|^2}\\
    \nonumber
    &\leq  2\norm{O} \,\sqrt{1-(\Re\braket{\Psi|\Psi'})^2}\\
    \nonumber
    & =    2\norm{O} \,\sqrt{1-\frac{1}{4}{\left(2-\norm{\ket{\Psi} - \ket{\Psi'}}_{2}^2\right)}^2}\\
    \nonumber
    & =    2\norm{O} \, \norm{\ket{\Psi} - \ket{\Psi'}}_{2} \, \sqrt{1-\frac{1}{4}\norm{\ket{\Psi} - \ket{\Psi'}}_{2}^2} \\
    \nonumber
    &\leq  2 \norm{O} \, \norm{\ket{\Psi} - \ket{\Psi'}}_{2},
    \nonumber
\end{align}
where in the first step we have used Hölder's inequality, in the second we have used the Fuchs-van de Graaf inequality and in the fourth step we have used that $\norm{\ket{\Psi} - \ket{\Psi'}}_{2}^2=2(1-\Re\braket{\Psi|\Psi'})$ which implies $\Re\braket{\Psi|\Psi'}=\frac{1}{2}(2-\norm{\ket{\Psi} - \ket{\Psi'}}_{2}^2)$. 
Applying the same derivation to the mixed states $\rho = \sum_k p_k \ketbra{\Psi_k}$ and $\rho' = \sum_k p_k \ketbra{\Psi'_k}$, we have
\begin{align}
    \abs{\tr{O \rho} - \tr{O\rho'}} &\leq \norm{O} \norm{\rho - \rho'}_1 \\
    \nonumber
    &\leq \norm{O}  \sum_k \norm{p_k \ketbra{\Psi_k}{\Psi_k} -  \ketbra{\Psi'_k}{\Psi'_k}}_1 \\
    \nonumber& \leq 2 \norm{O}\sum_k p_k \norm{\ket{\Psi_k} -  \ket{\Psi'_k}}_2. 
    \nonumber
\end{align}
Now using that $\sum_k p_k = 1$ and $p_k \geq 0$ for all $k$ shows that
\begin{align}
    \abs{\tr{O \rho} - \tr{O\rho'}}  \leq 2 \norm{O} \sup_{k} \norm{\ket{\Psi_k} -  \ket{\Psi'_k}}_2,
\end{align} 
which leads to the statement to be shown.
\end{proof}
\section{Stability of analog quantum simulation} \label{app:analog-worst-case}
In this appendix we prove our stability results for analog simulation under worst-case noise (i.e., Theorem~\ref{thm:wcas}). For this, we will use the notation and technical lemmas introduced in detail in Appendix~\ref{app:notation-preliminaries}.

\subsection{Proof of Theorem \ref{thm:wcas}}
\vspace{-6pt}
We now prove Theorem \ref{thm:wcas}, which we restate here for convenience.
\setcounter{theorem}{0}
\begin{theorem}[Restated, upper bound for worst case errors in analog simulators]
    Consider a perturbed analog time evolution $V(t)$ defined by the time-dependent Hamiltonian 
    \begin{align}
        H'(s)=\sum_\gamma H_\gamma'(s),
    \end{align}
    where $\norm{H'_\gamma(s) - H_\gamma} \leq \delta$ for all $s<t$ and all $\gamma\in\Gamma$. We further assume that each local term $H'_\gamma(s)$ always has the same support as the corresponding $H_\gamma$.
    
    Then, the error on the time-evolution of a local observable $O$ is at most
    \begin{align}
        \Delta \leq \BOO{t^{d+1} \delta}.
    \end{align}
\end{theorem}
\begin{proof}
    Considering the definition of $\Delta$ and applying Lemma~\ref{duhamel} gives
    \begin{align}
        \Delta = \norm{V^\dag(t) O V(t) - U^\dag(t) O U(t) } = \norm{\int_0^t \!ds \; V(t,s)^\dag \left[\sum_{\gamma} \tilde{H}_{\gamma}(s), O(s)\right] V(t,s)} \leq \int_0^t ds \sum_{\gamma} \norm{\left[\tilde{H}_{\gamma}(s), O(s)\right]},
    \end{align}
    where $\tilde{H}_{\gamma}(s)=H'_{\gamma}(s)-H_\gamma$ and $O(t)=U^\dag(t) O U(t)$. By assumption, each $\tilde{H}_{\gamma}(s)$ is supported on some fixed region of the lattice throughout its time evolution. We can thus apply a formulation of the Lieb-Robinson theorem (see for instance \cite{SpeedLimits}, Theorem 3.11) to bound each term $\norm{\left[\tilde{H}_{\gamma}(s), O(s)\right]}$ as
    \begin{align}
        \norm{[\tilde{H}_{\gamma}(s), O(s)]} &\leq \norm{O}\norm{\tilde{H}_{\gamma}(s)} \abs{\supp{O}}  \min\Big( \left(e^{\mu v s} - 1\right) e^{-\mu l_{\gamma}},1 \Big) \\\
        &\leq \delta \norm{O}\abs{\supp{O}} \min\Big( e^{-\mu(l_\gamma-vt)}, 1\Big) , \label{eq:delta-bound-wcas}
    \end{align}
    where $l_\gamma$ is the distance between the supports of $H_\gamma$ and $O$. Here we used that $\norm{\tilde{H}_\gamma(s)}\leq\delta$ and $s<t$. 
   
   We can now split this sum over $\gamma$ into two contributions. For this let us set $l\coloneqq vt$. If $\gamma\in\Theta_l$ then, by Definition~\ref{def:trunctated-Hamiltonian}, $l_{\gamma} \leq l = vt$ and thus
\begin{align}
    \sum_{\gamma\in\Theta_l} \min\Big( e^{-\mu(l_\gamma-vt)}, 1\Big) \leq \abs{\Theta_l}  \leq 2^d \Lambda_d (vt)^d,  \label{eq:first-sum_wcas}
\end{align}
where we applied Lemma~\ref{localterms}. If, on the other hand, $\gamma\notin \Theta_l$, then by Assumption~\ref{ass:locality} this corresponds at most to summing over all lattice sites at distance from $\supp{O}$ greater than $vt$. We then have
\begin{align}
    \sum_{\gamma\notin \Theta_l} e^{-\mu (l_\gamma-vt)} \leq \int \! d^d x \, \norm{x}^{d-1} \, e^{-\mu \norm{x}}  \leq K_d ,
\end{align}
for a constant $K_d \leq \Gamma(d)/\mu^d$.

Substituting this into Eq.~\eqref{eq:delta-bound-wcas}, the total error is bounded by
\begin{align}
    \Delta &\leq \delta \norm{O} \abs{\supp{O}} \int_0^t ds \left(2^d \Lambda_d v^d t^d + K_d\right) \\
    &=   \delta t^{d+1} \norm{O} \abs{\supp{O}} v^d \left(2^d \Lambda_d  +\frac{K_d}{(vt)^d}\right) \\ & \leq M  \delta t^{d+1}
\end{align}
with $M \coloneqq \norm{O} \abs{\supp{O}} v^d \left(2^d \Lambda_d  +K_d\right)$, assuming that $vt > 1$. This completes the proof.
\end{proof}
\begin{rem}[Analog simulation in finite size devices] \label{rem:wcas}
    In the previous Theorem~\ref{thm:wcas}, we have considered a noisy analog simulator evolution $V(t)$ implemented on the full lattice $\mathbb{Z}^d$. In a practical scenario, this will rather be implemented on a system truncated to a finite size $l$, leading to the simulator evolution $V_l(t)$.
    In this case, the error $\Delta$ can be split in two contributions (as we will discuss more in detail for example in the proof of Theorem~\ref{wcds}). One is the finite size perturbation term $\norm{V^\dag_l(t) O V_l(t) - U_l^\dag(t) O U_l(t) }$ and the other is the truncation error $\norm{U_l^\dag(t) O U_l(t) - U(t) O U(t)}$. With a reasoning analogous to the proof above, the first term can be seen to scale as $\BOO{\delta t \, l^d}$ if we choose $l\geq vt$ (indeed only the term~\eqref{eq:first-sum_wcas} will contribute to the sum over $\gamma$). The second term can be bounded with the help of Lemma~\ref{lem:trunc} by $\BOO{e^{-\mu(l-vt)}}$.

    It follows that a choice of truncation length $l = vt - \frac{1}{\mu} \log(\delta t^{d+1})$ balances these two error terms, achieving an optimal final scaling which, up to logarithmic factors, is equal to the one found in Theorem~\ref{thm:wcas}:
    \begin{align}
        \Delta \leq \BOO{\delta t^{d+1} \left(1-\frac{1}{\mu v t}\log{\delta t^{d+1}}\right)^d}\leq \BOO{\delta t^{d+1} \log^d{\left(\frac{1}{\delta t^{d+1}}\right)}}.
    \end{align}
\end{rem}

\section{Stability of analog quantum simulation under stochastic errors}\label{app:analog-average-case}
In this appendix we prove our stability results for analog quantum simulation under stochastic errors. We will first introduce some intermediate results on random Gaussian processes and on Ito calculus derivations for white noise processes. We then proceed to prove Theorems~\ref{acac}, \ref{ito-acas} and~\ref{acas}. For this, we will use the notation and preliminary lemmas introduced in 
detail in Appendix~\ref{app:notation-preliminaries}.

\subsection{Perturbations generated by Gaussian processes}
    As discussed in Section~\ref{sec:average-analog}, we consider an analog simulator where the the Hamiltonian is perturbed by a noise process of the form
    \begin{align}
        L_{\gamma}(t) = \sum_{a=1}^m \xi_{\gamma, a} (t) X_{\gamma, a}.
    \end{align}
    Here, for every $\gamma\in\Theta_l$, $\{X_{\gamma,a}\}_{a=1}^m$ is a set of $m$ Hermitian operators supported on $\supp{H_\gamma}$ and with $\norm{X_{\gamma, a}} \leq 1$. These operators could correspond, for instance, to a basis of the operators supported on $\supp{H_\gamma}$ (in which case $m=2^{2 \Lambda_d R^d -1}$). However in general we just assume that there is a constant number $m$ of them. In what follows, to simplify notation, we will combine the indices $(\gamma,a)$ into a single index $\sigma$, which then runs from $1$ to $m\abs{\Theta_l}$.

    We further assume $\xi_{\sigma} (t)$ to be uncorrelated stationary Gaussian processes with correlation function $D(t)$. That is we have 
\begin{align}
    \expect{\xi_{\sigma}(t)} = 0\,,  \hspace{20mm}  \expect{\xi_\sigma(t) \xi_{\sigma'}(s)} =  \delta_{\sigma\sigma'} D(t-s). 
\end{align} 

    With these definitions, we have that the states of the simulator system will evolve under the stochastic Schr\"odinger equation
    \begin{align}
        \label{st-sch}
        \frac{d}{dt} \ket{\psi(t)} = -i \left(H + \delta\sum_{\sigma=1}^{m\abs{\Theta_l}} \xi_\sigma(t) X_\sigma \right)\ket{\psi(t)}.
    \end{align}
    As a first step, we use a Dyson series expansion to find a simple expectation value perturbation bound for this time-evolution. 

\begin{lemma}[Variance of the expected error under Gaussian noise]
\label{gaussian_noise}
    Consider an initial state $\ket{\psi}$ and let $\ket{\psi(t)}$ be the evolved state under the stochastic Schr\"odinger equation~\eqref{st-sch}, while $\ket{\psi_0(t)}$ is the same state evolved under the unperturbed evolution generated by $H$.
    Then the variance of the distance between these states under Gaussian noise processes $\xi_\sigma(t)$ is bounded by:
      \begin{align}
        \label{res-stoch}
         \expect{\norm{\ket{\psi(t)} - \ket{\psi_0(t)}}_2^2} \leq  \exp\left[ \frac{\delta^{2}m\abs{\Theta_l}}{2}  \int_0^t ds\int_0^t ds' D(s-s')\right] -1.
    \end{align}
\end{lemma}
\begin{proof}
    The formal solution of the equation~\eqref{st-sch} for $\ket{\psi (t)}$ can be represented by a time-ordered exponential series
    \begin{align}
        \label{dyson}
        \ket{\psi(t)}_I &= \mathcal{T}\exp\left(-i\delta \sum_\sigma \int_0^t ds X^I_\sigma(s) \xi_\sigma(s)\right) \ket{\psi(0)} \\ & =\sum_{k=0}^\infty \frac{(-i)^k}{k!} \delta^k \sum_{\sigma_1, \dots \sigma_k} \int_0^t ds_1 \cdots \int_0^t ds_k  \prod_{j=1}^k\xi_{\sigma_j} (s_j) \; \mathcal{T}\prod_{j=1}^k X^I_{\sigma_j} (s_j)\;\ket{\psi(0)},
    \end{align} 
    where we used the interaction picture representation ($\ket{\psi(t)}_I=e^{iHt}\ket{\psi(t)}$ and $X^I_\sigma(t)=e^{iHt}X_\sigma e^{-iHt}$) and expanded in orders of $\delta$.

    Noting that inner products in the interaction picture are equivalent to the ones in the regular Schr\"odinger picture and that $\ket{\psi_0(t)}_I=\ket{\psi(0)}$, we thus have 
    \begin{align}
        \expect{\norm{\ket{\psi(t)} - \ket{\psi_0(t)}}_2^2} & = 2-2\Re{\expect{\braket{\psi_0(t)|\psi(t)}_I}} \\ 
        &\hspace{-20mm}= \sum_{k=1}^\infty \frac{\delta^k}{k!}  \sum_{\sigma_1, \dots \sigma_k}  \int_0^t ds_1 \cdots \int_0^t ds_k \; \expect{\prod_{j=1}^k \xi_{\sigma_j} (s_j)} \Re\left[ (-i)^k\braket{\psi(0)|\mathcal{T}\prod_{j=1}^k X^I_{\sigma_j}(s_j)|\psi(0)}\right]\\
        &\hspace{-20mm}\leq \sum_{k=1}^\infty \frac{\delta^k}{k!}  \sum_{\sigma_1, \dots \sigma_k}  \int_0^t ds_1 \cdots \int_0^t ds_k \; \expect{\prod_{j=1}^k \xi_{\sigma_j} (s_j)}.
    \end{align}
    where in the last step we used that $\norm{X_\sigma^I(s)}\leq1$.
    
    We can now apply Wick's Theorem to compute the expectation value of the product of Gaussian processes. First, we observe that $\expect{\prod_{j=1}^k \xi_{\sigma_j} (s_j)} = 0$ if $k$ is odd. Thus, only terms corresponding to an even $k=2q$ contribute. Each of these terms can be written as a sum of products of two point correlations, corresponding to all possible contractions of the sequence of $\xi_\sigma$'s. There are in total $\frac{(2q)!}{2^q q!}$ such contractions and each one leads to a term of the form
    \begin{align}
        \sum_{\sigma_1, \dots \sigma_{2q}}  \int_0^t ds_1 \cdots \int_0^t ds_{2q} \; &\expect{\xi_{\sigma_{j_1}}(s_{j_1})\, \xi_{\sigma_{j_2}}(s_{j_2})} \cdots \expect{\xi_{\sigma_{j_{2q-1}}}(s_{j_{2q-1}})\, \xi_{\sigma_{j_{2q}}}(s_{j_{2q}})} \\
        &=\sum_{\sigma_1, \dots \sigma_{2q}}  \int_0^t ds_1 \cdots \int_0^t ds_{2q} \; \delta_{\sigma_{j_1}\sigma_{j_2}} D(s_{j_1}-s_{j_2}) \cdots \delta_{\sigma_{j_{2q-1}}\sigma_{j_{2q}}} D(s_{j_{2q-1}}-s_{j_{2q}})\\
        &=\left[ \sum_{\sigma\sigma'} \delta_{\sigma\sigma'}\int_0^t ds\int_0^t ds' D(s-s')\right]^{q}\\
        &=\left[ m\abs{\Theta_l} \int_0^t ds\int_0^t ds' D(s-s')\right]^{q}
    \end{align}
    Putting this together, the expected deviation is 
    \begin{align}
        \expect{\norm{\ket{\psi(t)} - \ket{\psi_0(t)}}_2^2} \leq  \sum_{q=1}^\infty \frac{\delta^{2q}}{2^q q!} \left[m\abs{\Theta_l} \int_0^t ds\int_0^t ds' D(s-s')\right]^{q},
    \end{align}
    which leads to the final statement after resumming of the exponential series.
\end{proof}

To derive how the noisy evolution will concentrate around its expected value, we can use Lemma \ref{lem:cm-concentration}. To apply it we need to first compute the Cameron-Martin Lipschitz constant of the relevant observable function. This is done in the following Lemma.
\begin{lemma}[Cameron--Martin Lipschitz bound for the noisy expectation value]
\label{lem:cm-lip-F}
Let $\ket{\psi(t,\xi)}$ be the solution of the stochastic Schr\"odinger equation~\eqref{st-sch} for a fixed realisation $\xi(t)$ of the noise process and let $F_t(\xi)=\braket{\psi(t,\xi)|O|\psi(t,\xi)}$.
Then the function $F_t(\xi) $ is Lipschitz continuous with Lipschitz constant $L$  with respect to the Cameron-Martin norm:
\begin{align}
\abs{F_t(\xi+h) - F_t(\xi)} \leq L\,\norm{h}_{CM},
\qquad
L \coloneqq 2\,\norm{O}\,\delta\,\sqrt{m\abs{\Theta_l} t \sup_{s \in [0,t]} \int_0^t \abs{D(s-s')} ds'}
\end{align}
for all perturbations $h(t)$ with $\norm{h}_{CM}<\infty$.
\end{lemma}

\begin{proof}
Fix $h \in E_{CM}$ and consider Eq.~\eqref{st-sch} for the two noise processes $\xi$ and $\xi+h$. We define $U_\xi(t)$ and $U_{\xi+h}(t)$ as the time evolution operators corresponding to these two evolution equations and we further set $\Delta H(t) = H_{\xi+h}(t) - H_\xi(t) = \delta \sum_{\sigma} h_\sigma(t)\,X_\sigma$ to be the difference between the Hamiltonian operators appearing in the two evolutions. It follows that
$\ket{\psi(t,\xi)} = U_\xi(t)\ket{\psi}$ and $\ket{\psi(t,\xi+h)} = U_{\xi+h}(t)\ket{\psi}$ and we have
\begin{align}
    \abs{F_t(\xi+h) - F_t(\xi)}
    &= \abs{\bra{\psi(t,\xi+h)} O \ket{\psi(t,\xi+h)} - \bra{\psi(t,\xi)} O \ket{\psi(t,\xi)}} \\
    &
    \leq 2\,\norm{O}\,\norm{\ket{\psi(t,\xi+h)} - \ket{\psi(t,\xi)}}_2
    \\
    &\leq 2\,\norm{O}\,\norm{U_{\xi+h}(t) - U_\xi(t)}.
\end{align}
We apply Lemma \ref{duhamel} to each path of the stochastic process to expand the difference between $ U_{\xi+h}(t)$ and $U_\xi(t)$.\TG{The current statement of Lemma 7 is not compatible with the one below. Let's fix this without breaking other parts of the paper.}
\begin{equation}
    U_{\xi+h}(t) - U_\xi(t) = - i \int_0^t U_{\xi+h}(t,s)\,\Delta H(s)\,U_{\xi}(s)\,ds.
\end{equation}
Hence, by unitarity, $\norm{U_{\xi+h}(t) - U_\xi(t)} \leq \int_0^t \norm{\Delta H(s)} ds \leq \delta \int_0^t \sum_\sigma \abs{h_\sigma(s)} ds$.
Applying the Cauchy--Schwarz inequality first in $\mathbb{R}^{m\Theta_l}$ and then in $L^2([0,t])$ yields
\begin{align}
    \int_0^t \; \sum_{\sigma=1}^{m\abs{\Theta_l}} \abs{h_\sigma(s)} \, ds
    \leq \sqrt{m\abs{\Theta_l}} \int_0^t \,\left(\sum_{\sigma=1}^{m\abs{\Theta_l}} \abs{h_\sigma(s)}^2\right)^{1/2} \!\!\!ds
    \leq \sqrt{m\abs{\Theta_l} t}\left(\int_0^t \sum_{\sigma=1}^{m\abs{\Theta_l}} \abs{h_\sigma(s)}^2 ds\right)^{1/2}
    = \sqrt{m\abs{\Theta_l} t}\,\norm{h}_E.
\end{align}
Therefore $\abs{F_t(\xi+h) - F_t(\xi)} \leq 2\,\norm{O}\,\delta\,\sqrt{m\abs{\Theta} t}\,\norm{h}_E$. Using the Cameron--Martin embedding
$\norm{h}_E \leq \sqrt{\norm{C_D}}\,\norm{h}_{CM}$ (see Eq.~\eqref{eq:CM-vs-L2-norms}) and the Schur bound $\norm{C_D} \leq \sup_{s \in [0,t]} \int_0^t \abs{D(s-s')} ds'$
(Lemma~\ref{lem:schur}) gives the claimed Lipschitz constant $L$.
This completes the proof.
\end{proof}

\subsection{Proof of Theorem \ref{acac}}
We repeat the statement of Theorem \ref{acac} here. \TG{In the proof of the theorem mention explicitly when we are using Lemma 5.} \JRR{Done!}
\setcounter{theorem}{3}
\begin{theorem}[Restated, Average case bounds for errors in analog simulators with Gaussian perturbations]
    Consider a perturbed analog time evolution given by the Hamiltonian
    \begin{align}
        H'(t)=\sum_{\gamma\in\Gamma} \left(H_\gamma + \delta  \sum_{a=1}^m \xi_{\gamma,a}(t) X_{\gamma,a}\right)\,,
    \end{align}
    where $t\mapsto \xi_{\gamma,a}(t)$ are uncorrelated Gaussian noise processes with time correlation function given by~\eqref{eq:noise-time-correlation}. Assume that the initial state is a given pure state $\rho=\ket{\psi}\!\bra{\psi}$.
    Then, the error on the time-evolution of a local observable $O$, on average over the noise realizations, has expectation:
    \begin{align}
        \expect{\Delta (\rho)} \leq \BOO{\sqrt{\lambda}\,\delta t^{\frac{d+1}{2}}\log^{\frac{d}{2}}\left(\frac{1}{\delta t^{\frac{d+1}{2}}}\right)}\,,
    \end{align}
    and
   \begin{align}
        \prob{   \Delta(\rho)\geq  \BOO{s \, \sqrt{\lambda} \delta  t^{\frac{d+1}{2}} \log^{\frac{d}{2}}\left(\frac{1}{ \delta t^{\frac{d+1}{2}}}\right)}} \leq 2 e^{-s^2}. \label{eq:acac-prob-statement}
    \end{align}
    In the case the infinite correlation length (\ie $\lambda \rightarrow +\infty$) we instead have
        \begin{align}
        \expect{\Delta (\rho)} \leq \BOO{\sqrt{\lambda}\,\delta t^{\frac{d}{2}+1}\log^{\frac{d}{2}}\left(\frac{1}{\delta t^{\frac{d}{2}+1}}\right)}\,,
    \end{align}
    and
    \begin{align}
       \prob{\Delta(\rho) \geq \BOO{s\sqrt{\lambda} \delta t^{\frac{d}{2}+1}\left(1-\frac{\log \delta  t^{\frac{d}{2}+1} }{\mu v t}\right)^{\frac{d}{2}}}} \leq 2e^{-s^2}.
    \end{align}
\end{theorem}
\begin{proof}[Proof of Theorem \ref{acac}]
    Let $\rho'(t)=\ket{\psi'(t)}\!\bra{\psi'(t)}$ and $\rho(t)=U_l(t)\rho U^\dag_l(t)=\ket{\psi(t)}\!\bra{\psi(t)}$ be the states evolved respectively under the perturbed and unperturbed evolutions on the system of size $l$. We consider the definition of $\Delta(\rho)$ and split the error into two contributions. To the first term we apply Lemma~\ref{onlypure} and to the second term we apply Lemma~\ref{lem:trunc}:
    \begin{align}
        \Delta(\rho) &\leq \abs{\tr{O\rho'(t) - O\,  U(t)\rho U^\dag(t)}} \\
        &\leq \abs{\tr{O\rho'(t) - O \rho(t)}} + \abs{\tr{O\,U_l(t)\rho U^\dag_l(t) - O \, U(t)\rho U^\dag(t)}}
        \\ &\leq 2 \norm{O} \norm{\ket{\psi'(t)}-\ket{\psi(t)}}_2  +  \abs{\supp{O}} \norm{O} e^{-\mu(l- vt)}. \label{eq:Delta-stoch-noise}
    \end{align}

    The only stochastic term in $\Delta(\rho)$ is therefore $\norm{\ket{\psi'(t)}-\ket{\psi(t)}}_2$. We can compute the expectation value of this quantity by applying Jensen's inequality and Lemma \ref{gaussian_noise} (where we assume that the exponent of expression~\eqref{res-stoch} is small enough such that we can linearize it).
    \begin{align}
        \expect{\norm{\ket{\psi'(t)} - \ket{\psi(t)}}_2} &\leq \sqrt{\expect{\norm{\ket{\psi'(t)} - \ket{\psi(t)}}_2^2}}\\
        &\leq \sqrt{\frac{\delta^{2}m\abs{\Theta_l}}{2}  \int_0^t \!ds\!\int_0^t\!ds' \; D(s-s')}\\
        &\leq \delta l^{\frac{d}{2}} \sqrt{\frac{2^d m \Lambda_d}{2}} \sqrt{\int_0^t \!ds\!\int_0^t\!ds' \; D(s-s')}\,,
    \end{align}
    where in the last step we have used Lemma~\ref{localterms}.

    Now we combine this with equation~\eqref{eq:Delta-stoch-noise} and, in order to cancel the exponential scaling in $t$ in the last term, we choose $l=vt - \frac{1}{\mu }\log\varphi$, where $\varphi$ will be specified later:
    \begin{align}
        \expect{\Delta(\rho)} &\leq 2 \norm{O} \sqrt{\frac{2^d m \Lambda_d}{2}} v^{\frac{d}{2}} \: \delta  t^{\frac{d}{2}} \left(1-\frac{1}{\mu v t}\log\varphi\right)^{\frac{d}{2}}\sqrt{\int_0^t \!ds\!\int_0^t\!ds' \; D(s-s')}  +  \abs{\supp{O}} \norm{O} \,\varphi\,. \label{eq:Delta-acac}
    \end{align}

    We can now evaluate this quantity for different choices of the covariance function $D(t)$. In particular, if $D(t)=e^{-\frac{t^2}{2\lambda^2}}$ we have
    \begin{align}
        \int_{0}^t \!ds \!\int_0^t\! ds' D(s-s')=\int_{0}^t \!ds \!\int_0^t\! ds' e^{-\frac{(s-s')^2}{2\lambda^2}} \leq \int_{0}^t \!ds \!\int_{-\infty}^{+\infty}\! ds' e^{-\frac{(s')^2}{2\lambda^2}} = \sqrt{2\pi} \lambda t\,.
    \end{align}
    Substituting this into~\eqref{eq:Delta-acac} implies
    \begin{align}
        \expect{\Delta(\rho)} &\leq 2 \norm{O} \sqrt{\frac{2^d m \Lambda_d \sqrt{2\pi}\lambda }{2}} v^{\frac{d}{2}} \: \delta  t^{\frac{d+1}{2}} \left(1-\frac{1}{\mu v t}\log\varphi\right)^{\frac{d}{2}}  +  \abs{\supp{O}} \norm{O} \,\varphi\,.
    \end{align}
    Then, making a choice of system truncation size given by $\varphi=\delta  t^{\frac{d+1}{2}}$ leads to 
    \begin{align}
        \expect{\Delta(\rho)} &\leq \delta  t^{\frac{d+1}{2}} \left(1-\frac{\log \delta  t^{\frac{d+1}{2}} }{\mu v t}\right)^{\frac{d}{2}} \norm{O} \left[ 2\sqrt{\frac{2^{d} m \Lambda_d \sqrt{2\pi}\lambda}{2}}  v^{\frac{d}{2}} \:    +  \abs{\supp{O}}  \right]. 
    \end{align}
    If we consider the limit where $\delta  t^{\frac{d+1}{2}}$ is small enough and $\lambda$ is large, this shows that $\expect{\Delta(\rho)}\leq \BOO{\sqrt{\lambda} \delta  t^{\frac{d+1}{2}}\log^{\frac{d}{2}} (1/\delta  t^{\frac{d+1}{2}})}$. \\
    We note that $\sup_{s \in [0,t]} \int_0^t ds' e^{-\frac{(s-s')^2}{2\lambda^2}} \leq \min{\left[\sqrt{2\pi} \lambda, t \right]}$.  By comparing this to Lemmas \ref{lem:cm-concentration} and \ref{lem:cm-lip-F}, after substituting $l=vt - \frac{1}{\mu }\log\delta t^{\frac{d+1}{2}}$ and rescaling $s\rightarrow s \sqrt{2tm\abs{\Theta_l}}$ we find that:
     \begin{align}
        \prob{   \Delta(\rho)\geq  \delta t^{\frac{d+1}{2}}  \min{\left[\sqrt{2\pi \lambda}, t \right]} \left(1-\frac{\log \delta t^{\frac{d+1}{2}}}{\mu v t}\right)^{\frac{d}{2}} \left(2\norm{O} \sqrt{2^{d+1} m \Lambda_d v^d \, }s \,    +  \abs{\supp{O}} \norm{O} \right) } \leq 2 e^{-s^2}.
    \end{align}
    The statement~\eqref{eq:acac-prob-statement} follows from this for $s>1$.
    
    If we instead consider $D(t)=1$ (which corresponds to the limit $\lambda\rightarrow+\infty$) and we make the choice $\varphi=\delta  t^{\frac{d}{2}+1}$, then from~\eqref{eq:Delta-acac} we have 
    \begin{align}
        \expect{\Delta(\rho)} &\leq \delta  t^{\frac{d}{2}+1} \left(1-\frac{\log \delta  t^{\frac{d}{2}+1} }{\mu v t}\right)^{\frac{d}{2}} \norm{O} \left[ 2\sqrt{\frac{2^d m \Lambda_d}{2}} v^{\frac{d}{2}} \:    +  \abs{\supp{O}}  \right]\,,
    \end{align}
    which corresponds to $\expect{\Delta(\rho)}\leq \BOO{\delta  t^{\frac{d}{2}+1}\log^{\frac{d}{2}} (1/\delta  t^{\frac{d}{2}+1})}$. The probability tail bounds follow analogously via Lemma \ref{lem:cm-lip-F}. This completes the proof.
\end{proof}
\begin{rem}[Scaling of the error in dependence of the covariance function]
    In general the $t$-dependence in this proof comes from the integral of
    \begin{align}
        \abs{\Theta_l}\int_0^t \int_0^t ds du D_{a,b}(s,u).
    \end{align}
    This would give $t^{\frac{d}{2}+\alpha}$, where $\alpha$ is the leading exponent of $t$ in this integral. We found $\alpha = \frac{1}{2}$ in these calculations, but the question of what $\alpha$ can be minimally is interesting from this point of view. The first thing to note is that $\alpha = 0$ can be ruled out because
    $\int_0^t \int_0^t ds du D_{a,b}(s,u)$ cannot be independent of $t$ for all $t$. Hence, $\alpha \neq 0$. 
\end{rem}
Note also that the $\sqrt{t}$ vs $t$ in the scaling is a sign of an Ito process and that this is the difference between the infinitely correlated process and the finitely correlated one. In the one with finite $\lambda$, we note that this process corresponds to a Ornstein-Uhlenbeck process \cite{stochasticcal}. We analyze this behavior now.

\subsection{Proof of Theorem~\ref{ito-acas}}
The white noise version of Equation~\eqref{st-sch} is given by the following Ito stochastic differential equation
\begin{align}
    \label{ito-sde}
    d \ket{\psi_t} = \left(-iH -\frac{\delta^2}{2} \sum_{\sigma=1}^{m\abs{\Theta_l}} X_\sigma^2 \right)\ket{\psi_t}\,  dt -i\delta  \sum_{\sigma=1}^{m\abs{\Theta_l}} X_\sigma\ket{\psi_t} \,dW_\sigma(t) \,,
\end{align}
where $dW_\sigma(t)$ are standard Wiener process increments. For the evolution generated by this equation we can prove the following theorem.

\begin{theorem}[Restated, Average case bounds for errors in analog simulators with white noise perturbations]
    Consider a perturbed analog time evolution $\ket{\psi'_t}$ given by the evolution~\eqref{ito-sde}. Assume that the initial state is a given pure state $\rho=\ket{\psi}\!\bra{\psi}$.
    Then, the error on the time-evolution of a local observable $O$ is, on average over the noise realizations,
    \begin{align}
        \expect{\Delta(\rho)}&\leq\BOO{ \delta t^{\frac{d+1}{2}} \left(1-\frac{\log \delta t^{\frac{d+1}{2}}}{\mu v t}\right)^{\frac{d}{2}} }\\
        &\leq  \BOO{\delta t^{\frac{d+1}{2}} \log^{\frac{d}{2}}\left(\frac{1}{ \delta t^{\frac{d+1}{2}}}\right)}.
    \end{align}
    Additionally, 
    \begin{align}
        \prob{   \Delta(\rho)\geq  \BOO{s \, \delta t^{\frac{d+1}{2}} \log^{\frac{d}{2}}\left(\frac{1}{ \delta t^{\frac{d+1}{2}}}\right)}} \leq 2 e^{-s^2}.
    \end{align}
\end{theorem}
\begin{proof}
	As in the proof of Theorem~\ref{acac} we have
    \begin{align}
        \Delta(\rho) &\leq 2 \norm{O} \norm{\ket{\psi'_t}-\ket{\psi_t}}_2  +  \abs{\supp{O}} \norm{O} e^{-\mu(l- vt)}, \label{eq:Delta-Ito}
    \end{align}
    so we focus on the stochastic part $\norm{\ket{\psi'_t}-\ket{\psi_t}}_2$.

	Integrating Equation~\eqref{ito-sde} in the interaction picture, where $\ket{\psi_t'}_I=e^{iHt}\ket{\psi'_t}$ and $X^I_\sigma(t)=e^{iHt}X_\sigma e^{-iHt}$, we have
    \begin{align}
        \ket{\psi_t'}_I-\ket{\psi}=- \frac{\delta^2}{2}  \int_0^t\!dt'\,  \sum_{\sigma=1}^{m\abs{\Theta_l}} X_\sigma^I(t')^2\, \ket{\psi_{t'}'}_I -i\delta  \int_0^t\,\sum_{\sigma=1}^{m\abs{\Theta_l}} X_\sigma^I(t')\ket{\psi_{t'}'}_I \,dW_\sigma(t')\,.
    \end{align}
    Noting that inner products in the interaction picture are equivalent to the ones in the regular Schr\"odinger picture and that $\ket{\psi_t}_I=\ket{\psi}$, we can write
    \begin{align}
        \norm{\ket{\psi'_t}-\ket{\psi_t}}^2_2 &=  2-2\Re\braket{\psi|\psi_t'}_I\\
        &= \frac{\delta^2}{2}  \int_0^t\!dt'\,  \sum_{\sigma=1}^{m\abs{\Theta_l}} \Re\bra{\psi} X_\sigma^I(t')^2\, \ket{\psi_{t'}'}_I -i \delta  \int_0^t\,\sum_{\sigma=1}^{m\abs{\Theta_l}} \Re \bra{\psi}X_\sigma^I(t')\ket{\psi_{t'}'}_I \,dW_\sigma(t')
    \end{align}

    Note that $\expect{\ket{\psi_t} dW_t} = 0$ because the Wiener increment is independent of $\ket{\psi_t}$. Therefore, taking the expectation value of the expression above we have
    \begin{align}
        \expect{\norm{\ket{\psi'_t}-\ket{\psi_t}}^2_2} &\leq \frac{\delta^2}{2}  \int_0^t\!dt'\,  \sum_{\sigma=1}^{m\abs{\Theta_l}} \expect{\abs{\bra{\psi} X_\sigma^I(t')^2\, \ket{\psi'_{t'}}_I}}\\
        &\leq \frac{\delta^2t m\abs{\Theta_l} }{2} \,, \label{eq:ito-variance-bound}
    \end{align}
    where we have used that $\ket{\psi_t}$ stays on average normalized during the evolution and that $
    \norm{X_\sigma^I(t)} =\norm{X_\sigma}\leq 1$.

    As in the proof of Theorem~\ref{acac} we apply Jensen's inequality and substitute into~\eqref{eq:Delta-Ito} to find
    \begin{align}
        \expect{\Delta(\rho)}&\leq 2\norm{O} \sqrt{\frac{\delta^2t m\abs{\Theta_l} }{2}} +  \abs{\supp{O}} \norm{O} e^{-\mu(l- vt)}\\
        &\leq  \norm{O} \sqrt{2^{d+1} m \Lambda_d v^d\, t } \, \delta\left(t-\frac{1}{\mu v}\log\varphi\right)^{\frac{d}{2}}   +  \abs{\supp{O}} \norm{O}\, \varphi\,,
    \end{align}
    where in the last step we have also applied Lemma~\ref{localterms} and set $l=vt - \frac{1}{\mu }\log\varphi$. We see then that the optimal scaling can be achieved by choosing $\varphi=\delta t^{\frac{d+1}{2}}$, which gives
    \begin{align}
        \expect{\Delta(\rho)}&\leq \delta t^{\frac{d+1}{2}} \left(1-\frac{\log \delta t^{\frac{d+1}{2}}}{\mu v t}\right)^{\frac{d}{2}} \left(\norm{O} \sqrt{2^{d+1} m \Lambda_d v^d \, } \,    +  \abs{\supp{O}} \norm{O} \right)\\
        &\leq \BOO{ \delta t^{\frac{d+1}{2}} \log^{\frac{d}{2}}\left(\frac{1}{ \delta t^{\frac{d+1}{2}}}\right)}
    \end{align}

    To derive the concentration bound, we note that by Equation 1.2 of Ref.~\cite{ito-concentration} the following inequality holds
    \begin{align}
        \prob{ \int_0^t\,\sum_{\sigma=1}^{m\abs{\Theta_l}} \norm{X_\sigma(t')\ket{\psi_{t'}'}}_2 \,dW_\sigma(t') \geq s  } \leq 2 e^{-\frac{s^2}{2M^2}},
    \end{align}
    provided that one has
    \begin{align}
        \int_0^t\,\sum_{\sigma=1}^{m\abs{\Theta_l}} \norm{X_\sigma(t')\ket{\psi_{t'}'}}_2^2 \,dt' \leq M^2\,.
    \end{align}
    By an argument analogous to~\eqref{eq:ito-variance-bound}, we can take $M^2=tm\abs{\Theta_l}$. It then follows, by the Cauchy-Schwartz inequality, that 
    \begin{align}
        \prob{   \abs{\int_0^t\,\sum_{\sigma=1}^{m\abs{\Theta_l}} \Re\bra{\psi}X_\sigma^I(t')\ket{\psi_{t'}'}_I \,dW_\sigma(t')} \geq s  } \leq 2 e^{-\frac{s^2}{2tm\abs{\Theta_l}}}.
    \end{align}
    By comparing this to Equation~\eqref{eq:Delta-Ito} and rescaling $s\rightarrow s \sqrt{2tm\abs{\Theta_l}}$, after substituting $l=vt - \frac{1}{\mu }\log\delta t^{\frac{d+1}{2}}$ we find
    \begin{align}
        \prob{   \Delta(\rho)\geq  \delta t^{\frac{d+1}{2}} \left(1-\frac{\log \delta t^{\frac{d+1}{2}}}{\mu v t}\right)^{\frac{d}{2}} \left(2\norm{O} \sqrt{2^{d+1} m \Lambda_d v^d \, }s \,    +  \abs{\supp{O}} \norm{O} \right) } \leq 2 e^{-s^2}.
    \end{align}
    This gives the result if we consider $s>1$. 

\end{proof}

\subsection{Proof of Theorem \ref{acas}}
\setcounter{theorem}{5}
\begin{theorem}[Restated]
    Assume that the implemented Hamiltonian:
    \begin{align}
        H'_{\gamma} = H_{\gamma} + \delta L_{\gamma},
    \end{align}
    where $L_{\gamma}$ is chosen from an ensemble of Hermitian matrices with $\expect{L_{\gamma}} = 0$ and $\norm{L_{\gamma}} \leq 1$. Then
    \begin{align}
        \expect{\Delta} \leq C \delta t^{d+1}.
    \end{align}
\end{theorem}
\begin{proof}
    We apply Lemma~\ref{duhamel} to $\Delta$, so that
    \begin{align}
        \Delta \leq \int_0^t ds \norm{\sum_{\gamma} [\delta L_{\gamma}, O(s)]}.
    \end{align}
    Now we take expectation values on both sides to find
    \begin{align}
        \expect{\Delta} \leq \int_0^t ds \expect{\norm{\sum_{\gamma} [\delta L_{\gamma}, O(s)]}}.
    \end{align}
    Notice that $\expect{[\delta L_{\gamma}, O(s)]} = 0$, and $\norm{[\delta L_{\gamma}, O(s)]} \leq 2\norm{O} \delta$. Hence, we can apply Lemma~\ref{lem:azuma} in this case.
    For this, we compute the variance parameter $\sigma^2 = \norm{\sum_{\gamma} \expect{L_{\gamma}^2}} \leq \abs{\Theta_l}$. Similar to Theorem \ref{thm:wcas}, we split the error into two terms:
    \begin{align}
        \Delta \leq \delta \int_0^t ds \norm{\sum_{\gamma} [L_{\gamma}, O(s)]} \leq \delta \left( \int_0^t ds \norm{\sum_{\gamma \in \Theta_l} [L_{\gamma}, O(s)]}  + \norm{O} \supp{O} K_d t\right).
    \end{align}
    We can then apply (Theorem 5.1 in \cite{random3}) to this and find:
    \begin{align}
        \expect{\norm{\sum_{\gamma \in \Theta_l} [\delta L_{\gamma}, O(s)]}} \leq 2 \sqrt{C}\norm{O} l^d,
    \end{align}
    because we sum over at most $\Theta_l$ terms in this sum, and we assume that the dimension of the Hilbert space is $2^{C l^d}$, where $C$ is a constant. Thus, we obtain
    \begin{align}
        \expect{\Delta} \leq2 \sqrt{C}\norm{O} \delta t \abs{\Theta_l} + \norm{O} \supp{O} K_d t\delta \leq K \delta t^{d+1},
    \end{align}
    for sufficiently large $t$. This completes the proof.
    \end{proof}

\subsection{Proof of Theorem \ref{pertlind}}
A reason why white noise is physically interesting is because the averaged density matrix $\rho(t) = \expect{\ketbra{\psi_t}{\psi_t}}$, follows a Lindblad type time evolution. 
\begin{lemma}[Lindblad Evolution as Average over White Noise]
\label{whitenoise}
    Assume that the stochastic state vector $\ket{\phi}_t$ evolves under $H_t$ as
\begin{align}         
    \label{ket-sde}
    d\ket{\psi}_t = \left( \left( -iH_0 -\frac{1}{2} \sum_a  S_a^2 \delta^2 \right) dt - i\sum_a \delta S_a dW_a(t) \right) \ket{\psi_t},
    \end{align}
    where $S_a$ are Hermitian operators that span the operator algebra. Then, define the averaged density matrix $\rho_t \coloneqq \expect{\ketbra{\psi_t}{\psi_t}}$. The average evolution of the density matrix is then given by
    \begin{align}
        \label{itolind}
        d\rho_t = \left( -i[H_0, \rho_t] - \frac{1}{2}\sum_a \{ S_a^2 , \rho_t\} \delta^2 + \sum_{a,b} \delta^2  S_a \rho_t S_a\right) dt = \lind(\rho_t) dt .
    \end{align}
\end{lemma} 
\begin{proof}
    Assuming that the state vector $\ket{\psi_t}$ follows the SDE in Equation \eqref{ket-sde}, then the stochastic evolution of the state projector $\ketbra{\psi_t}{\psi_t}$ is given by:
    \begin{align}
        d\left({\ketbra{\psi_t} {\psi_t}}\right) &= \left(d \ket{\psi_t}\right)\bra{\psi_t} + \ket{\psi_t} \left(d\bra{\psi_t} \right) + \frac{1}{2} d\ket{\psi_t} d\bra{\psi_t} \\ &= \left[ \left( -iH_0 \right) dt - i\sum_a \delta S_a dW_a(t)  , \ketbra{\psi_t}{\psi_t} \right] - \left\{  \frac{1}{2} \sum_a  S_a^2 \delta^2 , \ketbra{\psi_t}{\psi_t}\right\} \\ &\hspace{20mm} - \frac{1}{2} \left\{\left( -iH_0 -\frac{1}{2} \sum_a  S_a^2 \delta^2 \right) dt - i\sum_a \delta S_a dW_a(t) \right\} \ketbra{\psi_t}{\psi_t} \\ & \hspace{60mm}\Bigg\{\left( iH_0 -\frac{1}{2} \sum_a  S_a^2 \delta^2 \right) dt + i\sum_a \delta S_a dW_a(t) \Bigg\} \nonumber
    \end{align}
    where we used Ito's Lemma (compare to Theorem 4.1 \cite{watanabe}). 
    Then, the expected evolution of $\expect{\ketbra{\psi_t}{\psi_t}}$ is
    \begin{align}
        \frac{d\expect{\left({\ketbra{\psi_t}{\psi_t}}\right)}}{dt} &= \expect{d \left({\ketbra{\psi_t} {\psi_t}}\right)} = \left[\left[-iH_0, \expect{\ketbra{\psi_t}{\psi_t}}\right]  - \frac{\delta^2}{2} \sum_a S_a \expect{\ketbra{\psi_t}{\psi_t}} S_a  - \left\{  \frac{1}{2} \sum_a  S_a^2 \delta^2 , \ketbra{\psi_t}{\psi_t}\right\} \right] \\ & \coloneqq \lind(\rho_t),
    \end{align}
    where we used that $\expect{dW_a dt} = \expect{dt^2} = 0$ and $\expect{dW_a dW_b} = \delta_{ab} dt$, for all $a,b$. This shows the claim.
\end{proof} 
\begin{rem}[Non-Hermitian jump operators via non-commutative stochastic calculus]
\vspace*{-\topsep}\vspace*{-\partopsep}
    We remark that in the case of Non-Hermitian Jump Operators, the above argument does not trivially extend. Instead, the correct SDE type unravelling into a stochastic 
    unitary $U_{t,s}$ is (compare to Section 2.2 in \cite{Benoist2022deviationbounds}):
    \begin{align}
        dU_{t,s} = \left\{ - iH_0 dt - \frac{\delta^2}{2} \sum_a L_a^\dag L_a dt + \delta \left(\sum_a L_a d \mathbf{W}_a^\dag (t) - d\mathbf{W}_{a} (t) L_a^\dag \right)  \right\} U_{t,s},
    \end{align}
    where $d\mathbf{W}_a$ denotes a quantum white noise process, defined with respect to an environment described by a bosonic Fock space. Then, one may recover the Lindblad generator as
    \begin{align}
        \lind[\rho_t] = -i[H, \rho_t] + \delta^2 \sum_a L_a \rho_t L_a^\dag - \frac{1}{2} \left\{L_a^\dag L_a, \rho_t\right\}.
    \end{align}
    Following an argument similar to Theorem 5.1 of \cite{Benoist2022deviationbounds}, one could derive a "quantum" analogue of Theorem \ref{ito-acas}.
\end{rem}
In the following theorem, we show that if a pure state evolution carries Lindbladian noise, we can find improvements over previous known bounds of this form:
\setcounter{theorem}{8}
\begin{theorem}[Perturbations of Lindbladians]
Assume now that:
\begin{align}
    \lind[\rho_t] = -i[H, \rho_t] + \delta^2 \sum_a L_a \rho_t L_a^\dag - \frac{\delta^2}{2} \left\{L_a^\dag L_a, \rho_t\right\}.
\end{align}
Let $\rho_t = e^{\lind t} (\ketbra{\psi_t}{\psi_t})$, then
\begin{align}
      \norm{\rho_t - \ketbra{\psi_0(t)}{\psi_0(t)}}_1 \leq\BOO{\delta t^{\frac{d+1}{2}}} + \BOO{\delta \sqrt{t} \log^{\frac{d}{2}} (\delta t^{\frac{d+1}{2}})}.
\end{align}
\end{theorem}
\setcounter{theorem}{6}
\begin{proof}
We can expand this, using the Fuchs-van-de-Graaf inequality between the trace distance and the fidelity:
\begin{align}
    \norm{\rho- \sigma}_1 \leq 2 \sqrt{1-F(\rho, \sigma)} = 2 \sqrt{1-\braket{\psi(t)|\rho|\psi(t)}},
\end{align}
where we inserted $\sigma = \ketbra{\psi(t)}{\psi(t)}$ (as $\ket{\psi(t)} = U(t) \ket{\psi(0)}$ is a unitary time-evolution). 
We insert $\rho_t = e^{\lind t}[\ketbra{\psi_0}{\psi_0}]$, and rewrite $F(\rho_t, \sigma_t)$:
\begin{align}
    \braket{\psi_0(t)|\rho_t|\psi_0(t)} = \braket{\psi_0|U(t)^\dag  e^{\lind t} [\ketbra{\psi_0}{\psi_0}] U(t)|\psi_0} = \braket{\psi(0)|\unital^\dag (t) \circ e^{\lind t }|\psi(0)},
\end{align}
where $\unital$ is the time evolution channel with respect to $U(t)$ and $\lind$ is the Lindblad super-operator defined in \eqref{itolind}. We expand the channel $\unital^\dag \circ e^{\lind t}$ in a power series and note that
\begin{align}
    \unital^\dag \circ e^{\lind t} = \mathcal{I} + it [H_0, \cdot]  - it[H_0, \cdot] + \frac{t}{2}\sum_a \{ L_a^\dag  L_a , \cdot\} \delta^2 + \sum_{a,b} \delta^2  L_a (\cdot) L_a^\dag + \dots.
\end{align}
We can then conclude that (and using the notation $\rho_0 = \ketbra{\psi(0)}{\psi(0)}$):
\begin{align}
      \norm{\ketbra{\psi_t}{\psi_t} - e^{\lind t} \ketbra{\psi(0)}{\psi(0)}}_1 &\leq 2\sqrt{1-\braket{\psi_0|\unital^\dag \circ e^{\lind t}|\psi_0}} \leq 2\sqrt{t\braket{\psi_0|\left\{- \frac{1}{2}\sum_a \{ L_a^\dag L_a , \rho_0\} \delta^2 + \sum_{a,b} \delta^2  L_a \rho_0 L_a^\dag \right\}|\psi_0}} \\ &\leq 2 \delta t \sqrt{\norm{\sum_a L_a^\dag L_a}_{\infty}} \leq 2^{d+1} \Lambda_d C_d \delta  \sqrt{t l^d} \leq  2^{d+1} \Lambda_d C_d \delta  \sqrt{t^{d+1} \left(1 + \frac{1}{\mu vt} \log \left[\varphi\right]\right)^d},
    \end{align}
where we used that $\braket{\psi(0)|\sum_a L_a^\dag L_a | \psi(0)} \leq \norm{\sum_a L_a^\dag L_a}$, and that $- \frac{\delta^2}{2}\sum_a \braket{\psi(0)| \{ L_a^\dag L_a , \rho_0\}| \psi(0)} \leq 0$.
We can make a choice of the parameter $\varphi = \delta t^{\frac{d+1}{2}}$ and conclude that:
\begin{align}
    \Delta(\psi) \leq \BOO{\delta t^{\frac{d+1}{2}}} + \BOO{\delta \sqrt{t} \log^{\frac{d}{2}}(\delta t^{\frac{d+1}{2}})},
\end{align}
which completes the proof.
\end{proof}

\section{Stability of digital quantum simulation by Suzuki-Trotter formulas} \label{app:digital-worst-case}
In this appendix we first introduce more in detail the notion of digital quantum simulation by Suzuki-Trotter product formulas. We review how to define them and how to evaluate the discretization error that one makes by using them to represent continuous time quantum dynamics. We then proceed to prove our stability results for digital simulation with these formulas under worst-case noise (i.e., Theorems~\ref{wcds} and~\ref{wccds}). For this, we will use the notation and technical lemmas introduced in detail in Appendix~\ref{app:notation-preliminaries}.

\subsection{Quantum simulation by Suzuki-Trotter formulas}
One common approach, which we focus on here is the product formula decomposition of local Hamiltonians. This is the most widely used method to decompose unitaries into products of local unitaries which can then be implemented in a quantum circuit.
We consider a local Hamiltonian $H = \sum_{\gamma} H_\gamma$ and choose to implement its evolution truncated to a system of size $l$. In what follows, we will,  therefore, consider only the local terms of the Hamiltonian with $\gamma\in\Theta_l$, where the set $\Theta_l$ is defined as in Definition~\ref{def:trunctated-Hamiltonian}.

In the simplest case we observe that the time evolution unitary can be approximated by a product of local terms according to
\begin{align} \label{eq:basic-trotter-forumla}
    e^{itH} = \prod_{\gamma} e^{itH_\gamma} + E^{(1)}, 
\end{align}
where the error is bounded by $\norm{ E^{(1)}} \leq \frac{t^2}{2}\sum_{\gamma_1 , \gamma_2 = 1}^{\Gamma} \norm{[H_{\gamma_1}, H_{\gamma_2}]}$ (see Ref.\ \cite{Chen_2024,Childs_2021}). By noticing that $e^{itH} = \left( e^{\frac{it}{n} H} \right)^n$ for any \textit{Trotter number} $n \in \mathbb{N}$, one can use the expression~\eqref{eq:basic-trotter-forumla} as an approximation of each term $e^{i\frac{t}{n} H}$ and obtain
\begin{align}
    e^{itH} = \left( \prod_{\gamma = 1}^\Gamma e^{i\frac{t}{n}H_{\gamma}} \right)^n + E^{(1)}_n
    =: U^{(1)}_{l,n} (t) + E^{(1)}_n .
\end{align}
This result is the \textit{first order Trotter formula}. The error term can be bounded as
\begin{align}
    \norm{E^{(1)}_n} = \norm{e^{itH} - \left( \prod_{\gamma = 1}^\Gamma e^{i\frac{t}{n}H_{\gamma}} \right)^n } \leq \frac{t^2}{2n}\sum_{\gamma_1 , \gamma_2 = 1}^{\Gamma} \norm{[H_{\gamma_1}, H_{\gamma_2}]},
\end{align}
where we have used Lemma \ref{telesc} in the last step.   If $\frac{t^2}{2n} \sum_{\gamma_1 , \gamma_2 = 1}^{\Gamma} \norm{[H_{\gamma_1}, H_{\gamma_2}]} \rightarrow 0$ as $n \rightarrow \infty$, the Trotter expansion is a better and better approximation of the unitary $e^{itH}$ as a larger $n$ is chosen.
One can generalize the above formula to a $p$-th order formula \cite{Childs_2021}
\begin{align}
    e^{itH} = U^{(p)}_{l,n}(t)+ E^{(p)}_{n},
\end{align}
such that $\norm{E^{(p)}_{n}} \leq \frac{t^{p+1}}{(p+1)! n^{p+1}}  \sum_{\gamma_1 , \dots, \gamma_{p+1} = 1}^{\Gamma} \norm{[H_{\gamma_{p+1}}, [, \cdots, [H_{\gamma_2}, H_{\gamma_1}]} $. For a generic order $p$, finding a suitable expression for $U^{(p)}_{l,n}(t)$ is not as straightforward as finding $U^{(1)}(t)$. 
However, such a $p$-th order formulas in general take the following form.
\begin{define}[Product-formula unitary \cite{Childs_2021}] \label{def:product-formula}
    Given a Hamiltonian $H = \sum_{\gamma} H_\gamma$,  the corresponding $p$-th order product unitary with Trotter number $n$ and truncated to a system of size $l$ has the form
    \begin{align}
        U^{(p)}_{l,n} (t) = \prod_{j = 1}^n \prod_{\upsilon=1}^\Upsilon \prod_{\gamma\in\Theta_l}  e^{-i\frac{t}{n}a_{\upsilon,\gamma} H_{\pi_\upsilon (\gamma)}},
    \end{align}
    where $a_{\upsilon, \gamma}$ are specific constants associated to higher order product formulas. The product is composed of a number $\Upsilon$ of so-called \emph{stages} (which depends on $p$). The index $\upsilon$ labels the stages. At each stage $\upsilon$ a different ordering of the Hamiltonian terms is used, as given by the permutation $\pi_\upsilon (\gamma)$. The number of distinct Hamiltonian terms appearing is $\abs{\Theta_l}$.
\end{define}

For even orders of $p = 2k$, we can use the methods developed by Suzuki \cite{Suz91} to construct such $2k$-th order formulas. From now on, in what follows we will always consider $p$-th order product unitaries $U_{l,n}^{(p)}(t)$ of even order and we will assume that they are constructed according to these Suzuki-Trotter formulas, which we now briefly introduce.

 To this end, one must first define the second order Suzuki-Trotter formula, which we indicate as $S_2(t)$. 
This can be derived as the product of two first order formulas where 
the local Hamiltonian terms appear in reversed orders
\begin{align}
    S_2(t) = \prod_{\gamma = \abs{\Theta_l}}^{1}e^{i\frac{t}{2}H_{\gamma}} \: \prod_{\gamma = 1}^{\abs{\Theta_l}} e^{i\frac{t}{2}H_{\gamma}}.
\end{align}
This can be shown to be an approximation of $e^{itH}$ of optimal error and gate count~\cite{_Anthony_Chen_2023}
\begin{align}
    \norm{e^{itH} - S_2(t)} \leq \frac{t^3}{2} \sum_{\gamma_1, \gamma_2, \gamma_3 \in\Theta_l} \norm{[H_{\gamma_3},[H_{\gamma_2} H_{\gamma_1}]]}.
\end{align}
From this, one can derive the even order Suzuki Trotter formulas $S_{2k}$ recursively from
\begin{align}
\label{suzuki}
    S_{2k+2}(t) = S_{2k} (P_k \, t)^2 \,  S_{2k}\!\left((1-4P_k)\,t\right) \, S_{2k}(P_k\,t)^2,
\end{align}
where $P_k = (4-4^{\frac{1}{2k+1}})^{-1}$. Note for $p \geq 4$, $1-4P_k = \left( 1-\frac{4}{4-4^{\frac{1}{3}}}\right) < 0$, so we also have backwards evolutions in higher order Suzuki-Trotter formulas. As stated and shown in \cite{Suz91}, there exists no higher order product formula without backwards evolution.
From Eq.~\eqref{suzuki} it follows that the number of stages of $S_2k$  is $\Upsilon = 2 \cdot 5^{k-1}$. For the Suzuki-Trotter formulas, the  error bound 
\begin{align}
    \norm{e^{itH} - S_p(t)} \leq \frac{t^{p+1}}{(p+1)!}  \sum_{\gamma_1 , \dots, \gamma_{p+1} \in \Theta_l} \norm{[H_{\gamma_{p+1}}, [, \cdots, [H_{\gamma_2}, H_{\gamma_1}]}
\end{align}
holds for all $t$ \cite{Suz91}.
Analogously to above, we can now use the  Suzuki-Trotter formulas $S_{2k}(t)$ to construct $p$-th order product unitaries for even $p$ as $U_{l,n}^{(p)}(t)\coloneqq S_p(\frac{t}{n})^n$. Using Lemma \ref{telesc} as before, the corresponding error can be expressed as follows. 
\begin{lemma}[Trotter error \cite{Childs_2021,Suz91}]
\label{fact-trotter}
    Let $H_l = \sum_{\gamma\in\Theta_l} H_\gamma$ be a local Hamiltonian truncated to a system of size $l$. Approximating the evolution $U_l(t) = e^{-itH_l}$ with a Suzuki-Trotter product unitary $U_{l,n}^{(p)}(t)$ for any evolution time $t$ and for any even order $p=2k$ gives an approximation error
    \begin{align}
         \norm{U_l(t)-U_{l,n}^{(p)}(t)} \leq \norm{\sum_{\gamma_1, \dots , \gamma_{p+1} \in \Theta_l} [H_{\gamma_{p+1}}, \dots , [H_{\gamma_2}, H_{\gamma_1}]]} \; \frac{t^{p+1}}{n^{p} (p+1)!}.
    \end{align}   
\end{lemma}

\subsection{Locality and Suzuki-Trotter products}
In the following, we will further evaluate the nested commutators appearing in Lemma \ref{fact-trotter}. We will do so using the geometrical locality of the $H_{\gamma}$ terms as stated in Assumption \ref{ass:locality}.

\begin{lemma}[Nested commutator scaling] \label{nested}
    We have that
    \begin{align}
       \norm{\sum_{\gamma_1,\dots, \gamma_p} [H_{\gamma_p}, [H_{\gamma_{p-1}}, \dots [H_{\gamma_2}, H_{\gamma_1}]]]}\leq 2^p \left(\Lambda_d 2^d R^d\right)^{p-1} [(p-1)!]^d \abs{\Theta_l}.
    \end{align}
\end{lemma}
\begin{proof}
    By Assumption~\ref{ass:locality}, we can associate each term $H_{\gamma_j}$ to a site $x_j$. For all $j=2,\dots, p$, let 
    \begin{align}
        C_{\gamma_1,\dots, \gamma_j} = [H_{\gamma_j}, [H_{\gamma_{j-1}}, [, 
        \dots [H_{\gamma_2}, H_{\gamma_1}]].
    \end{align}
     First we observe that $C_{\gamma_1,\dots, \gamma_j}$ can be non-zero only if $x_j\in B_{2(j-1)R}(x_1)$ and furthermore, for any such non vanishing term,
    \begin{align}
        \supp{C_{\gamma_1,\dots, \gamma_j}} \subseteq B_{(2j-1)R}(x_1)\,.
    \end{align} 
    We prove this by induction. For $j=2$, observe that $[H_{\gamma_2} , H_{\gamma_1}] = 0$ if $\supp{H_{\gamma_1}} \cap \supp{H_{\gamma_2}} = \emptyset$. Thus, 
    by Assumption~\ref{ass:locality}, the commutator can be non-vanishing only if $x_2$ is contained in a ball of center $x_1$ 
    and radius $2R$. Furthermore
    \begin{equation}
        \supp{[H_{\gamma_2} , H_{\gamma_1}]} \subseteq \supp{H_{\gamma_1}} \cup \supp{H_{\gamma_2}} \subseteq B_{3R}(x_1)\,.
    \end{equation}
    Assume now that the observation holds for the $j$-th nested commutator. 
    Reasoning like before, it is clear that $C_{\gamma_1,\dots, \gamma_{j+1}}=[H_{\gamma_{j+1}}, C_{\gamma_1,\dots, \gamma_{j}} ]$ can be non-zero only if $H_{\gamma_{j+1}}$ is associated to a site $x_{j+1}$ within a ball of radius $(2j-1)R+R=2jR$. Furthermore any non-vanishing term $[H_{\gamma_{p+1}}, C_{\gamma_1,\dots, \gamma_{p}} ]$ has support on a region contained in a ball of radius $(2p-1)R+2R=(2p+1)R$. This shows that the observation also holds for the $(j+1)$-th nested commutator. 
    
    In conclusion the sum can be restricted to those $\gamma_1,\dots,\gamma_p$ such that $x_j\in B_{2(j-1)R}(x_1)$. By Assumption~\ref{ass:locality} there are at most as many such local terms as there are sites within each of these balls, that is
    \begin{align}
        \norm{\sum_{\gamma_1, \dots, \gamma_p} C_{\gamma_1, \dots, \gamma_p}} &\leq \sum_{\gamma_1\in\Theta_l}\sum_{\substack{\gamma_2\ \mathrm{s.t.}\\x_2\in B_{2R}(x_1)}} \; \sum_{\substack{\gamma_p\ \mathrm{s.t.}\\x_p\in B_{2(p-1)R}(x_1)}} \norm{C_{\gamma_1, \dots, \gamma_p}} \\
        \nonumber
        &\leq \abs{\Theta_l} \abs{B_{2R}(x_1)} \cdots \abs{B_{2(p-1)R}(x_1)} \\
        & \leq  2^p \left(\Lambda_d 2^d R^d\right)^{p-1} [(p-1)!]^d \abs{\Theta_l}\,,
         \nonumber
    \end{align}
where we have used $\norm{C_{\gamma_1, \dots, \gamma_p}}\leq 2^p$ by the first point of Assumption~\ref{ass:locality}.
\end{proof}

By substituting the previous result into Lemma~\ref{fact-trotter} we immediately have
the following.

\begin{lemma}[Trotter error, with truncation]
    \label{lem:tr}
    Given a Hamiltonian $H_l = \sum_{\gamma\in\Theta_l} H_\gamma$, truncated to a system of size $l$ and  a $p$-th order product unitary with Trotter number $n$, the latter gives an approximation of the time evolution $U_l(t)=e^{-itH_l}$ up to an error
    \begin{align}
        \norm{U_l(t) - U^{(p)}_{l,n}\!\left( t\right)} &\leq \frac{t^{p+1}}{n^p(p+1)!}\norm{\sum_{\gamma_1, \dots , \gamma_{p+1}\in\Theta_l} [H_{\gamma_{p+1}}, \dots , [H_{\gamma_2}, H_{\gamma_1}]]}  \leq  K \abs{\Theta_l} \frac{t^{p+1}}{ n^p},
        \label{eq:trotter-bound}
    \end{align}    
    where $K=2^p\left(\Lambda_d 2^d R^d\right)^{p} \; \frac{[(p-1)!]^d}{(p+1)!}$.
\end{lemma}
\subsection{Proof of Theorem \ref{wcds}}
We now prove Theorem \ref{wcds}, whose formal statement we repeat here for convenience.

\setcounter{theorem}{1}
\begin{theorem}[Restated, upper bound for worst case errors in digital simulators with gate-dependent perturbations]
    Consider a perturbed Suzuki-Trotter product unitary of order $p=2k$, which takes the form
    \begin{align}
       V^{(p)}_{l,n} (t)  = \prod_{j=1}^n \prod_{\upsilon=1}^\Upsilon \prod_{\gamma\in\Theta_l} V_{\gamma, j, \upsilon}\,,
    \end{align}
    where each local gate is a perturbed version of the exact gate, satisfying $ \norm{ V_{\gamma, j, \upsilon} - e^{-i\frac{t}{n}a_{\upsilon,j} H_{\pi_v(\gamma)}}} \leq \delta\frac{t}
    {n}$. The product unitary has Trotter number $n$ and is implemented on a system of size $l$. Then, the error on the time-evolution of a local observable $O$ is at most
\begin{align}
        \Delta \leq \BOO{\delta t^{d+1}\left(1-\frac{1}{\mu v t}\log(\delta t^{d+1})\right)^d}\leq\BOO{\delta t^{d+1} \log^d(\frac{1}{\delta t^{d+1}})},
    \end{align}
    if the \textit{optimal choices} of $n_{\rm opt} \geq t/\delta^{\frac{1}{p}}$ and $l_{\rm opt}\geq vt-\frac{1}{\mu}\log(\delta t^{d+1})$ are made, where $v = e \Lambda_d R^{d+1} $, $\mu = \frac{1}{R}$ as in Lemma~\ref{lem:trunc}.
\end{theorem}
\begin{proof}\label{pr:wcds}
    We consider the definition of $\Delta$ and split it into three error contributions by applying the triangle inequality
    \begin{align}
        \Delta&= || V^{(p)}_{l,n}(t)^\dag O V^{(p)}_{l,n}(t) - U(t)^\dag O U(t) ||  \nonumber
        \\
        & \leq || V^{(p)}_{l,n}(t)^\dag O V^{(p)}_{l,n}(t) - U^{(p)}_{l,n}(t)^\dag O U^{(p)}_{l,n}(t) || \nonumber\\
        &\hspace{20mm}+ || U^{(p)}_{l,n}(t)^\dag O U^{(p)}_{l,n}(t) - U_l(t)^\dag O U_l(t)|| \nonumber\\
        &\hspace{40mm}+ || U_l(t)^\dag O U_l(t) - U(t)^\dag O U(t)||\,, \label{eq:three-terms_wcds}
    \end{align}
    where $U^{(p)}_{l,n} (t)$ is the noiseless 
    product unitary as in 
    Definition~\ref{def:product-formula}. We will now separately bound the three terms appearing in the 
    last inequality above.

    The first term reflects the contribution to the total error from the noisy gates and can be bounded by applying Corollary~\ref{cor:telescopic}:
    \begin{align}
        || V^{(p)}_{l,n}(t)^\dag O V^{(p)}_{l,n}(t) - U^{(p)}_{l,n}(t)^\dag O U^{(p)}_{l,n}(t) || \leq 2 ||O|| \sum_{j=1}^n \sum_{\upsilon=1}^\Upsilon \sum_{\gamma\in\Theta_l} \norm{ V_{\gamma, j, \upsilon} - e^{-i\frac{t}{n}a_{v,j} H_{\pi_v(\gamma)}}} \leq 2||O|| \Upsilon |\Theta_l| \delta \, t \,. \label{eq:wcds-term1}
    \end{align}
    
    The second term gives the Trotter decomposition error, which can be bounded as
    \begin{align}
        || U^{(p)}_{l,n}(t)^\dag O U^{(p)}_{l,n}(t) - U_l(t)^\dag O U_l(t)||\leq 2||O||\, || U^{(p)}_{l,n}(t) - U_l(t)|| \leq 2 ||O||K \abs{\Theta_l} \frac{t^{p+1}}{n^p}\,, \label{eq:wcds-term2}
    \end{align}
    where in the first step we have used Corollary~\ref{cor:telescopic} and in the second step we have used Lemma~\ref{lem:tr} and $K=2^p\left(\Lambda_d 2^d R^d\right)^{p} \; [(p-1)!]^d/(p+1)!$.
    
    The third term represents the error that we make by considering only the Hamiltonian terms within the truncation length $l$. It can be bounded using a Lieb-Robinson bound such as Lemma~\ref{lem:trunc}, which directly gives us
    \begin{align}
        \norm{U_l^\dag(t) O U_l(t) - U^\dag(t) O U(t)} \leq \abs{\supp{O}} \norm{O} \min \left( e^{-\mu l} \left( e^{\mu v t} - 1\right),1 \right)\,. \label{eq:wcds-term3}
    \end{align}
    
    So far the truncation length $l$ and the Trotter number $n$ are free parameters. The aim is now to choose them such that each of the error terms above scales in the same way with respect to $t$ and $\delta$. This corresponds to the choice that achieves the optimal trade-off between the various error sources.  
    First of all, in order to cancel the exponential scaling in $t$ in the last error term we choose $l=vt - \frac{1}{\mu }\log\varphi$, where $\varphi$ will be specified later. Notice that, according to Lemma~\ref{localterms}, for large enough $l$ (which we will see corresponds to large $t$ and small $\delta$) we have
    \begin{align}
        \abs{\Theta_l} \leq 2^d \Lambda_d l^d\,. 
    \end{align}
    Substituting this into Eqs.~\ref{eq:wcds-term1}, \ref{eq:wcds-term2} and~\ref{eq:wcds-term3} we have
    \begin{align}
        \Delta &\leq 2^{d+1}\norm{O}\Upsilon \Lambda_d v^d  \,  \left(t-\frac{1}{\mu v}\log\varphi\right)^d \delta \,t \nonumber \\
        &\hspace{30mm} + 2^{d+1} \norm{O}K \Lambda_d v^d  \left(t-\frac{1}{\mu v}\log\varphi\right)^d \frac{t^{p+1}}{n^p} \nonumber \\
        &\hspace{70mm} + \abs{\supp{O}} \norm{O} \varphi .
    \end{align}
    We now make a choice of $n$ and $\varphi$ which balances the scaling in $t$ and $\delta$ of all the three remaining terms (up to logarithmic factors), namely
    \begin{equation}
        n \geq \frac{t}{\delta^{1/p}}\,, \hspace{30mm} \varphi\leq\delta  t^{d+1}\,.
    \end{equation}
    This gives us
   \begin{align}
        \Delta &\leq \delta \,t^{d+1}\, \norm{O} \left[ 2^{d+1} \Upsilon \Lambda_d v^d  \, \left(1-\frac{1}{\mu v t}\log(\delta t^{d+1})\right)^d + 2^{d+1} K \Lambda_d v^d \,  \left(1-\frac{1}{\mu vt}\log(\delta t^{d+1})\right)^d + \abs{\supp{O}} \right] \\
        &\leq    \delta \, t^{d+1} \log^d(\frac{1}{\delta t^{d+1}})\, \norm{O} \left[ 2^{d+1} \Upsilon \Lambda_d v^d  \, \left(1+\frac{1}{\mu v t}\right)^d + 2^{d+1} K \Lambda_d v^d \,  \left(1+\frac{1}{\mu vt}\right)^d + \abs{\supp{O}} \right]    \,.
    \end{align}
    In the last step, we have recognised that, if $\delta\,t^{d+1}$ is small enough, then $\log(\frac{1}{\delta t^{d+1}})\geq 1$, which in particular means that
    \begin{equation}
        1-\frac{1}{\mu vt}\log(\delta t^{d+1}) = 1+\frac{1}{\mu vt} \log(\frac{1}{\delta t^{d+1}}) \leq \log(\frac{1}{\delta t^{d+1}}) \left(1+\frac{1}{\mu vt}\right)\,.
    \end{equation}
    We can, therefore, conclude that $\Delta \leq \BOO{\delta\, t^{d+1} \log^d(\frac{1}{\delta t^{d+1}})}$.
\end{proof}
\subsection{Proof of Theorem \ref{wccds}}
We now prove Theorem \ref{wccds}, whose formal statement we repeat here for convenience.

\begin{theorem}[Restated, upper bound for worst case errors in digital simulators with constant gate perturbations]
    Consider a perturbed Suzuki-Trotter product unitary of order $p=2k$, which takes the form
    \begin{align}
       V^{(p)}_{l,n} (t)  = \prod_{j=1}^n \prod_{\upsilon=1}^\Upsilon \prod_{\gamma\in\Theta_l} V_{\gamma, j, \upsilon}\,,
    \end{align}
    where each local gate is a perturbed version of the exact gate, satisfying $ \norm{ V_{\gamma, j, \upsilon} - e^{-i\frac{t}{n}a_{\upsilon,j} H_{\pi_v(\gamma)}}} \leq \delta$. The product unitary has Trotter number $n$ and is implemented on a system of size $l$. Then, the error on the time-evolution of a local observable $O$ is at most
\begin{align}
        \Delta \leq\BOO{\delta^{\frac{p}{p+1}} t^{d+1}\, \left(1-\frac{1}{\mu v t}\log(\delta^{\frac{p}{p+1}}t^{d+1})\right)^d} \leq \BOO{\delta^{\frac{p}{p+1}} t^{d+1} \log^d(\frac{1}{\delta^{\frac{p}{p+1}}t^{d+1}})},
    \end{align}
    if the \textit{optimal choices} of $n_{\rm opt} = t/\delta^{\frac{1}{p+1}}$ and $l_{\rm opt}\geq vt-\frac{1}{\mu}\log(\delta^{\frac{p}{p+1}} t^{d+1})$ are made, where $v = e \Lambda_d R^{d+1} $, $\mu = \frac{1}{R}$ as in Lemma~\ref{lem:trunc}.
\end{theorem}
\begin{proof}
    We consider the definition of $\Delta$ and we split it into three error contributions, as we did in the proof of Theorem~\ref{wcds} at line~\eqref{eq:three-terms_wcds}. These three terms can be bounded exactly like in the proof of that theorem, except for the first one, where instead of~\eqref{eq:wcds-term1} we find 
    \begin{align}
        || V^{(p)}_{l,n}(t)^\dag O V^{(p)}_{l,n}(t) - U^{(p)}_{l,n}(t)^\dag O U^{(p)}_{l,n}(t) || \leq 2 ||O|| \sum_{j=1}^n \sum_{\upsilon=1}^\Upsilon \sum_{\gamma\in\Theta_l} \norm{ V_{\gamma, j, \upsilon} - e^{-i\frac{t}{n}a_{v,j} H_{\pi_v(\gamma)}}} \leq 2||O|| n\Upsilon |\Theta_l| \delta \,. \label{eq:wccds-term1}
    \end{align}
    
    Again, the truncation length $l$ and the Trotter number $n$ are free parameters. The aim is now to choose them such that each of the error terms above scales in the same way with respect to $t$ and $\delta$. This corresponds to the choice that achieves the optimal trade-off between the various error sources.  
    As before, we choose $l=vt - \frac{1}{\mu }\log\varphi$, where $\varphi$ will be specified later.
    We now also consider that, according to Lemma~\ref{localterms}, for large enough $l$ (which we will see corresponds to large $t$ and small $\delta$) we have
    \begin{align}
        \abs{\Theta_l} \leq 2^d \Lambda_d l^d\,. 
    \end{align}
    Substituting all this into Eqs.~(\ref{eq:wccds-term1}), (\ref{eq:wcds-term2}) and~(\ref{eq:wcds-term3}), 
    we have
    \begin{align}
        \Delta &\leq 2^{d+1}\norm{O}\Upsilon \Lambda_d v^d  \,  \left(t-\frac{1}{\mu v}\log\varphi\right)^d \delta \,n \\
        &\hspace{30mm} + 2^{d+1} \norm{O}K \Lambda_d v^d  \left(t-\frac{1}{\mu v}\log\varphi\right)^d \frac{t^{p+1}}{n^p} \nonumber \\
        &\hspace{70mm} + \abs{\supp{O}} \norm{O} \varphi.
        \nonumber
    \end{align}
    We now make a choice of $n$ and $\varphi$ which balances the scaling in $t$ and $\delta$ of all the three remaining terms (up to logarithmic factors), namely
    \begin{equation}
        n = \frac{t}{\delta^{\frac{1}{p+1}}}\,, \hspace{30mm} \varphi\leq \delta^{\frac{p}{p+1}} t^{d+1}\,.
    \end{equation}
   This actually gives 
   us
   \begin{align}
        \Delta &\leq \delta^{\frac{p}{p+1}} t^{d+1}\, \norm{O} \left[ 2^{d+1} \Upsilon \Lambda_d v^d  \, \left(1-\frac{1}{\mu v t}\log(\delta^{\frac{p}{p+1}}t^{d+1})\right)^d + 2^{d+1} K \Lambda_d v^d \,  \left(1-\frac{1}{\mu vt}\log(\delta^{\frac{p}{p+1}}t^{d+1})\right)^d + \abs{\supp{O}} \right] \\
        &\leq    \delta^{\frac{p}{p+1}} t^{d+1} \log^d(\frac{1}{\delta^{\frac{p}{p+1}}t^{d+1}})\, \norm{O} \left[ 2^{d+1} \Upsilon \Lambda_d v^d  \, \left(1+\frac{1}{\mu v t}\right)^d + 2^{d+1} K \Lambda_d v^d \,  \left(1+\frac{1}{\mu vt}\right)^d + \abs{\supp{O}} \right]    \,.\nonumber
    \end{align}
    In the last step we have recognized that, if $\delta^{\frac{p}{p+1}} t^{d+1}$ is small enough, then $\log(\frac{1}{\delta^{\frac{p}{p+1}}t^{d+1}})\geq 1$, which in particular means that
    \begin{equation}
        1-\frac{1}{\mu vt}\log(\delta^{\frac{p}{p+1}}t^{d+1}) = 1+\frac{1}{\mu vt} \log(\frac{1}{\delta^{\frac{p}{p+1}}t^{d+1}}) \leq \log(\frac{1}{\delta^{\frac{p}{p+1}}t^{d+1}}) \left(1+\frac{1}{\mu vt}\right)\,.
    \end{equation}
    We can, therefore, conclude that $\Delta \leq \BOO{\delta^{\frac{p}{p+1}} t^{d+1} \log^d(\frac{1}{\delta^{\frac{p}{p+1}}t^{d+1}})}$.
\end{proof}

\section{Stability of digital quantum simulation under stochastic errors} \label{app:digital-average-case}
In this appendix we prove our stability results for digital quantum simulation under stochastic errors. We first introduce some intermediate technical results and then prove Theorems~\ref{fixed-states-ads}, \ref{fixed-states-2}, \ref{acdc} and~\ref{accds}. For this, we will use the notation and preliminary lemmas introduced in detail in Appendix~\ref{app:notation-preliminaries}.

\subsection{Sums of random matrices}
One important tool for this discussion is the analysis of the behaviour of the sum of mean-zero random perturbations. This can be studied by applying Pinelis' Lemma, which we introduced in Appendix~\ref{app:notation-preliminaries}. 

\begin{lemma}
    [Sum of mean-zero random perturbations]\label{randomsum}
    Consider a truncation length $l>0$ and a sequence of random Hermitian operators $L_J$ for $J=1,\dots,J_{\rm tot}$ with $J_{\rm tot}=n\Upsilon_p|\Theta_l|$. Assume that each of these operators has mean $\expect{L_J}=0$, bounded norm $\norm{L_J} \leq 1$ and support contained in the support of the truncated Hamiltonian $H_l$. Then, for any pure state $\ket{\psi}$ and  any unitary operator $V^+_J$ that only depends on the random variables $L_{J'}$ with $J'>J$, we have
    \begin{align}
        \expect{\norm{\sum_{J=1}^{J_{\rm tot}} (V_J^{+})^ \dag L_J V^+_J\ket{\psi}}_2} \leq \sqrt{2\pi \, n \Upsilon_p \abs{\Theta_l}},
    \end{align}
    and, furthermore,
    \begin{align}
        \prob{\norm{\sum_{J=1}^{J_{\rm tot}}  (V_J^{+})^ \dag L_J V^+_J\ket{\psi}}_2 > s} \leq 2 e^{-\frac{s^2}{2n \Upsilon_p \abs{\Theta_l}}}.
    \end{align}
\end{lemma}
\begin{proof}
    The proof is a simple application of Lemma~\ref{lem:pinelis}. Define $Y_J := (V_{(J_{\rm tot}-J+1)}^{+})^ \dag L_{(J_{\rm tot}-J+1)} V^+_{(J_{\rm tot}-J+1)}\ket{\psi}$ and $X_J:=L_{(J_{\rm tot}-J+1)}$ for every $J=1,\dots,J_{\rm tot}$ (where we reverse the order of the indices to be compatible with the notation of the Lemma). Then, given that $V^+_{J}$ are unitary, we have $\norm{Y_J}_2 \leq \norm{L_{(J_{\rm tot}-J+1)}}\leq 1$. Additionally, we observe that $V^+_{J}$ is independent of $L_{J'}$ for $J'\leq J$, which implies
    \begin{align}
        \mathbb{E}_{X_J, \dots, X_{J_{\rm tot}}}\left[Y_J\right] &= \mathbb{E}_{L_1, \dots, L_{(J_{\rm tot}-J+1)}}\left[ (V_{(J_{\rm tot}-J+1)}^{+})^ \dag L_{(J_{\rm tot}-J+1)} V^+_{(J_{\rm tot}-J+1)}\ket{\psi}\right] \\
        \nonumber
        &= (V_{(J_{\rm tot}-J+1)}^{+})^ \dag \; \mathbb{E}_{L_{(J_{\rm tot}-J+1)}}\left[L_{(J_{\rm tot}-J+1)}\right] \; V^+_{(J_{\rm tot}-J+1)}\ket{\psi} \\
         \nonumber
        &=0,
    \end{align}
    where in the last step we used $\mathbb{E}_{L_J}\left[L_J\right]=0$. Hence, we can apply Lemma~\ref{lem:pinelis} to $\sum_J Y_J$, to find
    \begin{align}
        \prob{ \norm{\sum_J Y_J}_2 > s} \leq 2 e^{-\frac{s^2}{2J_{\rm tot}}} = 2 e^{-\frac{s^2}{2n\Upsilon_p \abs{\Theta_l}}}.
    \end{align}
    Since $\norm{\sum_J Y_J}_2$ is a positive and bounded random variable, we can represent its expectation value by
    \begin{align}
        \expect{\norm{\sum_J     Y_J}_2} = \int_0^\infty \!ds\;  \prob{\norm{\sum_J Y_J}_2 > s} \leq \sqrt{2\pi \, n \Upsilon_p \abs{\Theta_l}}.
    \end{align}
    This completes the proof.
\end{proof}

A similar result can also be proven for sums of operators rather than vectors, although with a less optimal dimensional factor.
\begin{lemma}
    [Sum of mean-zero random operators]\label{randomsum-matrix}
    Consider a truncation length $l>0$ and a sequence of random Hermitian operators $L_J$ for $J=1,\dots,J_{\rm tot}$ with $J_{\rm tot}=n\Upsilon_p|\Theta_l|$. Assume that each of these operators has mean $\expect{L_J}=0$, bounded norm $\norm{L_J} \leq 1$ and support contained in the support of the truncated Hamiltonian $H_l$. Then, for any unitary operator $V^+_J$ that only depends on the random variables $L_{J'}$ with $J'>J$, we have
    \begin{align}
        \expect{\norm{\sum_{J=1}^{J_{\rm tot}} (V_J^{+})^ \dag L_J V^+_J}} \leq 2\sqrt{2^{d+1}\Lambda_d n \Upsilon_p \abs{\Theta_l} \,  l^d},
    \end{align}
    and, furthermore,
    \begin{align}
        \prob{\norm{\sum_{J=1}^{J_{\rm tot}}  (V_J^{+})^ \dag L_J V^+_J} > s} \leq 2 e^{-\frac{s^2}{2n\Upsilon_p \abs{\Theta_l}}+2^d\Lambda_d l^d}.
    \end{align}
\end{lemma}
\begin{proof}
We follow the same derivation of the previous Lemma, except that we now use Lemma~\ref{lem:azuma} to find 
\begin{align}
        \prob{ \norm{\sum_{J=1}^{J_{\rm tot}}  (V_J^{+})^ \dag L_J V^+_J}> s} \leq 2 De^{-\frac{s^2}{2J_{\rm tot}}} = 2^{2^d\Lambda_dl^d+1} e^{-\frac{s^2}{2n\Upsilon_p \abs{\Theta_l}}}\leq 2 e^{-\frac{s^2}{2n\Upsilon_p \abs{\Theta_l}}+2^d\Lambda_d l^d},
    \end{align}
    where we have used that the operators are defined on the truncated lattice $\Omega_l$, therefore their dimension can be assumed to be $D=2^{2^d\Lambda_dl^d}$ by Lemma~\ref{localterms}.

    As in the previous Lemma, we integrate this quantity to find a bound on the expectation value. We however now split the integral at $s_0=\sqrt{2J_{\rm tot}\log D}$ (similarly to the proof of Corollary 7.3.2 in \cite{tropp2015introductionmatrixconcentrationinequalities}):
    \begin{align}
        \expect{\norm{\sum_{J=1}^{J_{\rm tot}} (V_J^{+})^ \dag L_J V^+_J}} &\leq \int_0^{s_0} \!ds \, 1+  2 D\int_{s_0}^\infty \! ds \, \frac{s}{\sqrt{2J_{\rm tot}}} \,e^{-\frac{s^2}{2J_{\rm tot}}} \\
        &=s_0+\sqrt{2J_{\rm tot}}D e^{-\frac{s_0^2}{2J_{\rm tot}}}\\[2mm]
        &\leq 2\sqrt{2J_{\rm tot}\log D}\\[1mm]
        &\leq 2\sqrt{2 n \Upsilon \abs{\Theta_l} \, 2^d\Lambda_d l^d}\,,
    \end{align}
    where we have used that the integrand is upper-bounded by $1$ in the first integral and that $s/\sqrt{2J_{\rm tot}}\geq\sqrt{\log D}\geq 1$ in the domain of the second integral.
\end{proof}

\subsection{Bounds on perturbations of Trotter products}

We continue by proving the second main tool of this part of this work, namely an improved bound for local perturbations of the Trotter formula, which is better than the telescopic sum bound from Lemma \ref{telesc}. The main idea is to find a \textit{norm of sum} statement to which Lemma \ref{randomsum} can be applied.

\begin{lemma}[Perturbation Bound on Noisy Product Unitaries for fixed inputs]
\label{seriesbound_vec}
Let  $U^{(p)}_{l,n}$ be a product unitary according to Definition~\ref{def:product-formula}, with Trotter number $n$ and truncated to a system of size $l$. Let $V_{l,n}^{(p)}$ be a noisy version of this product unitary of the form~\eqref{eq:noisy-product-unitary-def}, that is
\begin{align}
    V^{(p)}_{l,n}(t) = \prod_{j=1}^n \prod_{\upsilon=1}^{\Upsilon_p} \prod_{\gamma\in\Theta_l} e^{-i\frac{t}{n} (a_{v,\gamma} H_{\pi_\upsilon (\gamma)}+ \delta L_{\gamma, \upsilon, j})}, 
\end{align}
with $\norm{L_{\gamma, \upsilon, j}}\leq 1$. Then, for any pure state $\ket{\psi}$, we have
\begin{align}
     \norm{{U_{l,n}^{(p)}\ket{\psi} - V_{l,n}^{(p)}}\ket{\psi}}_2 &\leq \frac{t}{n} \:\int_0^\delta d\delta'\,\norm{\sum_{\gamma, \upsilon, j} \left(V^{+}_{\gamma, \upsilon, j}(\delta')\right)^\dag L_{\gamma, \upsilon, j} V^+_{\gamma, \upsilon, j}(\delta')\ket{\psi}}_2 
        + \delta \Upsilon_p \abs{\Theta_l} \frac{t^2}{n}\,,
\end{align}
where $V^+_{\gamma, \upsilon, j}(\delta)$ is the product of all the terms in the product unitary $V_{l,n}^{(p)}$ appearing to the right of the term $(\gamma, \upsilon, j)$. 
\end{lemma}
\begin{proof}
    The operator $V_{l,n}^{(p)}$ is a product of unitaries, which for convenience we express as 
    \begin{align}
        V_{l,n}^{(p)}(\delta) = \prod_{j = 1}^n \prod_{\upsilon=1}^\Upsilon \prod_{\gamma\in\Theta_l}  e^{-i\frac{t}{n}a_{\upsilon,\gamma} H_{\pi_\upsilon (\gamma)} - i\delta \frac{t}{n}L_{\gamma, \upsilon, j}} = \prod_{J} e^{ Z_{J}}\,. \label{eq:V_trotter-perturbation-bounds}
    \end{align}
    Here, $Z_J=-i\frac{t}{n}a_{\upsilon,\gamma} H_{\pi_\upsilon (\gamma)} - i\delta \frac{t}{n}L_{\gamma, \upsilon, j}$ and $J$ is a multi-index which runs over all choices of $(\gamma, \upsilon, j)$ in the order in which they appear in the product formula (in total there are $J_{\rm tot}=n\Upsilon_p |\Theta_l|$ such choices). We have further made the dependence of $V_{l}^{(p)}$ on $\delta$ explicit. 
    
    Let us note that, for $\delta=0$, we have $U_{l,n}^{(p)} - V_{l,n}^{(p)}(0)=0$. Therefore, by the fundamental theorem of calculus\begin{align}
        \label{deriv_vec}
        \norm{V_l^{(p)} (\delta) - U_l^{(p)} \ket{\psi}}_2 \leq \int_0^\delta  d \delta' \norm{\frac{d}{d\delta}V_l^{(p)}(\delta')\ket{\psi}}_2.
    \end{align}
    We are now going to evaluate the derivative of $V_{l,n}^{(p)}(\delta')$ more in detail. For this, we use the known formula for the derivative of a matrix exponential
    \begin{align}
        \frac{d}{d\epsilon} e^{A(\epsilon)} = \int_{0}^1 ds \: \: e^{(1-s)A(\epsilon)}\,\left[\frac{d}{d\epsilon}A(\epsilon)\right] \;  e^{s A(\epsilon)}.
    \end{align}
    Applying this to~\eqref{eq:V_trotter-perturbation-bounds} gives
    \begin{align}
       \nonumber 
        \frac{d}{d\delta}V_{l,n}^{(p)}(\delta') &= -\frac{it}{n}\sum_{J} \left[\prod_{J'=1}^{J-1} e^{Z_{J'}} \int_{0}^1 ds \: e^{(1-s)Z_{J}} \,  L_{\gamma, \upsilon, j} \;e^{s Z_J} \prod_{J''=J+1}^{J_{\rm tot}} e^{Z_{J''}}\right]
        \\
        \nonumber 
        &=-\frac{it}{n}\, V_{l,n}^{(p)}(\delta')\,\sum_{J} \left(\prod_{J'=J+1}^{J_{\rm tot}} e^{Z_{J'}} \right)^{-1} \left(\int_{0}^1 ds \: e^{-s\, \mathrm{ad}_{Z_{J}}}[L_{\gamma, \upsilon, j}]\right) \left(\prod_{J'=J+1}^{J_{\rm tot}} e^{\mathrm{ad}_{Z_{J'}}} \right) \, \\
        &=-\frac{it}{n}\, V_{l,n}^{(p)}(\delta')\,\sum_{J} \left(V^+_J(\delta')\right)^\dag \left(\int_{0}^1 ds \: e^{-s\, \mathrm{ad}_{Z_{J}}}[L_{\gamma, \upsilon, j}]\right) V^+_J(\delta')\,, \label{eq:derivative-V}
    \end{align}
    where in the second step we have rearranged some terms and defined the superoperator $\mathrm{ad}_{Z}[\,\cdot\,] \coloneqq [Z,\,\cdot \,]$. We have also introduced the notation $V^+_J(\delta)$ for the product of all the terms in the product unitary $V_{l,n}^{(p)}$ appearing to the right of the $J$-th term.

    Now expanding the Taylor series of $e^{-s\, \mathrm{ad}_{Z_{J}}}$ and performing the integrals in $s$, we obtain
    \begin{align}
         \int_{0}^1 ds \: e^{-s\, \mathrm{ad}_{Z_{J}}}[L_{\gamma, \upsilon, j}] = L_{\gamma, \upsilon, j} + \sum_{m=1}^\infty \frac{(-1)^m}{(m+1)!} \mathrm{ad}_{Z_{J}}^m[L_{\gamma, \upsilon, j}].
    \end{align}
We can use the numerical bound
    \begin{align}
        \norm{ \sum_{m=1}^\infty \frac{(-1)^m}{(m+1)!} \mathrm{ad}_{Z_{J}}^m[L_{\gamma, \upsilon, j}]} \leq \frac{1}{2}\norm{\mathrm{ad}_{Z_J}[L_{\gamma, \upsilon, j}]} = \frac{t}{2n}\abs{a_{\upsilon,\gamma}} \norm{[H_{\pi_\upsilon (\gamma)},L_{\gamma, \upsilon, j}]} \leq \frac{t}{n}, \label{eq:bound-adZ}
    \end{align}
    where we have used that $\norm{H_{\pi_\upsilon (\gamma)}}$, $\norm{L_{\gamma, \upsilon, j}}$ and $\abs{a_{\upsilon,\gamma}}$ are all smaller than $1$. 
    Substituting this into Eq.~\eqref{eq:derivative-V} and considering that unitary operators such as $V_{l}^{(p)}(\delta')$ and $V_J^{+}(\delta')$ do not change the $2$-norm of a vector, we can write
    \begin{align}
        \norm{\frac{d}{d\delta}V_{l,n}^{(p)}(\delta')\ket{\psi}}_2 &\leq \frac{t}{n}\norm{\sum_{\gamma, \upsilon, j} \left(V^{+}_{\gamma, \upsilon, j}\right)^\dag L_{\gamma, \upsilon, j} V^+_{\gamma, \upsilon, j}(\delta')\ket{\psi}}_2 +\frac{t}{n}\sum_{\gamma, \upsilon, j} \frac{t}{n} \\
        &\leq \frac{t}{n}\norm{\sum_{\gamma, \upsilon, j} \left(V^{+}_{\gamma, \upsilon, j}\right)^\dag L_{\gamma, \upsilon, j} V^+_{\gamma, \upsilon, j}(\delta')\ket{\psi}}_2 
        +\Upsilon_p \abs{\Theta_l} \frac{t^2}{n} \,.
        \nonumber
    \end{align}
    The result then follows by substituting this into~\eqref{deriv_vec}.
\end{proof}

A similar result can be proven also directly for the operator norm of the product formulas.

\begin{lemma}[Operator Norm Bound on Noisy Product Unitaries]
\label{seriesbound-matrix}
Let  $U^{(p)}_{l,n}$ be a product unitary according to Definition~\ref{def:product-formula}, with Trotter number $n$ and truncated to a system of size $l$. Let $V_{l,n}^{(p)}$ be a noisy version of this product unitary of the form~\eqref{eq:noisy-product-unitary-def}, that is
\begin{align}
    V^{(p)}_{l,n}(t) = \prod_{j=1}^n \prod_{\upsilon=1}^{\Upsilon_p} \prod_{\gamma\in\Theta_l} e^{-i\frac{t}{n} (a_{v,\gamma} H_{\pi_\upsilon (\gamma)}+ \delta L_{\gamma, \upsilon, j})}, 
\end{align}
with $\norm{L_{\gamma, \upsilon, j}}\leq 1$. Then
\begin{align}
     \norm{{U_{l,n}^{(p)} - V_{l,n}^{(p)}}} &\leq \frac{t}{n} \:\int_0^\delta d\delta'\,\norm{\sum_{\gamma, \upsilon, j}\left(V^+_{\gamma, \upsilon, j}(\delta')\right)^\dag L_{\gamma, \upsilon, j} V^+_{\gamma, \upsilon, j}(\delta')} 
        +\Upsilon_p \abs{\Theta_l}  \frac{t^2}{n} \delta \,,
\end{align}
where $V^+_{\gamma, \upsilon, j}(\delta)$ is the product of all the terms in the product unitary $V_{l,n}^{(p)}$ appearing to the right of the term $(\gamma, \upsilon, j)$.
\end{lemma}
\begin{proof}
We proceed in the same way as in the previous Lemma~\ref{seriesbound_vec}, by observing that
\begin{align}
    \label{bound-delta}
    \norm{{ V_{l,n}^{(p)}(\delta) - U_{l,n}^{(p)}}} \leq \int_0^\delta d\delta' \norm{\frac{d}{d\delta} V_{l,n}(\delta')}.
\end{align}
The result follows by substituting the expression~\eqref{eq:derivative-V} and using the same bounds as in the proof of the previous Lemma.
    \end{proof}
\subsection{Proof of Theorem \ref{fixed-states-ads}}
We are now ready to prove Theorem \ref{fixed-states-ads}, whose formal statement we repeat here for convenience.
\setcounter{theorem}{6}
\begin{theorem}[Restated, average case errors in digital simulators with gate-dependent perturbations]
    Consider a perturbed Suzuki-Trotter product unitary of order $p =2k$, which takes the form
    \begin{align}
        V_{l,n}^{(p)}(t) = \prod_{j=1}^n \prod_{\upsilon =1}^\Upsilon \prod_{\gamma}^{\abs{\Theta_l}} e^{i\frac{t}{n} (H_\gamma a_{\gamma, \upsilon} + \delta L_{\gamma, \upsilon, j})},
    \end{align}
    where $L_{\gamma, \upsilon, j}$ are random perturbations, drawn independently from a distribution of Hermitian operators with bounded norm $\norm{L_{\gamma, \upsilon, j}}\leq 1$ and vanishing mean $\expect{L_{\gamma, \upsilon, j}}=0$. The product unitary has Trotter number $n$ and is implemented on a system of size $l$. Assume that the initial state is a given pure state $\rho=\ket{\psi}\!\bra{\psi}$.
    
    Then, for any $\varepsilon>0$, there exists a choice of $n\geq \BOO{\frac{t^{d+2}}{\varepsilon^2}\log^d\!\left(\frac{1}{\varepsilon}\right)}$ and $l\geq vt-\frac{1}{\mu}\log\BOO{\varepsilon}$ such that the error on time evolution of a local observable $O$ is on average
    \begin{align}
        \expect{\Delta(\rho)} \leq \varepsilon\,.
    \end{align}
    Additionally, for the same choices, we have
\begin{align}
    \prob{\Delta(\rho) > s\,\varepsilon} \leq 2e^{-s^2}\,.
\end{align}
Here, $v = e \Lambda_d R^{d+1} $, $\mu = \frac{1}{R}$ as in Lemma~\ref{lem:trunc}.
\end{theorem}
\begin{proof}
We consider the total error for the given input state and divide it into three contributions, analogously to what we did in the proof of Theorem~\ref{wcds}.  Using Lemma \ref{lem:trunc} and Lemma \ref{lem:tr}, 
we then find 
\begin{align}
    \Delta(\rho) & \leq 2\norm{O}\norm{(V_{l,n}^{(p)}(t) - U_{l,n}^{(p)}(t)) \ket{\psi}}_2 + 2\norm{O}\norm{(U_{l,n}^{(p)}(t) - U_l(t))}  + \norm{(U_l^\dag(t) O U_l(t)-U^\dag(t) O U(t)) } \\ 
    \nonumber
    & \leq 2\norm{O}\norm{(V_{l,n}^{(p)}(t) - U_{l,n}^{(p)}(t)) \ket{\psi}}_2 +  \norm{O} \abs{\supp{O}} e^{-\mu l} \left( e^{\mu v t} - 1\right) + 2\norm{O} K \abs{\Theta_l} \frac{t^{p+1}}{n^p}\,,
\end{align}
for $K=2^p\left(\Lambda_d 2^d R^d\right)^{p} \; [(p-1)!]^d/(p+1)!$.
By applying Lemma~\ref{seriesbound_vec} to the first term we then have
\begin{align}
    \Delta(\rho) & \leq 2\norm{O}\frac{t}{n} \:\int_0^\delta d\delta'\,\norm{\sum_{\gamma, \upsilon, j} \left(V^{+}_{\gamma, \upsilon, j}(\delta')\right)^\dag L_{\gamma, \upsilon, j} V^+_{\gamma, \upsilon, j}(\delta')\ket{\psi}}_2   \nonumber \\ 
    &\hspace{10mm}+ 2\norm{O} \abs{\Theta_l} \left( \delta \Upsilon_p  \frac{t^2}{n} +  K \frac{t^{p+1}}{n^p}\right) +\norm{O} \abs{\supp{O}}e^{-\mu l} \left( e^{\mu v t} - 1\right)\,.\label{eq:Delta-fixed-state-acds}
\end{align}

Note here that $V^+_{\gamma, \upsilon, j}(\delta)$ is the product of all the terms in the product unitary $V_{l,n}^{(p)}$ appearing to the right of the term $(\gamma, \upsilon, j)$. Therefore we can apply Lemma \ref{randomsum} to bound the expectation value of $\Delta(\rho)$
\begin{align}
    \expect{\Delta(\rho)}&\leq 2\norm{O}\frac{t}{n} \:\int_0^\delta d\delta'\,\expect{\norm{\sum_{\gamma, \upsilon, j} \left(V^{+}_{\gamma, \upsilon, j}(\delta')\right)^\dag L_{\gamma, \upsilon, j} V^+_{\gamma, \upsilon, j}(\delta')\ket{\psi}}_2}  \nonumber\\
    &\hspace{20mm}+ 2\norm{O} \abs{\Theta_l} \left( \delta \Upsilon_p  \frac{t^2}{n} +  K \frac{t^{p+1}}{n^p}\right) +\norm{O} \abs{\supp{O}}e^{-\mu l} \left( e^{\mu v t} - 1\right) \\[4mm]
    &\leq 2\norm{O}\sqrt{2\pi\Upsilon_p \abs{\Theta_l}} \delta \frac{t}{\sqrt{n}} \nonumber\\
    &\hspace{20mm} + 2\norm{O} \abs{\Theta_l} \left( \delta \Upsilon_p  \frac{t^2}{n} +  K \frac{t^{p+1}}{n^p}\right) +\norm{O} \abs{\supp{O}}e^{-\mu l} \left( e^{\mu v t} - 1\right) \\[4mm]
    &\leq 2\norm{O} \sqrt{2\pi\Upsilon_p 2^d\Lambda_d v^d}  \delta \left(t-\frac{1}{\mu v}\log\varphi\right)^{\frac{d}{2}} \frac{t}{\sqrt{n}} \nonumber\\
    &\hspace{20mm} + 2^{d+1} \norm{O} \Lambda_d v^d\left(t-\frac{1}{\mu v}\log\varphi\right)^d\left( \delta \Upsilon_p  \frac{t^2}{n} +  K \frac{t^{p+1}}{n^p}\right) +\norm{O} \abs{\supp{O}}\varphi\,, \label{eq:expect-Delta-acds}
\end{align}
where we have also applied Lemma~\ref{localterms} and chosen $l=vt - \frac{1}{\mu }\log\varphi$, for a $\varphi$ that will be specified later.

We now observe that, by choosing a large enough $n$ and a small enough $\varphi$, this last quantity can be made arbitrarily small. In particular, for any $\varepsilon>0$, we can pick 
\begin{align}
    \varphi\leq\varepsilon, \hspace{20mm} n\geq \frac{t^{d+2}}{\varepsilon^2} \,\log^d\!\left(\frac{1}{\varepsilon}\right)
\end{align}
such that 
\begin{align}
    \expect{\Delta(\rho)}&\leq \varepsilon\left[ 2\norm{O}\sqrt{2\pi\Upsilon_p 2^d\Lambda_d v^d}+ 2^{d+1} \norm{O} \Lambda_d v^d\left(  \Upsilon_p   +  K \right) +\norm{O} \abs{\supp{O}} \right]\,,
\end{align}
where we have used that $\varepsilon\leq 1$, $\delta \leq 1$ and that, for small enough $\varepsilon$ and large enough $t$, $(1/\log (\frac{1}{\varepsilon})+1/\mu v t)^d\leq 1$. The first statement of the theorem then follows by rescaling $\varepsilon$ by an appropriate constant.

Similarly, we can observe from expression~\eqref{eq:Delta-fixed-state-acds} that if $\norm{\sum_{\gamma, \upsilon, j} \left(V^{+}_{\gamma, \upsilon, j}(\delta')\right)^\dag L_{\gamma, \upsilon, j} V^+_{\gamma, \upsilon, j}(\delta')\ket{\psi}}_2\leq s$, then 
\begin{align}
    \Delta(\rho) & \leq 2\norm{O}\frac{t}{n} \delta s  + 2\norm{O} \abs{\Theta_l} \left( \delta \Upsilon_p  \frac{t^2}{n} +  K \frac{t^{p+1}}{n^p}\right) +\norm{O} \abs{\supp{O}}e^{-\mu l} \left( e^{\mu v t} - 1\right)\,.
\end{align}
This implies the following probabilistic statement, to which we can apply Lemma~\ref{randomsum}:
\begin{align}
    &\prob{\Delta(\rho) > 2\norm{O}\frac{t}{n} \delta s  + 2\norm{O} \abs{\Theta_l} \left( \delta \Upsilon_p  \frac{t^2}{n} +  K \frac{t^{p+1}}{n^p}\right) +\norm{O} \abs{\supp{O}}e^{-\mu l} \left( e^{\mu v t} - 1\right)} \nonumber\\
    &\hspace{5mm} \leq\prob{\norm{\sum_{\gamma, \upsilon, j} \left(V^{+}_{\gamma, \upsilon, j}(\delta')\right)^\dag L_{\gamma, \upsilon, j} V^+_{\gamma, \upsilon, j}(\delta')\ket{\psi}}_2>s}\\
    &\hspace{5mm} \leq 2\exp\left(-\frac{s^2}{2n \Upsilon_p \abs{\Theta_l}}\right)\,.
    \nonumber
\end{align}
Now, by rescaling $s \rightarrow s \sqrt{2n \Upsilon_p \abs{\Theta_l}}$ and using as before $\abs{\Theta_l}\leq 2^d \Lambda_d (vt-\frac{1}{\mu}\log\varphi)^d$, we have
\begin{align}
    &\mathrm{Prob}\Bigg[ \Delta(\rho) > 2\norm{O}\sqrt{2\Upsilon_p 2^d\Lambda_d v^d}  \delta \left(t-\frac{1}{\mu v}\log\varphi\right)^{\frac{d}{2}} \frac{t}{\sqrt{n}} \; s\nonumber\\
    &\hspace{30mm} + 2^{d+1} \norm{O} \Lambda_d v^d\left(t-\frac{1}{\mu v}\log\varphi\right)^d\left( \delta \Upsilon_p  \frac{t^2}{n} +  K \frac{t^{p+1}}{n^p}\right) +\norm{O} \abs{\supp{O}}\varphi \Bigg] \leq 2e^{-s^2}
\end{align}
By making the same choices for $n$ and $\varphi$ as before and choosing $s>1$, we arrive at the second statement of the theorem. 
\end{proof}

\begin{rem}[Optimal scaling of $n$ in Theorem \ref{fixed-states-ads}]
    Note that in the previous theorem we have made some relatively loose estimates of the value of $n$ needed to achieve a given precision. A more precise analysis, taking into account the potential dependence of $n$ on $\delta$, would be the following. 
    From expression~\eqref{eq:expect-Delta-acds}, we see that, in order to suppress the terms coming from the gate noise, we need to choose
    \begin{equation}
        n\geq\BOO{\frac{\delta\,  t^{d+2} \,\log^d(\frac{1}{\varepsilon})}{\varepsilon^2}}\,.
    \end{equation}
    At the same time, to suppress the Trotter error term, we need
    \begin{equation}
        n\geq\BOO{\frac{\delta\,  t^{\frac{d+2}{p}} \,\log^{\frac{d}{p}}(\frac{1}{\varepsilon})}{\varepsilon^{\frac{1}{p}}}} \,.
    \end{equation}
    The required scaling will be given by the largest of the two quantities above. Which one will be the dominating term will depend on the relative value of $t$ and $\delta$ in the setting we are considering. 
\end{rem}

\subsection{Proof of Theorem \ref{fixed-states-2}}
\begin{theorem}[Restated, average case errors in digital simulators with constant gate perturbations]
    Consider a perturbed Suzuki-Trotter product unitary of order $p =2k$, which takes the form
    \begin{align}
        V_{l,n}^{(p)}(t) = \prod_{j=1}^n \prod_{\upsilon =1}^\Upsilon \prod_{\gamma}^{\abs{\Theta_l}} e^{i\frac{t}{n} H_\gamma a_{\gamma, \upsilon} + i\delta L_{\gamma, \upsilon, j}}\,,
    \end{align}
    where $L_{\gamma, \upsilon, j}$ are random perturbations, drawn independently from a distribution of Hermitian operators with bounded norm $\norm{L_{\gamma, \upsilon, j}}\leq 1$ and vanishing mean $\expect{L_{\gamma, \upsilon, j}}=0$. The product unitary has Trotter number $n$ and is implemented on a system of size $l$. Assume that the initial state is a given pure state $\rho=\ket{\psi}\!\bra{\psi}$.
    Then, the error on time evolution of a local observable $O$ is on average
    \begin{align}
        \expect{\Delta(\rho)} \leq \BOO{\left(1+\frac{1}{\mu v t}\log{\frac{1}{\delta^{\frac{2p}{2p+1}} t^{\frac{2}{3}(d+1)}}}\right)^d \delta^{\frac{2p}{2p+1}} t^{\frac{2}{3}(d+1)}} \leq \BOO{\log^d\left(\frac{1}{\delta^{\frac{2p}{2p+1}} t^{\frac{2}{3}(d+1)}}\right) \delta^{\frac{2p}{2p+1}} t^{\frac{2}{3}(d+1)}}\,,
    \end{align}
    if the optimal choices $n_{\rm opt}=\delta^{-\frac{2}{2p+1}}\,t^{\frac{d+4}{3}}$ and $l_{\rm opt}=vt-\frac{1}{\mu}\log\left( \delta^{\frac{2p}{2p+1}} t^{\frac{2}{3}(d+1)}\right)$ are made. Additionally, for the same choices, we have
\begin{align}
    \prob{\Delta(\rho) > \BOO{s\log^d\left(\frac{1}{\delta^{\frac{2p}{2p+1}} t^{\frac{2}{3}(d+1)}}\right) \delta^{\frac{2p}{2p+1}} t^{\frac{2}{3}(d+1)}}} \leq 2e^{-s^2}\,.
\end{align}
Here, $v = e \Lambda_d R^{d+1} $, $\mu = \frac{1}{R}$ as in Lemma~\ref{lem:trunc}.
\end{theorem}
\begin{proof}
The proof follows the same steps as the one of the previous Theorem~\ref{fixed-states-ads}, except that a slight variation of Lemma~\ref{seriesbound_vec} holds in this case. In particular, we claim that
\begin{align}
    \label{new-seriesbound}
     \norm{{U_{l,n}^{(p)}\ket{\psi} - V_{l,n}^{(p)}}\ket{\psi}}_2 &\leq \:\int_0^\delta d\delta'\,\norm{\sum_{\gamma, \upsilon, j} \left(V^{+}_{\gamma, \upsilon, j}\right)^\dag L_{\gamma, \upsilon, j}  V^+_{\gamma, \upsilon, j}(\delta')\ket{\psi}}_2\nonumber\\
     &\hspace{20mm}+ \frac{t}{2n}\int_0^\delta d\delta'\,\norm{\sum_{\gamma, \upsilon, j} \left(V^{+}_{\gamma, \upsilon, j}\right)^\dag  [H_{\pi_\upsilon(\gamma)}, L_{\gamma, \upsilon, j}]  V^+_{\gamma, \upsilon, j}(\delta')\ket{\psi}}_2 \nonumber\\
     &\hspace{40mm}+\frac{2}{3}\Upsilon_p \,\delta^2t\abs{\Theta_l} +\frac{2}{3}\Upsilon_p \,\delta\frac{ t^2}{n}\abs{\Theta_l} \,.
\end{align}
The proof of this is in essence the same as that of Lemma~\ref{seriesbound_vec}, except that we now define $Z_J=-i\frac{t}{n}a_{\upsilon,\gamma} H_{\pi_\upsilon (\gamma)} - i\delta L_{\gamma, \upsilon, j}$ and we expand the Taylor series to the second order. This implies that Eq.~\eqref{eq:derivative-V} must be adapted to 
\begin{align}
    \frac{d}{d\delta}V_{l,n}^{(p)}(\delta') &=-i\, V_{l,n}^{(p)}(\delta')\,\sum_{J} \left(V^+_J(\delta')\right)^\dag \left(L_{\gamma, \upsilon, j}-\frac{it}{2n} a_{\upsilon,\gamma} [H_{\pi_\upsilon (\gamma)},L_{\gamma, \upsilon, j}] +\mathcal{R}\right) V^+_J(\delta')\,,
\end{align}
where we can bound the remainder term as 
\begin{align}
     \norm{\mathcal{R}}=\norm{\sum_{m=2}^\infty \frac{(-1)^m}{(m+1)!}\mathrm{ad}^m_{Z_J} [L_J] } \leq \frac{1}{3!} \norm{\Big[\frac{t}{n}a_{\upsilon,\gamma} H_{\pi_\upsilon (\gamma)} +\delta L_{\gamma, \upsilon, j}, \Big[\frac{t}{n}a_{\upsilon,\gamma} H_{\pi_\upsilon (\gamma)}, L_{\gamma, \upsilon, j}\Big]\Big] } \leq \frac{2t^2}{3n^2} + \delta \frac{2t}{3n},
\end{align}
which leads to expression~\eqref{new-seriesbound}.

Note now that the first two terms in~\eqref{new-seriesbound} can both be addressed using Lemma \ref{randomsum}. Indeed, the random variables $\frac{1}{2}[H_{\pi_\upsilon(\gamma)}, L_{\gamma, \upsilon, j}]$ also have bounded norm and mean zero, similarly to $L_{\gamma, \upsilon, j}$. Thus, proceeding like in Theorem~\ref{fixed-states-ads} we find
\begin{align}
    \expect{\Delta(\rho)}&\leq 2\norm{O}\sqrt{2\pi\Upsilon_p 2^d\Lambda_d v^d}  \; \delta \left(t-\frac{1}{\mu v}\log\varphi\right)^{\frac{d}{2}} \left(\sqrt{n}+\frac{t}{\sqrt{n}}\right) \nonumber\\
    &\hspace{20mm} + 2^{d+1} \norm{O} \Lambda_d v^d\left(t-\frac{1}{\mu v}\log\varphi\right)^d\left( \frac{2}{3} \Upsilon_p \delta^2 t+\frac{2}{3} \Upsilon_p \delta \frac{t^2}{n} +  K \frac{t^{p+1}}{n^p}\right) +\norm{O} \abs{\supp{O}}\varphi\,.
\end{align}

Notice now that the resulting expression contains terms that depend on $n$ and a term that scales as $\BOO{\delta^2t^{d+1}}$ independently of $n$. We treat the former terms as we did in Theorem~\ref{wccds}, choosing the optimal scaling of $n$ to balance the various error contributions. This leads to an optimal scaling of these terms of $\BOO{\delta^{\frac{2p}{2p+1}}t^{\frac{2}{3}(d+1)}}$. More specifically, with the choice
\begin{align}
    n = \delta^{-\frac{2}{2p+1}}\,t^{\frac{d+4}{3}} ,\hspace{20mm} \varphi = \delta^{\frac{2p}{2p+1}} t^{\frac{2}{3}(d+1)}\,,
\end{align}
we find 
\begin{align}
    \expect{\Delta(\rho)} &\leq \BOO{\left(1+\frac{1}{\mu v t}\log{\frac{1}{\delta^{\frac{2p}{2p+1}} t^{\frac{2}{3}(d+1)}}}\right)^d \delta^{\frac{2p}{2p+1}} t^{\frac{2}{3}(d+1)}} +  \BOO{\left(1+\frac{1}{\mu v t}\log{\frac{1}{\delta^{\frac{2p}{2p+1}} t^{\frac{2}{3}(d+1)}}}\right)^d \delta^{2} t^{d+1}}\,.
\end{align}
Which one of these two terms above will be the dominating one will depend on the relation between $t$ and $\delta$. In particular, the first term will dominate if $\delta\leq t^{-\frac{1}{3}\frac{2p+1}{2p+2}(d+1)}$. Notice, however that in order for the bounds above to be meaningful we must in any case also assume $\delta\leq t^{-\frac{2}{3}\frac{2p+1}{2p}(d+1)}$, otherwise for large enough $t$ the bounds will be larger than $1$ which is clearly trivial. So in general it will also be true that the first term is the dominating one. We can thus conclude
\begin{align}
    \expect{\Delta(\rho)} &\leq \BOO{\left(1+\frac{1}{\mu v t}\log{\frac{1}{\delta^{\frac{2p}{2p+1}} t^{\frac{2}{3}(d+1)}}}\right)^d \delta^{\frac{2p}{2p+1}} t^{\frac{2}{3}(d+1)}} \leq \BOO{\log^d\left(\frac{1}{\delta^{\frac{2p}{2p+1}} t^{\frac{2}{3}(d+1)}}\right) \delta^{\frac{2p}{2p+1}} t^{\frac{2}{3}(d+1)}}\,.
\end{align}

Under these assumptions, by proceeding like in Theorem~\ref{fixed-states-ads}, we also find
\begin{align}
    &\mathrm{Prob}\Bigg[ \Delta(\rho) > \Big(\; 4\norm{O}\sqrt{2\Upsilon_p 2^d\Lambda_dv^d}   \; s +2^{d+1} \norm{O} \Lambda_d v^d\left( \frac{2}{3} \Upsilon_p +\frac{2}{3} \Upsilon_p  +  K\right)\nonumber\\
    &\hspace{40mm}  +\norm{O} \abs{\supp{O}} \Big)\left(1+\frac{1}{\mu v t}\log{\frac{1}{\delta^{\frac{2p}{2p+1}} t^{\frac{2}{3}(d+1)}}}\right)^d \delta^{\frac{2p}{2p+1}} t^{\frac{2}{3}(d+1)} \Bigg] \leq 2e^{-s^2}\,,
\end{align}
which by picking $s\geq 1$ leads to the second statement of the theorem.
\end{proof}

\subsection{Average scaling of the state-independent error}
Analogously to what we did in Theorem~\ref{acas} for analog simulation, we discuss here the average scaling of the quantity $\Delta$. As discussed in the main text (Section~\ref{sec:average-case}), this only provides an upper bound on the quantity $\expect{\Delta(\rho)}$ for a fixed state. The following results show that this is in fact a loose bound. We present these statements nonetheless, as they show some useful methods for bounding average values of operator norm quantities, which may be of independent interest. 

Under the same assumptions of Theorem~\ref{fixed-states-ads}, one can prove the following.
\setcounter{theorem}{10} \begin{theorem}[Average error upper-bound in digital simulators with gate-dependent perturbations] \label{acdc}
    Consider a perturbed Suzuki-Trotter product unitary of order $p =2k$, which takes the form
    \begin{align}
        V_{l,n}^{(p)}(t) = \prod_{j=1}^n \prod_{\upsilon =1}^\Upsilon \prod_{\gamma}^{\abs{\Theta_l}} e^{i\frac{t}{n} (H_\gamma a_{\gamma, \upsilon} + \delta L_{\gamma, \upsilon, j})},
    \end{align}
    where $L_{\gamma, \upsilon, j}$ are random perturbations, drawn independently from a distribution of Hermitian operators with bounded norm $\norm{L_{\gamma, \upsilon, j}}\leq 1$ and vanishing mean $\expect{L_{\gamma, \upsilon, j}}=0$. The product unitary has Trotter number $n$ and is implemented on a system of size $l$. 
    
    Then, for any $\varepsilon>0$, there exists a choice of $n\geq \BOO{\frac{t^{2d+2}}{\varepsilon^2} \,\log^{2d}\!\left(\frac{1}{\varepsilon}\right)}$ and $l\geq vt-\frac{1}{\mu}\log\BOO{\varepsilon}$ such that the error on time evolution of a local observable $O$ is on average
    \begin{align}
        \expect{\Delta} \leq \varepsilon\,.
    \end{align}
    Additionally, for the same choices, we have
\begin{align}
    \prob{\Delta > s\,\varepsilon} \leq 2e^{-s^2}\,.
\end{align}
Here, $v = e \Lambda_d R^{d+1} $, $\mu = \frac{1}{R}$ as in Lemma~\ref{lem:trunc}.
\end{theorem}
\begin{proof}
The proof follows the same steps as the one of Theorem~\ref{fixed-states-ads}. We consider the definition of $\Delta$ and divide it into three contributions, which after applying Lemma \ref{lem:trunc} and Lemma \ref{lem:tr} are
\begin{align}
    \Delta & \leq 2\norm{O}\norm{V_{l,n}^{(p)}(t) - U_{l,n}^{(p)}(t)} +  \norm{O} \abs{\supp{O}} e^{-\mu l} \left( e^{\mu v t} - 1\right) + 2\norm{O} K \abs{\Theta_l} \frac{t^{p+1}}{n^p}\,.
\end{align}
By applying Lemma~\ref{seriesbound-matrix} to the first term we then have
\begin{align}
    \Delta(\rho) & \leq 2\norm{O}\frac{t}{n} \:\int_0^\delta d\delta'\,\norm{\sum_{\gamma, \upsilon, j} \left(V^{+}_{\gamma, \upsilon, j}(\delta')\right)^\dag L_{\gamma, \upsilon, j} V^+_{\gamma, \upsilon, j}(\delta')}   \nonumber \\ 
    &\hspace{10mm}+ 2\norm{O} \abs{\Theta_l} \left( \delta \Upsilon_p  \frac{t^2}{n} +  K \frac{t^{p+1}}{n^p}\right) +\norm{O} \abs{\supp{O}}e^{-\mu l} \left( e^{\mu v t} - 1\right)\,.\label{eq:Delta-acds}
\end{align}

Noticing that $L_{\gamma, \upsilon, j}$ and $V^{+}_{\gamma, \upsilon, j}(\delta')$ satisfy the assumptions of Lemma~\ref{randomsum-matrix}, we can bound the expectation value of $\Delta$ as
\begin{align}
    \expect{\Delta}&\leq 4\norm{O}\sqrt{2^{d+1}\Lambda_d\Upsilon_p l^d\abs{\Theta_l}} \delta \frac{t}{\sqrt{n}} \nonumber\\
    &\hspace{20mm} + 2\norm{O} \abs{\Theta_l} \left( \delta \Upsilon_p  \frac{t^2}{n} +  K \frac{t^{p+1}}{n^p}\right) +\norm{O} \abs{\supp{O}}e^{-\mu l} \left( e^{\mu v t} - 1\right)\\
    &\leq2^{d+2} \norm{O} \Lambda_d v^d \sqrt{2\Upsilon_p} \delta \left(t-\frac{1}{\mu v} \log \varphi\right)^d \frac{t}{\sqrt{n}} \nonumber\\
    &\hspace{20mm} + 2^{d+1}\norm{O} \Lambda_d v^d \left(t-\frac{1}{\mu v} \log \varphi\right)^d \left( \delta \Upsilon_p  \frac{t^2}{n} +  K \frac{t^{p+1}}{n^p}\right) +\norm{O} \abs{\supp{O}}\, \varphi\,,
\end{align}
where we have also applied Lemma~\ref{localterms} and chosen $l=vt - \frac{1}{\mu }\log\varphi$. It is then clear that this quantity can be made arbitrarily small by choosing
\begin{align}
    \varphi\leq\varepsilon, \hspace{20mm} n\geq \frac{t^{2d+2}}{\varepsilon^2} \,\log^{2d}\!\left(\frac{1}{\varepsilon}\right)\,.
\end{align}

Similarly to Theorem~\ref{fixed-states-ads}, we can also apply Lemma~\ref{randomsum-matrix} to obtain the probabilistic statement
\begin{align}
    \prob{\Delta>2\norm{O} \delta \frac{t}{n} s + 2\norm{O} \abs{\Theta_l} \left( \delta \Upsilon_p  \frac{t^2}{n} +  K \frac{t^{p+1}}{n^p}\right) +\norm{O} \abs{\supp{O}}e^{-\mu l} \left( e^{\mu v t} - 1\right)}\leq 2e^{-\frac{s^2}{2n\Upsilon_p \abs{\Theta_l}}+2^d\Lambda_d l^d}\,,
\end{align}
which after the substitution $s\rightarrow \sqrt{2n\Upsilon_p\abs{\Theta_l}(2^d+\Lambda_d l^d+s^2)}$ is equivalent to (for large enough $t$ and $s$)
\begin{align}
    &\mathrm{Prob}\!\big[\Delta>2^{d+2} \norm{O} \Lambda_d v^d \sqrt{2\Upsilon_p} \delta \left(t-\frac{1}{\mu v} \log \varphi\right)^d \frac{t}{\sqrt{n}} \nonumber\\
    &\hspace{20mm} + 2^{d+1}\norm{O} \Lambda_d v^d \left(t-\frac{1}{\mu v} \log \varphi\right)^d \left( \delta \Upsilon_p  \frac{t^2}{n} +  K \frac{t^{p+1}}{n^p}\right) +\norm{O} \abs{\supp{O}}\, \varphi \big] \leq 2 e^{-s^2}\,.
\end{align}
With the same substitutions as before this shows the second statement of the theorem.
\end{proof}

Under the same assumptions as Theorem~\ref{fixed-states-2}, we find the following.

\begin{theorem}[Average error upper-bound in digital simulators with constant gate perturbations] \label{accds}
    Consider a perturbed Suzuki-Trotter product unitary of order $p =2k$, which takes the form
    \begin{align}
        V_{l,n}^{(p)}(t) = \prod_{j=1}^n \prod_{\upsilon =1}^\Upsilon \prod_{\gamma}^{\abs{\Theta_l}} e^{i\frac{t}{n} H_\gamma a_{\gamma, \upsilon} + i\delta L_{\gamma, \upsilon, j}}\,,
    \end{align}
    where $L_{\gamma, \upsilon, j}$ are random perturbations, drawn independently from a distribution of Hermitian operators with bounded norm $\norm{L_{\gamma, \upsilon, j}}\leq 1$ and vanishing mean $\expect{L_{\gamma, \upsilon, j}}=0$. The product unitary has Trotter number $n$ and is implemented on a system of size $l$. Then, the error on time evolution of a local observable $O$ is on average
    \begin{align}
        \expect{\Delta} \leq \BOO{\left(1+\frac{1}{\mu v t}\log{\frac{1}{\delta^{\frac{2p}{2p+1}} t^{d+\frac{2}{3}}}}\right)^d \delta^{\frac{2p}{2p+1}} t^{d+\frac{2}{3}}} \leq \BOO{\log^d\left(\frac{1}{\delta^{\frac{2p}{2p+1}} t^{d+\frac{2}{3}}}\right) \delta^{\frac{2p}{2p+1}} t^{d+\frac{2}{3}}}\,,
    \end{align}
    if the optimal choices $n_{\rm opt}=\delta^{-\frac{2}{2p+1}}\,t^{\frac{4}{3}}$ and $l_{\rm opt}=vt-\frac{1}{\mu}\log\left( \delta^{\frac{2p}{2p+1}} t^{d+\frac{2}{3}}\right)$ are made. Additionally, for the same choices, we have
\begin{align}
    \prob{\Delta(\rho) > \BOO{s\log^d\left(\frac{1}{\delta^{\frac{2p}{2p+1}} t^{d+\frac{2}{3}}}\right) \delta^{\frac{2p}{2p+1}} t^{d+\frac{2}{3}}}} \leq 2e^{-s^2}\,.
\end{align}
Here, $v = e \Lambda_d R^{d+1} $, $\mu = \frac{1}{R}$ as in Lemma~\ref{lem:trunc}.
\end{theorem}
\begin{proof}
   The proof follows the same steps of Theorem \ref{fixed-states-2}, except that we now use Lemmas~\ref{randomsum-matrix} and~\ref{seriesbound-matrix} (similarly to what we did in the previous Theorem~\ref{acdc}). This leads to the expression
   \begin{align}
    \expect{\Delta}&\leq 4\norm{O}\sqrt{2\Upsilon_p} 2^d\Lambda_d v^d \; \delta \left(t-\frac{1}{\mu v}\log\varphi\right)^{d} \left(\sqrt{n}+\frac{t}{\sqrt{n}}\right) \nonumber\\
    &\hspace{20mm} + 2^{d+1} \norm{O} \Lambda_d v^d\left(t-\frac{1}{\mu v}\log\varphi\right)^d\left( \frac{2}{3} \Upsilon_p \delta^2 t+\frac{2}{3} \Upsilon_p \delta \frac{t^2}{n} +  K \frac{t^{p+1}}{n^p}\right) +\norm{O} \abs{\supp{O}}\varphi\,.
\end{align}
We now make a choice of $n$ and $\varphi$ that optimally balances the scaling of the terms in this expression. More specifically, this corresponds to the choice
\begin{align}
    n = \delta^{-\frac{2}{2p+1}}\,t^{\frac{4}{3}} ,\hspace{20mm} \varphi = \delta^{\frac{2p}{2p+1}} t^{d+\frac{2}{3}}\,,
\end{align}
which leads to the scaling
\begin{align}
    \expect{\Delta(\rho)} &\leq \BOO{\left(1+\frac{1}{\mu v t}\log{\frac{1}{\delta^{\frac{2p}{2p+1}} t^{d+\frac{2}{3}}}}\right)^d \delta^{\frac{2p}{2p+1}} t^{d+\frac{2}{3}}} +  \BOO{\left(1+\frac{1}{\mu v t}\log{\frac{1}{\delta^{\frac{2p}{2p+1}} t^{\frac{2}{3}(d+1)}}}\right)^d \delta^{2} t^{d+1}}\,.
\end{align}
As before, we find that in the conditions in which these bounds are meaningful, the first term is the dominating one. We can thus conclude
\begin{align}
    \expect{\Delta(\rho)} &\leq \BOO{\left(1+\frac{1}{\mu v t}\log{\frac{1}{\delta^{\frac{2p}{2p+1}} t^{d+\frac{2}{3}}}}\right)^d \delta^{\frac{2p}{2p+1}} t^{d+\frac{2}{3}}} \leq \BOO{\log^d\left(\frac{1}{\delta^{\frac{2p}{2p+1}} t^{d+\frac{2}{3}}}\right) \delta^{\frac{2p}{2p+1}} t^{d+\frac{2}{3}}}\,.
\end{align}
A bound on the concentration of probability around this average scaling can be derived by applying Lemma~\ref{randomsum-matrix} as before.
\end{proof} 
\subsection{Brownian random walk} \label{browniantrotter}

    Let $(X_i)_{i \in I}$ be a sequence of \emph{identically and independently distributed} random variables (i.i.d.) with mean $0$ and variance $1$. Define
    \begin{align}
        S_n \coloneqq \sum_{i=1}^{n} X_i,
    \end{align}
     where $S$ is known as the random walk. Define the stochastic process
     \begin{align}
         W_n (t) \coloneqq \frac{S_{\lfloor nt\rfloor}}{\sqrt{n}} \: \: \: t \in [0,1].
     \end{align}
     Then Donsker's Theorem states that in the $n \rightarrow \infty$ limit it converges in distribution to the Wiener limit. One can generalize this to the setting we consider here to a type of discrete stochastic process which resembles the Wiener process we constructed in the continuous time case.
     A model of a Brownian circuit is presented in Ref.~\cite{chen2021concentrationotocliebrobinsonvelocity}, which is
     \begin{align}
         U^T = \prod_{j=1}^T \exp(iH_{X}^j \xi), 
     \end{align}
     where $X$ is a region of a lattice on which the Hamiltonian is supported on and $\xi$ is the 'Brownian time' defined by $\lim_{\xi \rightarrow 0} \xi^2 T = \tau$ is fixed, where $T$ is the depth of the circuit. In this model, like in the above random walk, each of the $H_X$ is chosen at random from an ensemble of Hermitian operators of mean $0$ and variance $1$. 
     
     In the assumptions of Theorem \ref{acdc}, we assumed a model of Hamiltonian perturbations which is given by $\frac{t}{n} \left( H_\gamma + \delta L_{\gamma, j}\right)$. The depth of the Trotter circuit in the case of geometrically local Hamiltonians is $D = O(n)$, the parameter $\frac{t}{n} \delta D = O(\delta t)$, which was fixed in the $n \rightarrow \infty$ limit. In Brownian time $\xi$,
     \begin{align}
        \label{browniannoise}
         V_{\gamma, j} = \exp(i\frac{t}{n} H_{(\gamma, j)} + i\delta \sqrt{\frac{t}{n}} L_{(\gamma, j)}),
     \end{align}
     where we have now used that $\xi^2 \delta D = O(\delta^2 t)$ holds. The limit is well-defined, as discussed in Ref.\ \cite{onorati_mixing_2017} and converges in distribution.  
     
     Indeed when we take $\sqrt{\frac{t}{n}} \rightarrow 0$, the Trotter error goes to zero (see Ref.~\cite{chen2021concentrationotocliebrobinsonvelocity}).
     Thus, if $H_{(\gamma, j)}$ and $L_{(\gamma, j)}$ 
     do not commute, since the Trotter is of sub-leading order, the non-commutativity is no longer a problem in the limit.  
     This noise model gives rise to a Lindblad type model, as we show here.
     
     \begin{lemma}[Brownian Limit of Trotter Circuit]
     \label{lem: brownian-trotter}
     Let $t > 0$ and let 
     \begin{align}
         \lind = i[H, \cdot] + \sum_a \delta^2 L_a^\dag (\cdot) L_a -  \frac{\delta^2}{2} \{L_a^\dag L_a, \cdot\}
     \end{align}
     be a Lindbladian super-operator, where $L_a$ are Hermitian operators (which form a basis of the local Hermitian operators). 
     Let $\ket{\psi}$ be an initial state vector and $V_l^{(p)} (t)$ the unitary as above. Define the averaged density operator $\rho(t) = \expect{V_l^{(p)} \ketbra{\psi}{\psi} V_l^{(p),\dag}}$. Then in the $n \rightarrow \infty$ limit
     \begin{align}
         \partial_t \rho(t) = \lind{\rho(t)}.
     \end{align} 
     \end{lemma}
     \begin{proof}
     The partial derivative of $t\mapsto \rho(t)$ is
     \begin{align}
         \partial_t \rho(t) = \expect{\sum_J \left[ V_J^{-} V_J \left(\int_0^1 du e^{i\mathrm{ad}_{Z_J} }\left[\frac{dZ_J}{dt}\right]\right) (V_J^{-} V_J)^\dag, \ketbra{\psi(t)}{\psi(t)}\right]}.
     \end{align}
     We can apply Taylor's theorem to
     \begin{align}
         V_J \left(\int_0^1 du e^{i\mathrm{ad}_{Z_J} }\left[\frac{dZ_J}{dt}\right]\right) V_J^\dag  = \frac{dZ_J}{dt} +  [Z_J, \frac{dZ_J}{dt}] + \cdots.
     \end{align}
     We can compute
     \begin{align}
        \label{derivative-brown}
         \expect{\frac{dZ_J}{dt}} = iH \hspace{10mm} \expect{[Z_J, \frac{dZ_J}{dt}]} = -\frac{\delta^2}{2n} \expect{\left[\sum_{a, j}  l_{a, j} L_{a} , \sum_{a, j} l_{a, j} L_{a}\right]},
     \end{align}
     where we used that $\expect{L_J} = 0$ and that we can write each $L_{\gamma, j} = \sum_{a, j} l_{a, j} L_a$, where $L_a$ is a local basis. In the $n \rightarrow \infty$ limit all higher order terms vanish in this expansion, thus we can drop them. 
     
     Since $[V_J, V_{J'}] \rightarrow 0$, $[Z_J, V_{J'}] \rightarrow 0$, and $[\frac{dZ_J}{dt}, V_{J'}] \rightarrow 0$ as $n \rightarrow \infty$, 
     \begin{align}
         \partial_t \rho(t) = \expect{\left[\left(\frac{dZ_J}{dt} + \left[Z_J, \frac{dZ_J}{dt}\right]\right), \ketbra{\psi (t)}{\psi(t)}\right]}.
     \end{align}
     Inserting Equation \eqref{derivative-brown} into this, completes the proof.
     \end{proof}
     
 \subsection{Proof of Theorem \ref{ito-acdc}}
 \setcounter{theorem}{9}
     \begin{theorem}[Restated, discrete-Ito]
         Given a noise model as described in Eq.~\eqref{browniannoise}, then
         the following holds. 
         \begin{itemize}
             \item The expected error behaves as $\expect{\Delta} \leq C \delta t^{d+\frac{1}{2}} + \BOO{\frac{1}{\sqrt{n}}}$.
             \item For a fixed input state, the $\expect{\Delta(\psi)} \leq C \delta t^{\frac{d+1}{2}} + \BOO{\frac{1}{\sqrt{n}}}$.
         \end{itemize}
     \end{theorem}
     \begin{proof}
         The proof is straightforward and follows the same steps as in Theorem \ref{fixed-states-ads} and Theorem \ref{acdc}, we thus omit here. 
     \end{proof}
     At this point, a remark is in order. 
     
     \begin{rem}[Comparison to Theorem \ref{pertlind}]
     As noted in Lemma \ref{whitenoise} and in Proposition \ref{pertlind}, if we considered the averaged state $\rho_t = \expect{\ketbra{\psi_t}{\psi_t}}$, we would find the perturbation bound
     \begin{align}
         \norm{\rho_t - \ketbra{\psi_0(t)}{\psi_0(t)}}_1 \leq \BOO{\delta t^{\frac{d+1}{2}}},
     \end{align}
     where $\ket{\psi_0(t)}$ captures the ideal state vector evolution. Since $\rho_t$ is evolved under a Lindbladian (see Lemma \ref{whitenoise}), the limit has the same scaling we already derived. 
     \end{rem}
     Given a fixed state vector $\ket{\psi}$, 
     we find that
     \begin{align}
         \expect{\Delta(\ket{\psi})} \leq 2\norm{O} \left(\frac{\delta \sqrt{t}}{\sqrt{n}}  \sqrt{n \abs{\Theta_l}} + \frac{t^{d+2}}{n} + \varepsilon \abs{\supp{O}} \right) = 2\norm{O} \left(\delta t^{\frac{d+1}{2}} + \frac{t^{d+2}}{n} + \varepsilon \abs{\supp{O}} \right) .
     \end{align}
     After choosing $n = \BOO{\frac{t^\frac{d+2}{2}}{\delta}}$ and $\varepsilon = e^{-vt}$ as 
     before, we can find that the 
     optimal accuracy in this case is
     \begin{align}
         \expect{\Delta(\ket{\psi})} \leq \BOO{\delta t^{\frac{d+1}{2}}}.
     \end{align}
\end{document}